\documentclass{jfm_ase_dec2010}

\DeclareGraphicsExtensions{.pdf,.eps,.jpg,.jpeg,.png}

\graphicspath{{Graphics/}}

\usepackage{graphicx}
\usepackage{xcolor}
\usepackage{color}
\usepackage{fancyhdr}
\usepackage{graphics}
\usepackage{amsmath}
\usepackage{amsbsy}
\usepackage{natbib}
\usepackage{amssymb}
\usepackage{mathrsfs}
\usepackage{subfig}
\usepackage{enumerate}
\usepackage{makeidx}
\usepackage{nomencl}
\usepackage{index}
\usepackage[colorlinks=true,urlcolor=black,linkcolor=blue]{hyperref}

\title{Structure, molecular dynamics, and stress in a linear polymer under dynamic strain}
\titlegraphic{\includegraphics[width=0.8\textwidth]{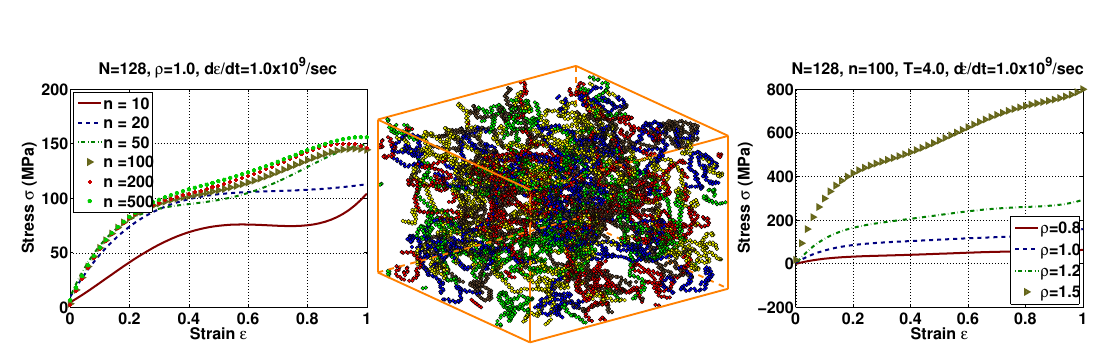}}
\crest{\includegraphics[width=4cm]{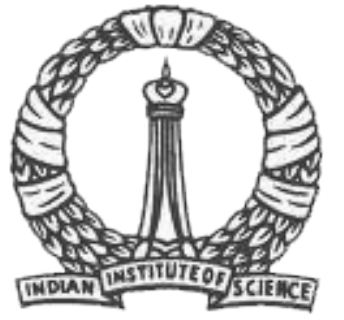}}
\author{Prashant Kumar Srivastava}
\degree{Graduate Student}
\advisor{Kartik Venkatraman}
\advdesignation{Associate Professor}
\date{\today}
\reportnumber{ASELAB-TR-2010-DEC-27}
\collegeordept{\href{http://www.aero.iisc.ernet.in}{Department of Aerospace Engineering}}
\university{\href{http://www.iisc.ernet.in}{Indian Institute of Science}}
\lab{\href{http://www.aero.iisc.ernet.in/kartik}{Aeroservoelasticity Laboratory}}
\address{Bangalore 560012, India}

\makenomenclature

\begin{document}

\maketitle

\begin{abstract}

The structural properties of a linear polymer and its evolution in time have a strong bearing on its anisotropic stress response. The mean-square bond length and mean bond angle are the critical parameters that influence the time-varying stress developed in the polymer. The bond length distribution along the chain is uniform without any abrupt changes at the ends. Among the externally set parameters such as density, temperature, strain rate, and chain length, the density as well as the chain length of the polymer have a significant effect on the stress. At high density values, changes in mean-square bond length dominates over changes in radius of gyration and end-to-end length. In other words, bond deformations dominate as opposed to changes in size and shape. Also, there is a large change in the mean-square bond length that is reflected as a jump in the stress. Beyond a particular value of the chain length, $n = 50$, called the entanglement length, stress-response is found to have distinctly different behavior which we attribute to the entanglement effects. Short chain polymers more or less behave like rigid molecules. There is no significant change in their internal structure when loaded. Further, temperature and rate of loading have a very mild effect on the stress. Besides these new results, we can now explain well known polymeric mechanical behavior under dynamic loading from the point of view of the evolution of the molecular dynamics and the derived structural properties. This could possibly lead to polymer synthesis with desired mechanical behavior.
 
\end{abstract}

\tableofcontents
\newpage
\addcontentsline{toc}{section}{List of figures}
\listoffigures
\newpage
\addcontentsline{toc}{section}{List of tables}
\listoftables
\newpage
\addcontentsline{toc}{section}{Nomenclature}
\printnomenclature

\newcommand{\Cite}[1]{\citeauthor{#1} (\citeyear{#1})}
\newcommand{\vecr}{\mathbf{r}}
\newcommand{\force}{\mathbf{f}}
\newcommand{\stress}{\boldsymbol{\tau}}
\newcommand{\PiolaT}{\mathbf{T}}
\newcommand{\PiolaS}{\mathbf{S}}
\newcommand{\Hamiltonian}{\mathscr{H}}
\newcommand{\Lagrangian}{\mathscr{L}}
\newcommand{\Defgrad}{\mathbf{F}}
\newcommand{\Kinpar}{\mathscr{A}}
\newcommand{\Force}{\mathscr{F}}
\newcommand{\Cellscale}{\mathbf{H}}
\newcommand{\Mom}{\mathbf{p}}
\newcommand{\Pos}{\mathbf{q}}
\newcommand{\refconf}{\mathbf{X}}
\newcommand{\curconf}{\mathbf{x}}
\newcommand{\trace}{\textrm{Tr}}
\newcommand{\boltz}{k_BT}
\newcommand{\etab}{\boldsymbol{\eta}}
\newcommand{\zetab}{\boldsymbol{\zeta}}
\newcommand{\Jb}{\mathbf{J}}
\newcommand{\scoor}{\mathbf{s}}
\newcommand{\dhhi}{\dot{\mathbf{H}}\mathbf{H}^{-1}}
\newcommand{\bond}{\mathbf{b}}

\pagenumbering{arabic}

\section{Introduction}

Polymers exhibit a wide variety of properties and behavior under different environmental conditions. Polymers are very large molecules made-up of smaller repeating units called monomers. Due to their flexibility in assuming random configurations they are capable of undergoing nonlinear and large elastic deformations when subjected to external loads. Also, because of the relaxation in internal structure they show viscoelastic behavior as well. Further, these materials can be doped with foreign particles to increase their strength and stiffness as well as exhibit field sensitive properties. 

At present many models exist to describe the behavior of polymers that can be broadly classified as continuum models and molecular models. In continuum models there are those that are based on primitive mechanical elements such as mechanical springs and dampers, and are used in different combination to describe the behavior of polymers \cite{Rajagopal}.  Other continuum models are based on the free energy of the system, where an arbitrary form of the free energy is chosen to fit the experimental data, and the behavior of the polymer is described in terms of this functional form of the free energy using the laws of thermodynamics \cite{Thien, Kankanala, Brigadnov}. The molecular models of polymers are too many. A few that come to mind are the freely jointed chain model, phantom model, Rouse model, Zimm model, reptation model, network model, tube model, work-like chain model, etc \cite{Doi}. None of these models describe the complete behavior of polymers in general but are only applicable in some limiting cases because of the simplifying assumptions that go into the model framework. Further, many of these models suffer from assumptions that could be physically unjustifiable. For instance, the phantom network model does not have any constraints on bonds crossing each other and is not physically plausible. Many of the advanced statistical mechanics models have been used to describe the behavior of polymers \cite{James1947, Boggs, Freed}. Studies have been performed using some of these models and their ranges of validity established \cite{Gao1992c, Rickayzen, Kremer1988}. 

Molecular dynamic (MD) and Monte Carlo simulations too have helped in understanding the deformation behavior of polymers.  Gao and Weiner \cite{Gao1984, Gao1987a, Gao1989b, Gao1994a} used a freely jointed monomer chain model of the polymer and came to the conclusion that stress anisotropy is primarily an effect of excluded volume interactions. The covalent bond force in freely-jointed models of a polymer network contributes predominantly to pressure. As the volume fraction increases, contribution of covalent bond interactions to the stress anisotropy decreases \cite{Gao1989a, Gao1991a, Gao1991b, Weiner1994}.

\citet{Bower2004} studied the role of pressure on the mechanical behavior of polymers. They show using molecular simulation on an idealized elastomer, a relation between the difference stress and pressure. Difference stress is that which resists extension or stretch and pressure resists volume change. If the scaled density is less than one the difference stress is purely entropic. When the normalized density exceeds this value, then at lower temperatures difference stress contains an additional energy component. The role of monomer packing fraction on rubber elasticity was studied by \citet{Bower2006}. They demonstrate that non-bonded interactions play a central role in controlling the mechanical response of rubber-like solids even under standard conditions. In their simulation, glass transition does not occur for scaled density less than 1, while at a scaled density value 1.2, they observe glass transition. They find that stress levels were considerably higher for higher densities as compared to lower densities. Finally, they conclude that short range excluded volume interaction is an active deformation mechanism contributing to difference stress whereas long range attractive interactions contribute only to mean stress or pressure. 

\citet{Weiner1989} introduced the concept of intrinsic chain stress that generalizes the concept of axial entropic force for the case of inter- and intra-chain non-covalent interactions for dense systems. They show that the macroscopic stress tensor in a network can be expressed as a sum of individual chain stress contributions. They also find that the interchain non-covalent contribution to stress is non-hydrostatic in the deformed network and varies with deformation due to variation in chain orientation. They again use the concept of intrinsic stress to study the Mooney effect in swollen networks where they show that due to non-hydrostatic contribution to stress, because of interchain non-covalent interaction, model polymer exhibits strain softening or Mooney effect characteristic of dry rubber solids \cite{Weiner1990}. 

Molecular dynamics simulation of polymer entanglement was performed by \citet{kremer1990}. They observed that the Rouse model provides an excellent description of short chain polymers while the dynamics of long chains can be described by the reptation model. They map the polymer to the experimental polymeric liquids and determine the entanglement length uniquely. They analyze the motion of the primitive chain to show the possibility of visualizing the confinement of motion of the chain within the tube. 

Stress anisotropy due to excluded volume interaction is quite well established in literature on polymeric stress-strain behavior. the previous works as mentioned earlier. In the present work, we address the effect of internal structure on stress anisotropy in a polymer melt under dynamic strain. A generic polymeric system is modeled as an NVT ensemble. We study the stress-strain response of the polymeric system in combination with various micro-structural parameters, namely, mean-square bond length, mean bond angle, mean-square end-to-end length, mean-square radius of gyration, mass-ratios, mean chain angle. The mean-square bond length and mean bond angle have significant effect on the stress anisotropy. Besides, bond deformation and conformational change becomes more significant when the chains are strongly aligned in the loading direction. With reference to external parameters such as density, temperature, chain length and rate of loading, we observe that the stress levels are highly sensitive to density and chain length. In contrast, the temperature and loading rate are only mildly influential. These influences are correlated in terms of the micro-structure changes in the polymeric system.

\section{Mathematical model}
\label{sec:model}

We consider a system consisting of $N$ polymer chains each having $n$ united atoms. The extended variables of the system are $\{\scoor_{ij},\, s\}$, where, $\scoor_{ij}$ is the scaled position vector of the atom $j$ in chain $i$ with respect to simulation cell dimension. $s$ is an extended dynamical variable introduced in the Lagrangian to control the temperature \cite{Nose}.

\nomenclature[A]{$N$}{number of chains}
\nomenclature[A]{$n$}{number of atoms in a chain}
\nomenclature[A]{$\mathbf{s}_{ij}$}{scaled coordinate of atom $j$ in chain $i$}
\nomenclature[A]{$s$}{dynamical variable to control temperature}

\begin{equation}
\vecr_{ij} = \Cellscale\scoor_{ij},
\end{equation}

\nomenclature[A]{$\mathbf{r}_{ij}$}{position vector of atom $j$ in chain $i$}

where

\[
\Cellscale= \begin{bmatrix}
L_1 & 0 & 0 \\
0 & L_2 & 0 \\
0 & 0 & L_3 
\end{bmatrix}.
\]

Alternately

\begin{equation}
r_{ijl} = L_l s_{ijl},\ l = 1,2,3\ (\textrm{no sum on}\ l).
\end{equation}

\nomenclature[A]{$L_l$}{cell dimension in the direction $\mathbf{e}_l$, $l = 1, 2, 3.$}
\nomenclature[S]{$i$}{index of polymer chain, $i = 1, 2, \ldots N$}
\nomenclature[S]{$j$}{index of atom in a polymer chain, $j = 1, 2, \ldots n$}
\nomenclature[S]{$l$}{coordinate directions, $l = 1, 2, 3$}

Here $l = 1,\,2,\,3$ refers to coordinate directions $e_1,\,e_2,\,e_3$ respectively. $\vecr_{ij}$ is the position of united atom $j$ in chain $i$. Thus, $s_{ij1},\, s_{ij2}\ \textrm{and}\ s_{ij3} \in [0,1)$, and the dimensions of the cell in the directions $e_1$, $e_2$ and $e_3$ are $L_1$, $L_2$ and $L_3$,  respectively. The dynamics of the system is controlled by the Lagrangian

\begin{equation}
\Lagrangian = \sum_{i=1}^N\sum_{j=1}^n\frac{1}{2}m\dot\vecr_{ij}^2-U(\vecr_{ij}) + \frac{w_s}{2}\frac{\dot s^2}{s^2} - f\boltz\ln s,
\end{equation}

which, when written in terms of scaled coordinates and scaled time, reads

\begin{equation}
\label{eq:Lag}
\Lagrangian = \sum_{i = 1}^N\sum_{j = 1}^n\frac{1}{2}ms^2\left(\scoor'_{ij}\cdot\Cellscale^t\Cellscale\scoor'_{ij}\right)-U(\Cellscale\scoor_{ij}) + \frac{w_s}{2}s'^2 - f\boltz\ln s.
\end{equation}

\nomenclature[X]{$\mathscr{L}$}{Lagrangian}
\nomenclature[A]{$m_{ij}$}{mass of the atom $j$ in chain $i$, In this case $m_{ij} = m$}
\nomenclature[A]{$U$}{potential energy}
\nomenclature[A]{$T$}{temperature}
\nomenclature[A]{$k_B$}{Boltzmann's constant}
\nomenclature[A]{$f$}{number of degrees of freedom}
\nomenclature[G]{$\tau$}{scaled time}
\nomenclature[A]{$t$}{real time}

In the above relations, $f$ denotes the number of degree of freedom of the system. $\dot{(\,)}$ and $(\,)'$ refers to the derivative with respect to the real time $t$ and scaled time $\tau\ (d\tau = dt/s)$ respectively. The momentum of the particle in terms of the Lagrangian is given by

\begin{equation}
\boldsymbol{\Pi}_{ij} = \frac{\partial \mathscr{L}}{\partial \mathbf{s}'_{ij}} = ms^2\mathbf{G}\scoor'_{ij},
\end{equation}

\nomenclature[G]{$\boldsymbol{\Pi}_{ij}$}{scaled momenta of atom $j$ in chain $i$}
\nomenclature[A]{$\mathbf{H}$}{matrix consisting of edge vectors of the simulation cell}
\nomenclature[A]{$\mathbf{G}$}{$\mathbf{H^tH}$}

where $\mathbf{G} = \Cellscale^t\Cellscale$ and $\boldsymbol{\Pi}_{ij}$ denotes the scaled momenta.

The equation of motion of a particle in a system whose Lagrangian is $\Lagrangian$, is given by

\begin{equation}
\frac{d}{dt}\frac{\partial \Lagrangian}{\partial \dot{\Pos}_i}-\frac{\partial \Lagrangian}{\partial \Pos_i} = 0.
\end{equation}

Therefore, the equations of motion of the particles in the scaled coordinates can be written as

\begin{subequations}
\setlength{\jot}{3pt}
\begin{equation}
\label{eq:extend}
\scoor''_{ij}  =  \displaystyle\frac{1}{ms^2}\Cellscale^{-1}\force_{ij}-2\frac{s'}{s}\scoor'_{ij}-\mathbf{G}^{-1}\mathbf{G}'\scoor'_{ij},
\end{equation}

and 

\begin{equation}
\label{eq:extends}
w_ss'' =  \displaystyle s\left[\sum_{i = 1}^N\sum_{j = 1}^nm\scoor'_{ij}\cdot\mathbf{G}\scoor'_{ij}-\frac{f\boltz_{ext}}{s^2}\right]
\end{equation}
\end{subequations}

\nomenclature[A]{$w_s$}{inertia parameter corresponding to the extended variable $s$}

The equations of motion in real variables and real time now take the form

\begin{subequations}
\setlength{\jot}{3pt}
\begin{equation}
\label{eq:motion}
\ddot\vecr_{ij}  = \displaystyle\frac{\force_{ij}}{m} - \frac{\dot s}{s}\frac{\Mom_{ij}}{m} + \ddot\Cellscale\Cellscale^{-1}\vecr_{ij} + \{(\dot\Cellscale\Cellscale^{-1})^t-\dot\Cellscale\Cellscale^{-1}\}(\dot\vecr_{ij}-\dot\Cellscale\Cellscale^{-1}\vecr_{ij}),
\end{equation}

\begin{equation}
\label{eq:motions}
\ddot{s}  = \displaystyle \frac{{\dot{s}}^2}{s}+\frac{s}{w_s}\left[\sum_{i=1}^N\sum_{j=1}^n\frac{\Mom_{ij}^2}{m}-f\boltz_{ext}\right].
\end{equation}
\end{subequations}

\nomenclature[A]{$\mathbf{f}_{ij}$}{force on united atom}
\nomenclature[A]{$T_{ext}$}{external temperature imposed on the system}
\nomenclature[X]{$\dot{(\,)}$}{derivative with respect to real time $t$}
\nomenclature[X]{$(\,)'$}{derivative with respect to scaled time $\tau$}
\nomenclature[A]{$\mathbf{p}_{ij}$}{momentum of the united atom $j$ in chain $i$ in real coordinates}
\nomenclature[G]{$\lambda$}{stretch ratio}

The momentum, $\Mom_{ij}$, of the atom $j$ in chain $i$ in term of real coordinates and real time is given by

\begin{equation}
\Mom_{ij} = \frac{\Cellscale^{-t}\boldsymbol{\Pi}_{ij}}{s} = m(\dot\vecr_{ij} - \dot\Cellscale\Cellscale^{-1}\vecr_{ij}).
\end{equation}

The expression for the particle momentum in this ensemble is not the same as that in the NVE ensemble. There are some extra terms in the above expression because of variation in the cell dimensions. In the case when the system is held at a constant strain, its Hamiltonian remains conserved throughout the simulation. But when the system is subject to a time-varying strain, this is no longer valid. This accounts for the extra terms. While applying the strain on the system, all atom positions are scaled in the same proportion as that of the cell dimensions. Here we assume the material to be incompressible. So, if stretch ratio in the $x$ direction is $\lambda$, then the stretch ratio in $y$ and $z$ direction would be $\frac{1}{\sqrt{\lambda}}$ because of symmetry. 

All the physical properties are defined in terms of momentum of the atoms thus defined. For example the kinetic energy is now defined as $\Mom^2/2m$ instead of $\frac{1}{2}m\dot\vecr^2$.

The calculation of stress is done by using the \emph{virial theorem} \cite{Greiner} of statistical mechanics. For a system of atoms interacting with two body potentials, the expression for the stress is given by

\begin{equation}
\boldsymbol{\tau} = -\frac{1}{V}\left[(Nk_BT)\mathbf{I}-\left<\sum_{\alpha>\beta}\frac{1}{|\mathbf{r}|_{\alpha \beta}}\frac{\partial U}{\partial |\mathbf{r}|_{\alpha\beta}}\mathbf{r}_{\alpha\beta}\otimes\mathbf{r}_{\alpha\beta}\right>\right]; \ \alpha ,\beta = 1, 2, \cdots, Nn,
\end{equation}

\nomenclature[G]{$\boldsymbol{\tau}$}{virial stress tensor}
\nomenclature[A]{$\mathbf{I}$}{identity tensor}
\nomenclature[A]{$V$}{volume of the simulation cell}

where $\boldsymbol{\tau}$ is the virial stress tensor, $k_B$ is the Boltzmann constant, $T$ is the temperature, $\mathbf{I}$ is unit tensor and $V$ is the current volume of the simulation cell. Various other parameters representing internal degrees of freedom and overall shape of the polymer chain considered in this study are briefly defined below.

The mean-square end-to-end length of the polymer chains give information on the compact or open configuration acquired by chains on an average. It is defined as

\begin{equation}
 \left<R^2\right> = \left<|\mathbf{r}_n-\mathbf{r}_1|^2\right>,
\end{equation}

\nomenclature[A]{$R$}{end to end length of the polymer chain}

where $\mathbf{r}_n$ and $\mathbf{r}_1$ are the position vectors of the last and first monomer or united atom, respectively, in a chain. 

The radius of gyration of a polymer chain denotes the entire mass distribution of the chain and plays a central role in interpreting light scattering and viscosity measurements. Radius of gyration for a polymer chain of length $n$ is given by

\begin{equation}
\left<R_g^2\right> = \frac{1}{n}\left<\sum_{i=1}^n|\mathbf{r}_i-\bar{\mathbf{r}}|^2\right>
\end{equation}

\nomenclature[A]{$R_G$}{radius of gyration of the polymer chain}

In any configuration of the chain, the mean spatial distribution of the chain mass need not be spherical. The moment of inertia of the monomer mass distribution in a sense signifies the shape of the chain. Elements of the tensor describing the mass distribution have the form

\begin{equation}
 G_{xy} = \frac{1}{n}\sum_{i=1}^n(r_{ix}-\bar{r}_x)(r_{iy}-\bar{r}_y).
\end{equation}

\nomenclature[A]{$G_{xy}$}{element of mass distribution tensor}
\nomenclature[A]{$g_k$}{eigen values of the mass distribution tensor, $k=1,\,2,\,3$}

Eigenvectors of the above tensor give the principle directions in which mass is distributed and the corresponding eigenvalues are a measure of the distribution. If the eigenvalues are arranged in a decreasing order, the two mass ratios of interest are the ratios of the second and third eigenvalues with respect to the first eigenvalue. These ratios give the relative measure of the mass distribution in the two transverse directions to the strain loading direction.

Chain angle is the angle which an end-to-end vector of the chain makes with the direction of the loading. This is also a measure of alignment of the chains in the loading direction. It is given by

\begin{equation}
\psi = \cos^{-1} \left(\frac{\mathbf{R}\cdot\mathbf{e}_l}{|\mathbf{R}|}\right)=\cos^{-1} \left(\frac{R_l}{|\mathbf{R}|}\right),
\end{equation}

\nomenclature[G]{$\psi$}{chain angle, angle between end to end vector of chain and loading direction}

where, $\mathbf{e}_l$ is the loading direction and  $l=1,\,2,\,3$.

In MD simulations, the equations of motion, Equation \ref{eq:motion}, need to be integrated to determine the new positions of the united atoms. The integration method that is quite popular in use is the Verlet algorithm \cite{Rapaport}. However, in this simulation since we are constraining the temperature, the accelerations are no longer only functions of position but also functions of momenta. This imposes considerable difficulty in integrating the equations of motion. A variant of the Verlet algorithm known as the velocity Verlet algorithm is used in this case since we need the current particle velocities too. This algorithm is very similar to the Verlet algorithm except for the fact that velocity rather than position at the previous time step is used to integrate the equations of motion. Due to application of temperature control, equations of motion of the particles depend on the current value of $s$ and its first derivative. Again, the second derivative of $s$ depends on the current particle momenta. This makes the equations of motion interdependent. This prevents the direct application of the velocity Verlet algorithm for integrating the equations of motion.  \citet{Fox} gave a procedure to modify the velocity Verlet algorithm to overcome this difficulty without affecting the accuracy of the method. In this procedure, $s$ and its derivatives are approximated by using the position and momenta at the previous time step. These approximate values are then used to integrate the equation of motion of the particles. The new values of the position and momenta thus obtained are used to make better approximation of $s$ and its derivative which are used in the next time step. We also use the same method to integrate the equations of motion.

We outline the principle numerical integration steps in this procedure.

 \begin{subequations}
\begin{equation}
 \label{eq:int1}
 x(t+dt) = x(t) + dt\dot{x}(t)+\frac{1}{2}(dt)^2\ddot{x}(t),
\end{equation}
\begin{equation}
 \label{eq:int2}
 \dot{x}\left(t + \frac{1}{2}dt\right) = \dot{x}(t)+\frac{1}{2}dt\ddot{x}(t),
\end{equation}
\begin{equation}
 \label{eq:int3}
 \dot{x}(t+dt) = \dot{x}\left(t+\frac{1}{2}dt\right)+\frac{1}{2}dt\ddot{x}(t+dt).
\end{equation}
\end{subequations}

Now the equations of motion are integrated in the following manner.

\begin{enumerate}[(1) ]

\item  \label{integration1} Assume that the values $\vecr(t), \dot\vecr(t),\, \ddot\vecr(t)$ and $s(t)$ are known at time $t$. With these values,  evaluate $\vecr(t+dt)$ and $s(t+dt)$ using Equation~(\ref{eq:int1}).

\item The velocities $\dot{\vecr}(t+\frac{1}{2}dt)$ and $\dot{s}(t+\frac{1}{2}dt)$ are next determined in the next half time-step using Equation~(\ref{eq:int2}).

\item Using Equation~(\ref{eq:motions}), determine $\ddot{s}(t+dt)$ using the $\dot{s}(t+\frac{dt}{2}),\, s(t+dt)$ and momenta $\mathbf{p}_{ij}(t)$. Since we are using the value of the temperature control variable $s$, its derivative and linear momenta from the previous time-steps to obtain $\ddot{s}(t+dt)$, we refer to this value as $\ddot{s}_{app}(t+dt)$.  This value in substituted in Equation (\ref{eq:int3}) to get the approximate value $\dot{s}_{app}(t+dt)$.

\item Since the position of the particles is known at step $t+dt$, the force on each particle can be calculated from the gradient of the potential energy at this instant.

\item \label{integration5} From Equation (\ref{eq:motion}) we observe that acceleration at time $t+dt$ is a function of current particle velocity. Further, from Equation (\ref{eq:int3}), particle velocity at time $t+dt$ is dependent on the acceleration at the current time step. This implies that acceleration and momenta are interdependent. Substituting the expression of accelerations from Equation (\ref{eq:motion}) into Equation (\ref{eq:int3}) we get

\begin{equation}
\begin{split}
\dot\vecr_{ij}(t+dt)  &= \left[\left(1 + \frac{dt}{2}\frac{\dot s}{s}\right)\mathbf{I}+\frac{dt}{2}\left(\dhhi-(\dhhi)^t\right)\right]^{-1}\left\{\dot\vecr_{ij}\left(t + \frac{dt}{2}\right) + \frac{dt}{2}\frac{\force_{ij}}{m} \right. \\
& \left. +  \frac{dt}{2}\left[\ddot\Cellscale\Cellscale^{-1} + \frac{\dot s}{s}\dhhi+\left(\dhhi-(\dhhi)^t\right)\dhhi\right]\vecr_{ij}\right\}.
\end{split}
\end{equation}

Substituting this expression for the velocity at the current time in Equation (\ref{eq:motion}), the acceleration at the current time step is obtained as

\begin{equation}
\begin{split}
\ddot\vecr_{ij}(t+dt) & = \left[\left(1 + \frac{dt}{2}\frac{\dot s}{s}\right)\mathbf{I}+\frac{dt}{2}\left(\dhhi-(\dhhi)^t\right)\right]^{-1} \\
& \left[\frac{\force_{ij}}{m} + \left\{\ddot\Cellscale\Cellscale^{-1}+\frac{\dot s}{s}\dhhi+\left(\dhhi-(\dhhi)^t\right)\dhhi\right\}\vecr_{ij} \right. \\
&\left. -\left\{\frac{\dot s}{s}\mathbf{I}+\left(\dhhi-(\dhhi)^t\right)\right\}\dot\vecr_{ij}\left(t + \frac{dt}{2}\right)\right].
\end{split}
\end{equation}

When the deformation is symmetric, $\dhhi-(\dhhi)^t$ vanishes. In this case, the above expressions change to

\begin{equation}
\dot\vecr_{ij}(t+dt) = \left[1 + \frac{dt}{2}\frac{\dot s}{s}\right]^{-1}\left[\dot\vecr_{ij}\left(t + \frac{dt}{2}\right) + \frac{dt}{2}\frac{\force_{ij}}{m} +
\frac{dt}{2}\left\{\ddot\Cellscale\Cellscale^{-1} + \frac{\dot s}{s}\dhhi\right\}\vecr_{ij}\right],
\end{equation}

\begin{equation}
\ddot\vecr_{ij}(t + dt) = \left[1 + \frac{dt}{2}\frac{\dot s}{s}\right]^{-1}\left[\frac{\force_{ij}}{m} + \left\{\ddot\Cellscale\Cellscale^{-1} +\frac{\dot s}{s}\dhhi\right\}\vecr_{ij} - \frac{\dot s}{s}\dot\vecr_{ij}\left(t + \frac{dt}{2}\right)\right].
\end{equation}

\item More accurate value of $\ddot{s}(t+dt)$ and $\dot{s}(t+dt)$ is obtained using the new values of position, momenta and acceleration. These values are again used in step (\ref{integration5}) to yield more accurate position, momenta and acceleration of the particles. These iterations are terminated once the values converge.

\item Finally, we increment the time-step and go back to step (\ref{integration1}).

\end{enumerate}

\section{Simulation}
\label{sec:sim}

\begin{figure}
\begin{center}
\includegraphics[scale=0.6]{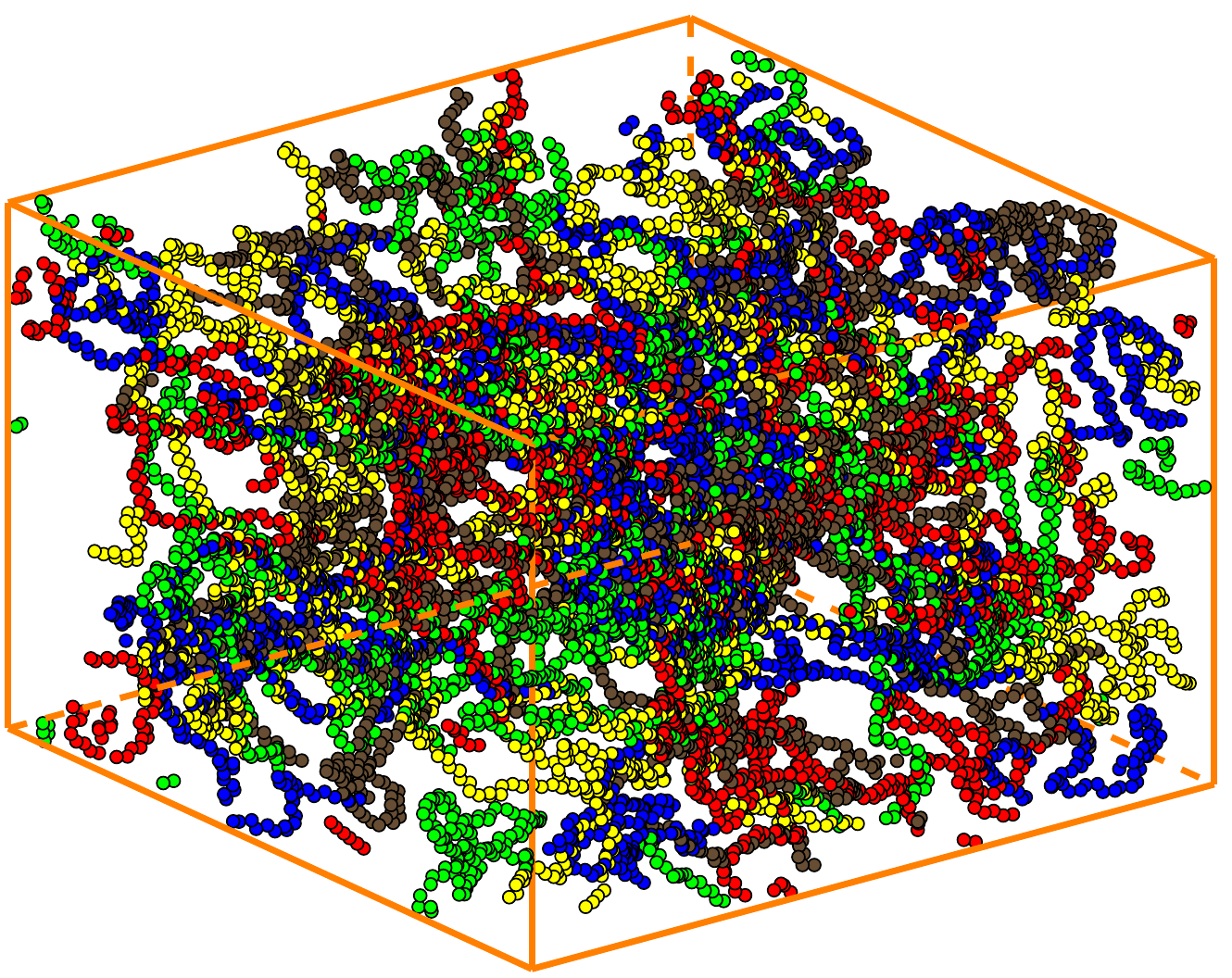}
\caption{Initial configuration}
\label{fig:init}
\end{center}
\end{figure}

Molecular dynamics simulation is performed on a linear polymeric system made up of polymer chains. The starting point of the simulation is to generate the initial equilibrated melt of the polymer chains. The cell is divided into a grid of uniformly spaced points and the center of mass of each polymer chain is placed at these grid points.  Chain growth takes place by generating a series of random bond vectors.  A random bond vector in three dimensions is generated by using a random dihedral angle between $0$ to $360^\circ$ while keeping the bond angle and bond length of the chain at their equilibrium values. This helps in generating an initial structure of the polymeric system very close to the one corresponding to the minimum energy configuration. Position of the new monomer is obtained by adding these bond vectors to the previously generated sites. While chain growth takes place, a check is performed for the occupancy of the newly generated site. In other words, if the new generated position for the monomer is already occupied we go back one step and generate a new site position. A partial overlap of the monomers is allowed during the chain growth which is later removed by subjecting the polymeric system to a soft repulsive potential to ensure that all the monomers are separated by a distance confirming the absence of overlap. \ref{fig:init} shows the initial structure of the polymer system generated by the above described method.

Once the initial structure is ready, we perform the molecular dynamic simulation by subjecting the system to the forces and moments due to bonded and non-bonded potentials which describe the inter-atomic interaction. To model the van dar Walls forces, we use the shifted and truncated Lennard-Jones potential which acts between all pairs of the united atoms 

\begin{equation}
U_{LJ}(r) = \begin{cases}
4\epsilon_{LJ}\left[\left(\frac{\sigma}{r}\right)^{12}-\left(\frac{\sigma}{r}\right)^6-\left(\frac{\sigma}{r_c}\right)^{12}+\left(\frac{\sigma}{r_c}\right)^6\right], & \text{if} \quad r < r_c = 2.5\sigma \\
0, & \text{otherwise}.
\end{cases}
\end{equation} 

\nomenclature[G]{$\sigma$}{length parameter in Lennard-Jones potential}
\nomenclature[G]{$\epsilon$}{energy parameter in Lennard-Jones potential}
\nomenclature[A]{$r_c$}{cut-off distance for short range potentials}

Here $\sigma = 0.398$~nm is the Lennard-Jones length parameter and $\epsilon = 0.477$~kJ/mol is the Lennard-Jones energy parameter that is related to the well-depth of the Lennard-Jones potential. Other kinds of bonded interaction used are bond stretching, bond bending and bond torsional potentials. 

\begin{figure}
\centering
\includegraphics[scale = 0.8]{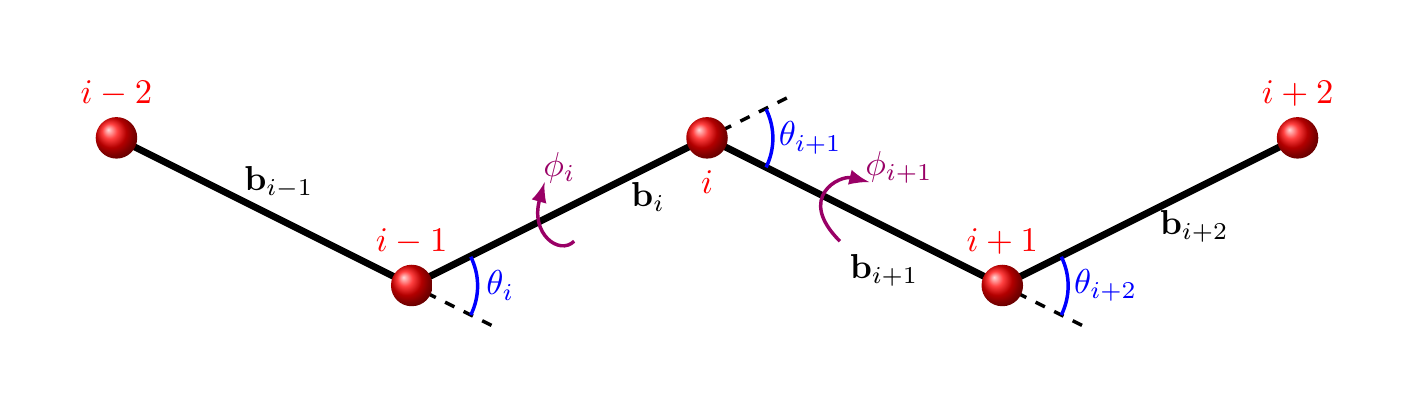}
\caption{Chain structure}
\label{fig:chain}
\end{figure}

Stretching of the bond is modeled through the finitely extended nonlinear elastic (FENE) potential

\begin{equation}
U_{FENE}(r) = -\frac{1}{2}KR_0^2\ln\left[1-\left(\frac{r}{R_0}\right)^2\right].
\end{equation}

\nomenclature[A]{$K$}{stiffness parameter in FENE potential}
\nomenclature[A]{$R_0$}{maximum allowable bond length}

Here $K = 30\epsilon/\sigma^2$ is the stiffness of the FENE potential.  $R_0 =1.5\sigma$. This potential acts between the monomers which are directly bonded to each other in a chain.

The bond bending potential is a function of the bond angle $\theta$ such that

\begin{equation}
\label{eq:bbend}
U_{bending}(\theta) = \frac{1}{2}K_{\theta}(\cos\theta-\cos\theta_0)^2,
\end{equation}

\nomenclature[A]{$K_{\theta}$}{bond bending stiffness}
\nomenclature[G]{$\theta_0$}{mean bond angle}
\nomenclature[G]{$\theta$}{bond angle}

where $K_\theta = 520$~kJ/mol is the bond bending stiffness and the equilibrium bond bending angle $\theta_0 = 110^\circ$. The bond angle $\theta$ is defined as the angle between the two contiguous covalent bonds

\[
\cos\theta_i = \frac{\bond_{i-1}\cdot\bond_i}{|\bond_{i-1}||\bond_i|},
\]

where $\bond_i = \vecr_i - \vecr_{i-1}$ is the bond vector between the atoms $i-1$ and $i$.

\nomenclature[A]{$\mathbf{b}_i$}{Bond vector between the united atoms $i-1$ and $i$}

The torsional potential is a four body potential which is a harmonic function of the dihedral angle $\phi$. That is

\begin{equation}
U_{torsion}(\phi) = \displaystyle\sum_{l=0}^{3} K_{\phi l} \cos^{l}\phi,
\end{equation}

\nomenclature[A]{$K_{\phi l}$}{bond torsion stiffness parameters, $l = 0,\ 1,\ 2,\ 3$}
\nomenclature[G]{$\phi$}{dihedral angle} 

where $K_{\phi 0} = 14.477\ \mathrm{kJ/mol},\ K_{\phi 1} = -37.594\ \mathrm{kJ/mol},\ K_{\phi 2} = 6.493\ \mathrm{kJ/mol}\ \mathrm{and}\ K_{\phi 3} = 58.499\ \mathrm{kJ/mol}$. The dihedral angle is defined as the angle between the plane formed by first two bond vectors and plane formed by the last two bond vectors from the set of three consecutive bonds

\[
\cos\phi_i = \frac{(\bond_{i-1}\times\bond_i)\cdot(\bond_i\times\bond_{i+1})}{|\bond_{i-1}\times\bond_i||\bond_i\times\bond_{i+1}|}.
\]

During the simulation, the following non-dimensional parameters are used which are described below.

\begin{table}
\setlength{\itemsep}{0cm}%
\setlength{\parskip}{0.5cm}%
\begin{center}
\begin{tabular}{ll}
\hline
time & $\bar{t}$ = $\displaystyle\frac{t}{\sigma}\left(\frac{\epsilon}{m}\right)^\frac{1}{2}$ \\[0.5cm]
number density & $\bar\rho$ = $\displaystyle\sigma^3\rho$ \\[0.5cm]
pressure & $\bar{p}$ = $\displaystyle\frac{\sigma^3p}{\epsilon}$ \\[0.5cm]
temperature & $\bar{T}$ = $\displaystyle\frac{k_BT}{\epsilon}$ \\
\hline
\end{tabular}
\end{center}
\caption{Scaled parameters}
\label{tb:ndp}
\end{table}

In Table~\ref{tb:ndp}, $m$ is the mass of the united atom---in this case $14\ \mathrm{amu}$. $\sigma$ and $\epsilon$ denote length and energy scale parameters, respectively, and their values are the same as for the Lennard-Jones potential. From here onwards, we drop the bar used for scaled symbols, and symbols without bar should be considered as scaled parameters. In order to study variations and their effect on the system response, we choose a reference system with $N=128$ polymer chains, each chain having $n = 100$ united atoms. The temperature of this reference system is maintained at $T = 4.0$ in a thermal bath. The simulation cell has a volume consistent with a number density $\rho = 1.0$. This reference model was strained at a strain rate of $\dot\epsilon = 1.0\times 10^9/\mathrm{sec}$. In the present simulation we use a time-step of $0.001$ for the numerical integration of equations of motion which is sufficient for the stability of the integration scheme and maintaining the accuracy of the same. In order to simulate the constant strain rate uniaxial deformation, the dimension of the simulation cell in the $\mathbf{e}_1$ direction - $L_1$ - was increased at a fixed rate. We assume that the polymer is incompressible and apply the appropriate contraction in the transverse direction of the simulation cell. The polymeric system considered by us is free of cross-linkages between chains. The position of the atoms are also scaled in the same proportion as the dimension of the simulation cell. The average of the property under consideration is taken over $200$ steps of the simulation.

\section{Stress-strain and micro-structure}
\label{sec:stress_strain_resp}

First we subject the polymer system to a constant rate loading and subsequent unloading at the same rate - a triangular loading - to validate our MD code with the results of \citet{Bergstrom}. The results obtained from our MD simulation show similar qualitative trend with those presented in Bergstrom and Boyce's study. Quantitative difference in the results are due to difference in the choice of model parameters, such as spring constants and mean bond-angles of the bond potentials, used in the respective simulations.

The stress response of the polymer is correlated with its micro-structural properties such as bond length, bond angle, dihedral angle, mean-square bond length, mean-square end-to-end length, and radius of gyration distributions. The effect of the constant strain-rate loading on the mass ratio, mean bond angle and mean chain angle is also discussed.

\begin{figure}
\centering
\subfloat[Stress-strain curve]{\label{fig:fit_stress_strain}\includegraphics[width=0.5\textwidth]{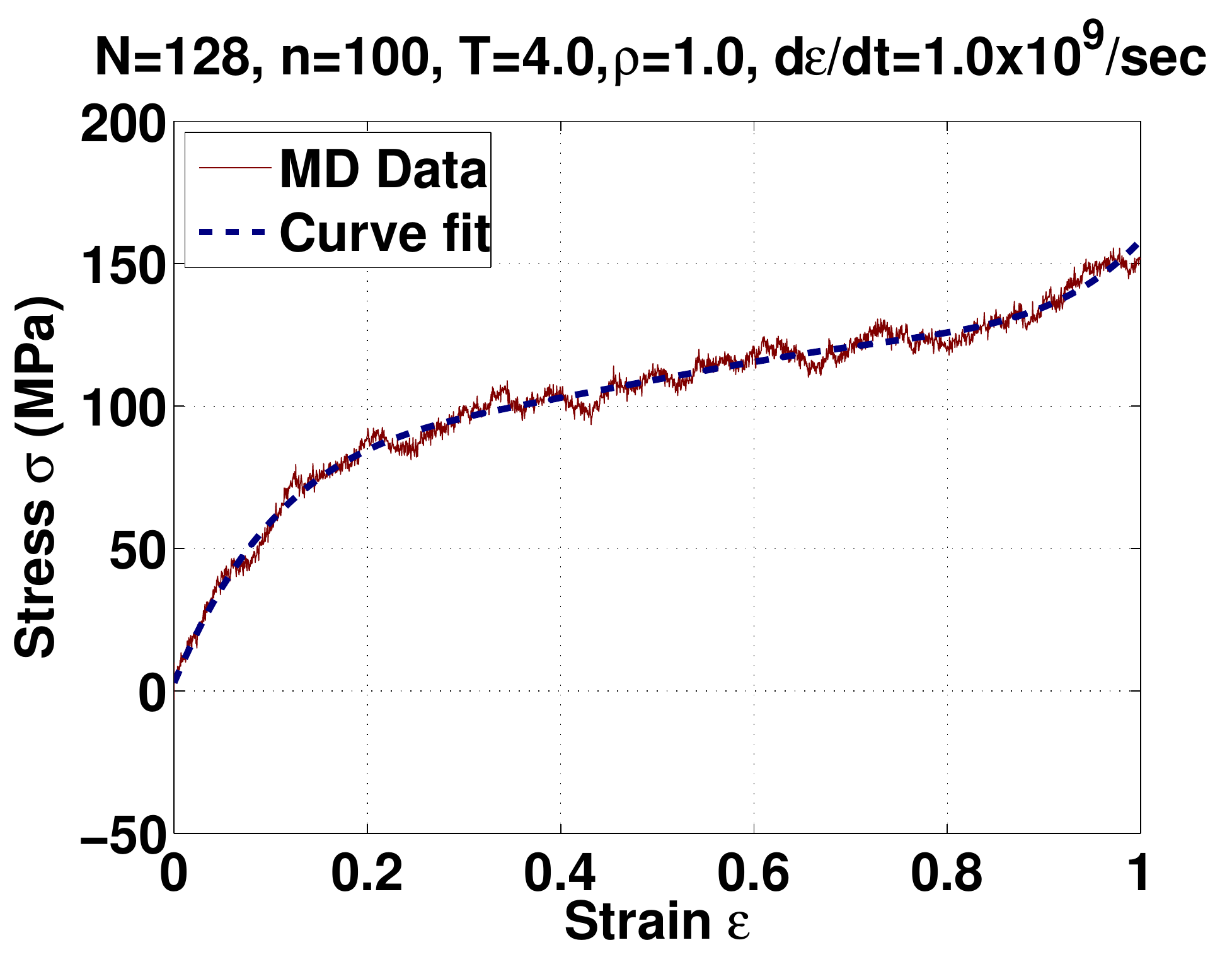}}
 \subfloat[Modulus vs. strain]{\label{fig:modulus}\includegraphics[width=0.5\textwidth]{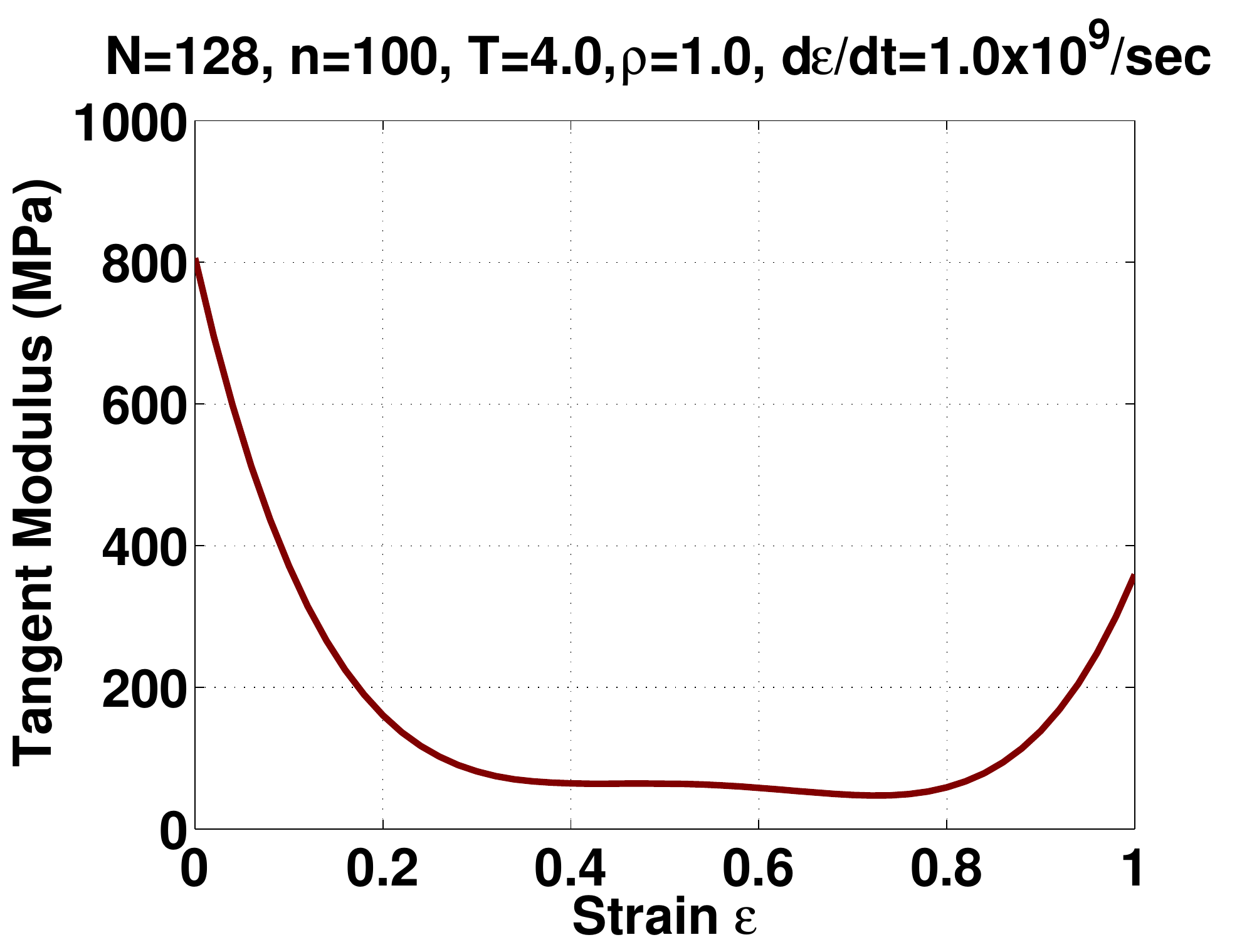}}
\caption{Stress response under uniaxial constant strain rate loading}
\label{fig:StressStrainMod}
\end{figure}

\begin{figure}
 \centering
 \subfloat[Mean-square bond length]{\label{fig:MeanBondLenStrain}\includegraphics[width=0.5\textwidth]{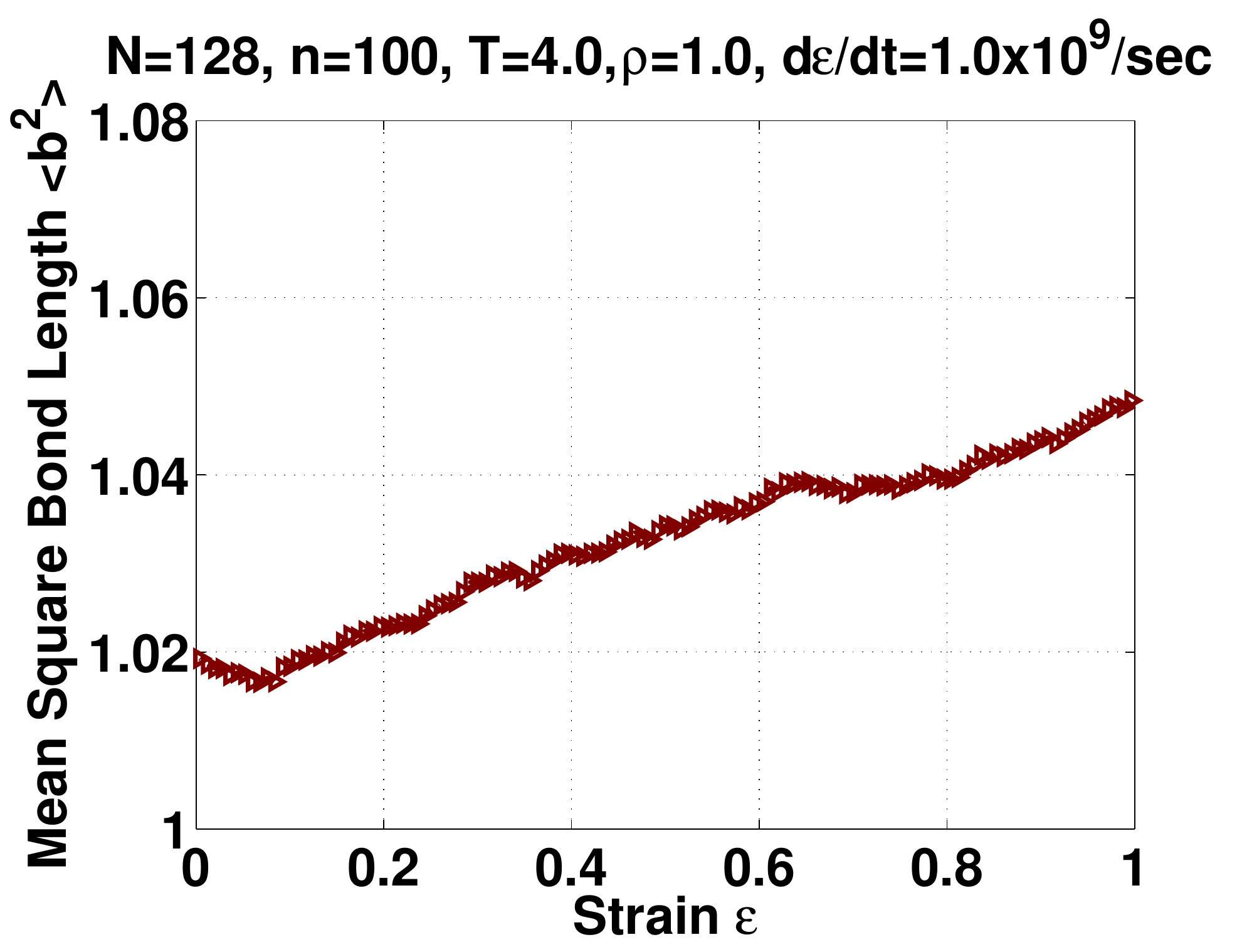}}
 \subfloat[End-to-end length]{\label{fig:EndToEnd}\includegraphics[width=0.5\textwidth]{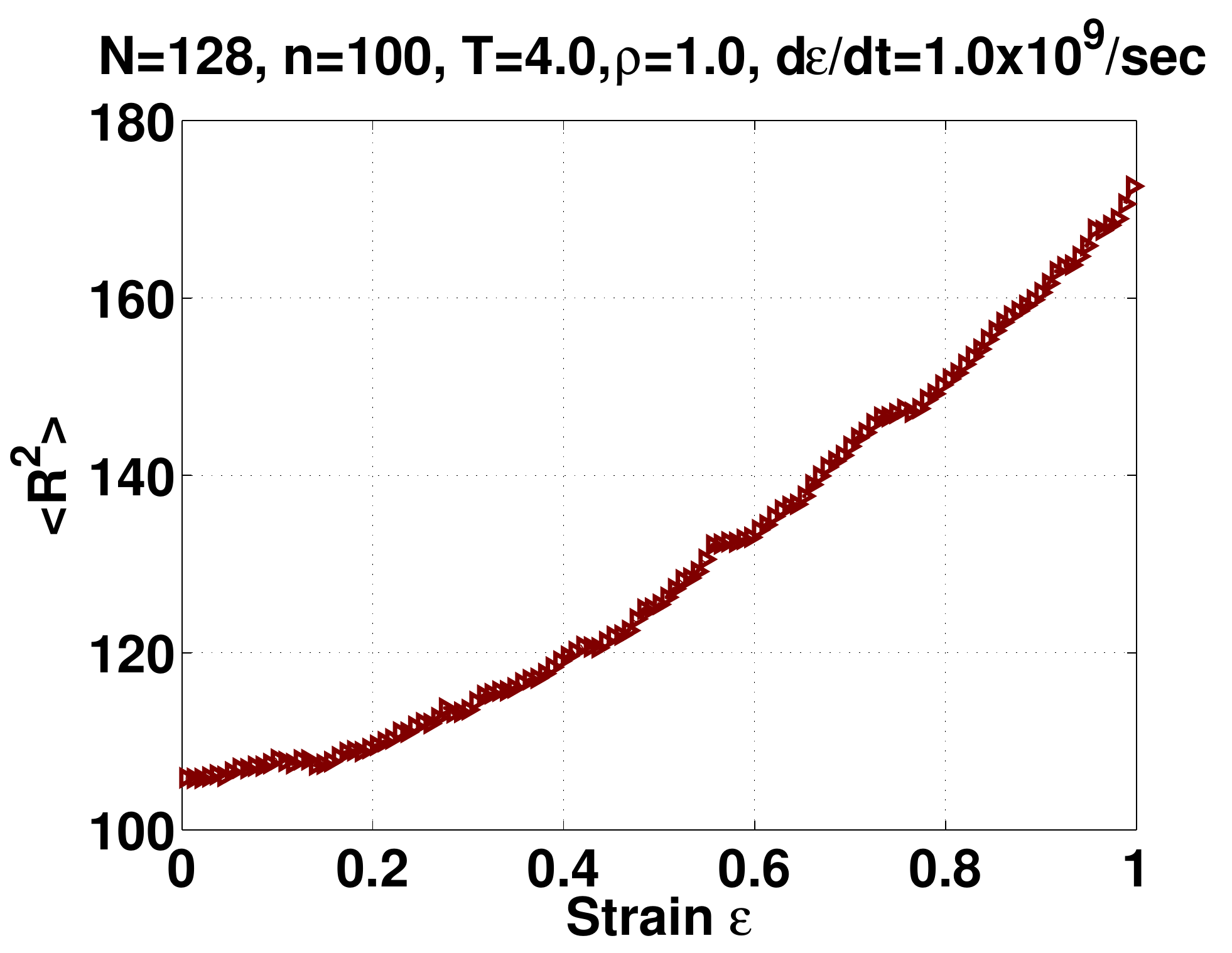}}\\
 \subfloat[Radius of gyration]{\label{fig:RadGyr}\includegraphics[width=0.5\textwidth]{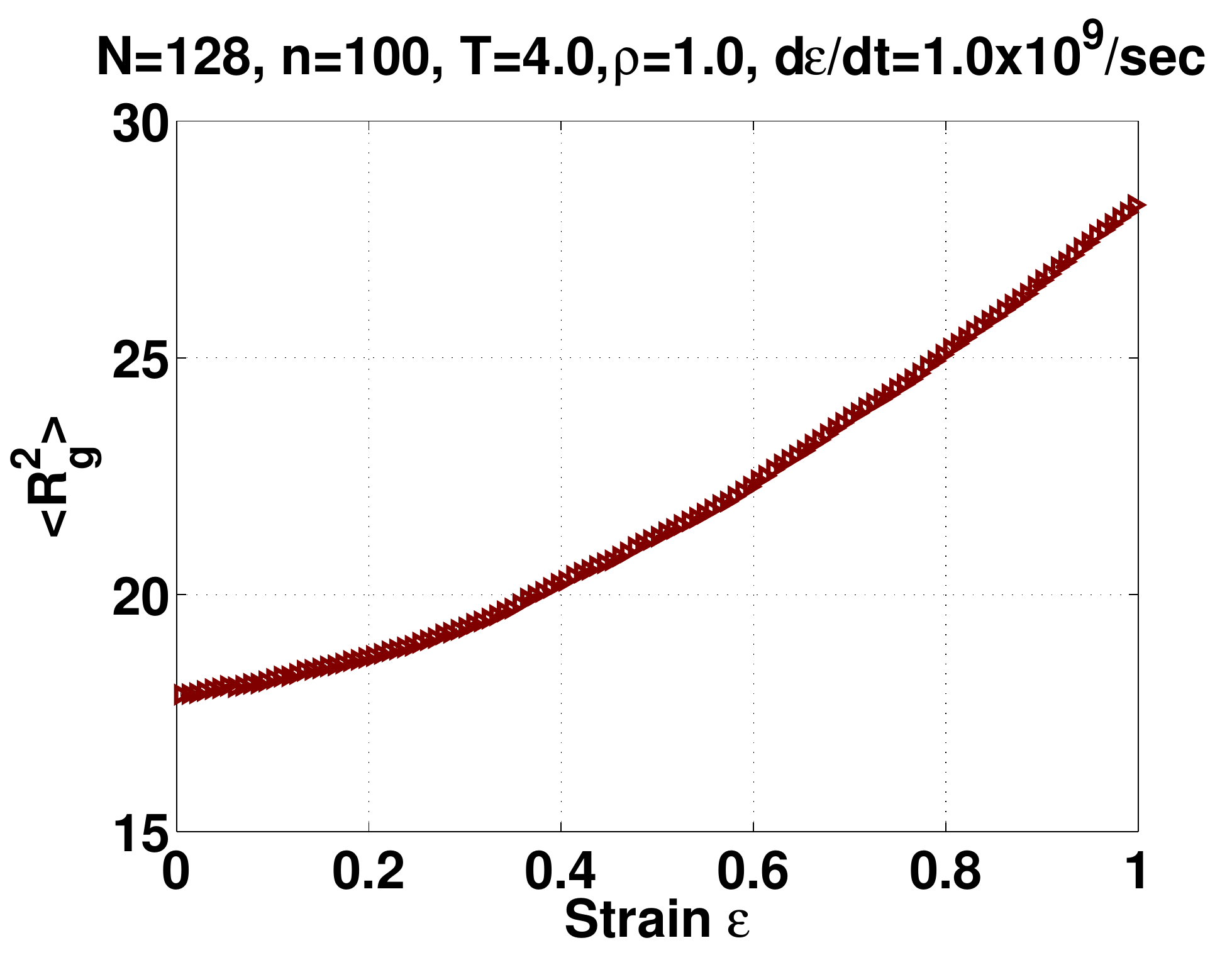}}
 \subfloat[Mass ratio]{\label{fig:MassRatio}\includegraphics[width=0.5\textwidth]{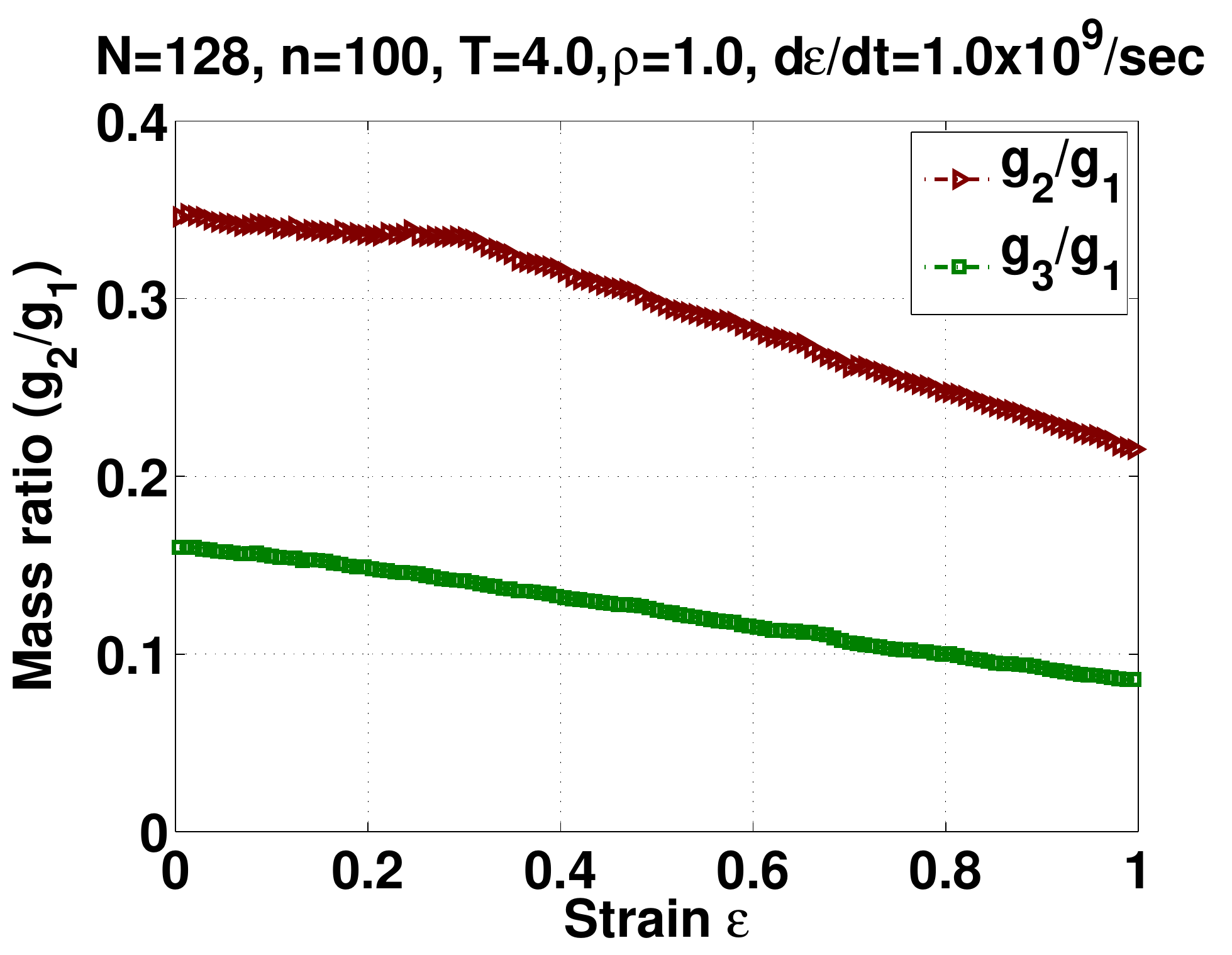}}\\
 \subfloat[Mean bond angle]{\label{fig:BondAngle}\includegraphics[width=0.5\textwidth]{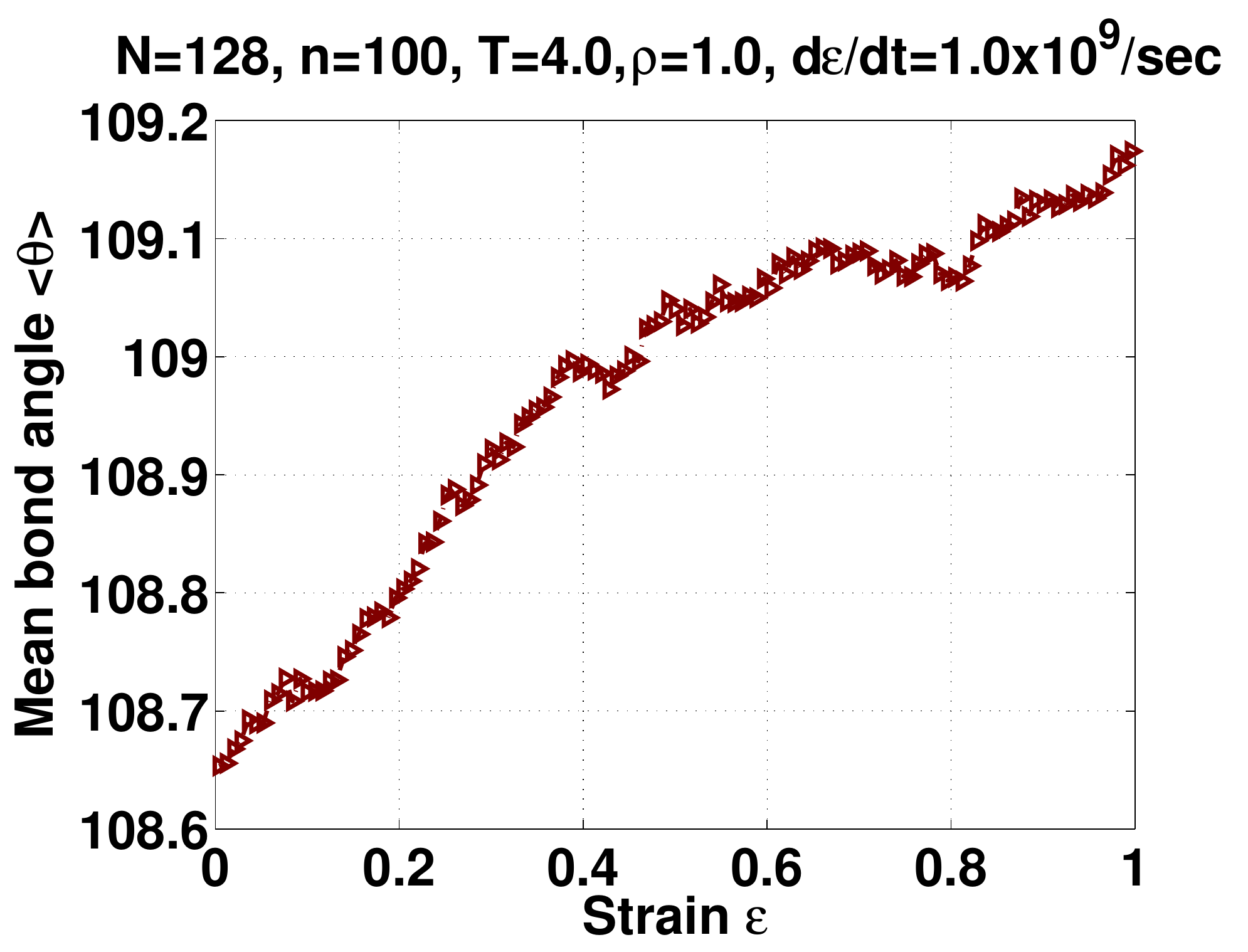}}
  \subfloat[Chain angle]{\label{fig:ChainAngle}\includegraphics[width=0.5\textwidth]{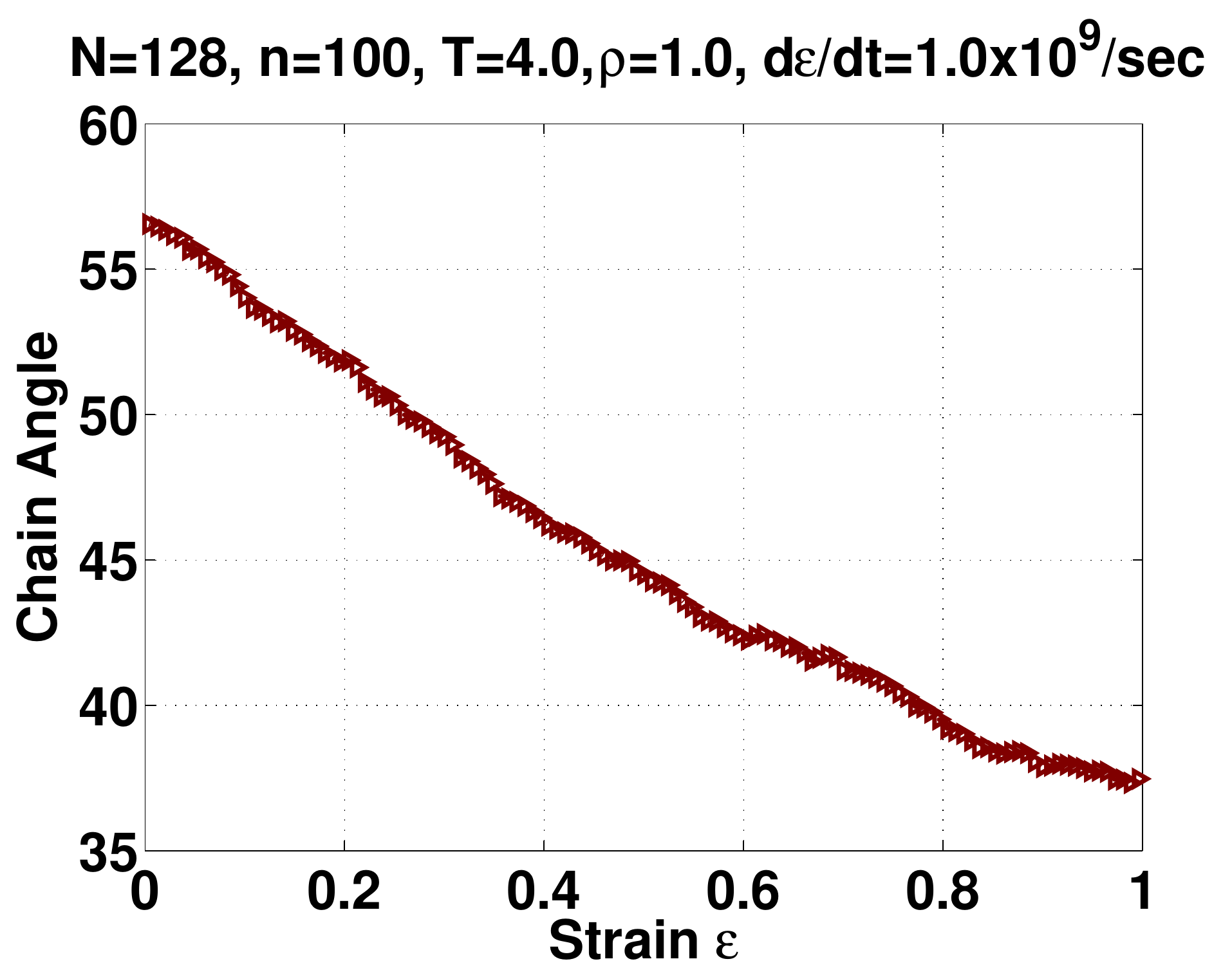}}
 \caption{Variation of structural properties for uniaxial constant strain rate loading}
 \label{fig:PropertiesA}
\end{figure}

Figure~\ref{fig:fit_stress_strain} shows the variation of the axial stress as the system evolves in time. It is observed that the axial stress increases as a nonlinear function of the imposed strain loading. A curve fit of the stress-strain data obtained from the MD simulation is also shown in Figure~\ref{fig:fit_stress_strain}.  From this curve fit we calculate the variation of the tangent modulus with strain. This is shown in Figure~\ref{fig:modulus}. The tangent modulus initially starts with a very high value for small values of strain and decreases to almost zero with further increase in strain. In the second phase there is not much change in the modulus. However for strains greater than $90$\%, the modulus increases at a high rate. This increment is due to the rapid change in the mean-square bond length and mean bond angle as shown in Figure~\ref{fig:MeanBondLenStrain} and ~\ref{fig:BondAngle}, respectively.   

Initially the chains take the folded configuration at equilibrium. When this system is subjected to a constant strain rate, the linear polymer deforms due to unfolding of the chains as well as stretching of the bonds. Hence the tangent modulus starts with a high value and decreases as more and more number of chains unfold. During the second phase, the deformation occurs mainly due to the unfolding of the chains, and hence in this phase there is very little change in the modulus. In the third phase, chains almost align themselves in the stretch direction and hence any further strain directly results in bond stretching which is a very stiff mode of deformation. Therefore, we observe the increase in the modulus at high values of strain.  

We now present results on the variation of the micro-structure parameters such as mean-square bond length, radius of gyration, mean-square end-to-end bond length, and mass ratio of the polymer as a function of the time-history of the constant strain rate loading on the polymer.

Figure~\ref{fig:MeanBondLenStrain} shows the variation of the mean-square bond length with strain. The mean-square bond length gives an estimate of the contribution to the deformation coming from bond deformation. We observe that the mean bond length decreases initially for very small strains but subsequently increases uniformly with strain till approximately $60\%$ strain. After this, the bond length remains constant till $80\%$ strain and subsequently increases rapidly between $80-100\%$ strain. If the chains are highly oriented in the direction of the loading, the predominant mechanisms of deformation are bond extension and bond conformation \cite{Simoes2004}. This is the primary reason for the increase in the modulus beyond a strain value of $90\%$. This shows that the deformation of the bond is strongly correlated with the stress. A sudden change in the bond length is due to the alignment of the chains in the loading direction. The change in the bond length contributed partially to the deformation, and thereby the stress, in the polymer. 

Figure~\ref{fig:EndToEnd} shows the behavior of the end-to-end length when the polymer is stretched. The end-to-end bond length estimates the degree of coiling of the polymer chain. If the chain is fully uncoiled, the end-to-end bond length will approach a value very close to the mean bond length times the number of bonds. In general, its value is much lower than this except at highly strained conditions. We observe that initially the end-to-end length grows linearly with strain, but in the later part of the loading, the rate of change of the end-to-end length grows non-linearly. This is due to the fact that in the later part of the loading many of the chains are open to an extent that they do not offer much resistance to the further opening of the chains. Hence a large and nonlinear increase in the mean-square end-to-end length.

The variation of radius of gyration with strain is shown in Figure~\ref{fig:RadGyr}. The radius of gyration is a measure of the coiling of the polymer chain. If the chain is highly coiled, the radius of gyration will take a low value, whereas if it is uncoiled, its value will be higher. During the initial phase of loading, the change in the value of the radius of gyration is less, but it slowly grows in the later part of the loading as resistance offered for uncoiling reduces as the chains open up. But if we keep straining the material, a stage is reached when most of the chains are completely uncoiled and there will not be any significant change in the end-to-end length and radius of gyration.
 
In Figure~\ref{fig:MassRatio}, the variation of the mass ratio to strain loading is shown. The mass ratio is indicative of the shape of the chains during the deformation. The mass ratio is a parameter that gives a rough estimate of the orientation of the chain and distribution of the mass in the chain. If the mass ratio is different from unity, the  distribution is non-spherical. We find that as the system is strained, the mass ratios in both the transverse directions to the strain loading, namely directions $2$ and $3$, decrease. Moreover, in direction $3$, the distribution of the mass is very low. Hence, one can conclude that the mass of the polymer is distributed more or less in a plane. It means that the chains become more and more like a flattened cigar when it is strained.

Now we take a look at the variation of the mean bond angle as the system is stretched. This is shown in Figure~\ref{fig:BondAngle}. We find that the bond angle, during the initial part of the loading, increases very rapidly. Subsequent to this, the rate of change decreases, and we find that from $\epsilon = 0.6$ to $\epsilon = 0.8$, there is no change in the mean bond angle. Towards the end of the loading, that is beyond $\epsilon = 0.8$, there is again a sudden increase in the bond angle. From Equation~(\ref{eq:bbend}) this results in an increase in the bond bending potential. Consequently, this leads to an increase in the stress towards the end of the loading, as observed in Figure~\ref{fig:fit_stress_strain}.

Figure~\ref{fig:ChainAngle} shows the variation of the mean chain angle. The chain angle is the angle between the loading direction and the end-to-end vector of the chain. We observe that the chain angle decreases uniformly, indicative of alignment of the chains in the loading direction.

\begin{figure}
 \centering
 \subfloat[Bond length]{\label{fig:LenBondLen}\includegraphics[width=0.5\textwidth]{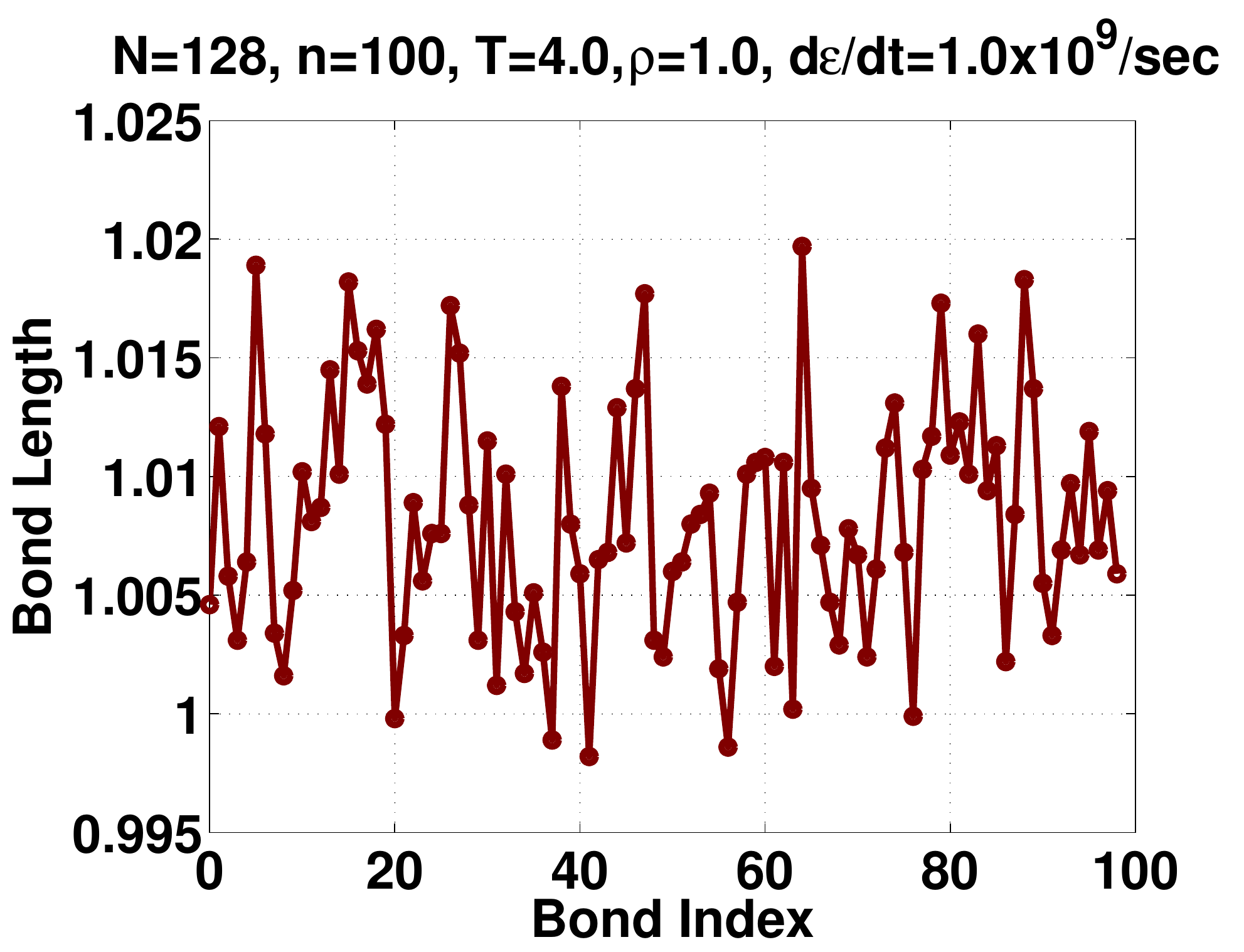}}
 \subfloat[Bond angle]{\label{fig:LenBondAng}\includegraphics[width=0.5\textwidth]{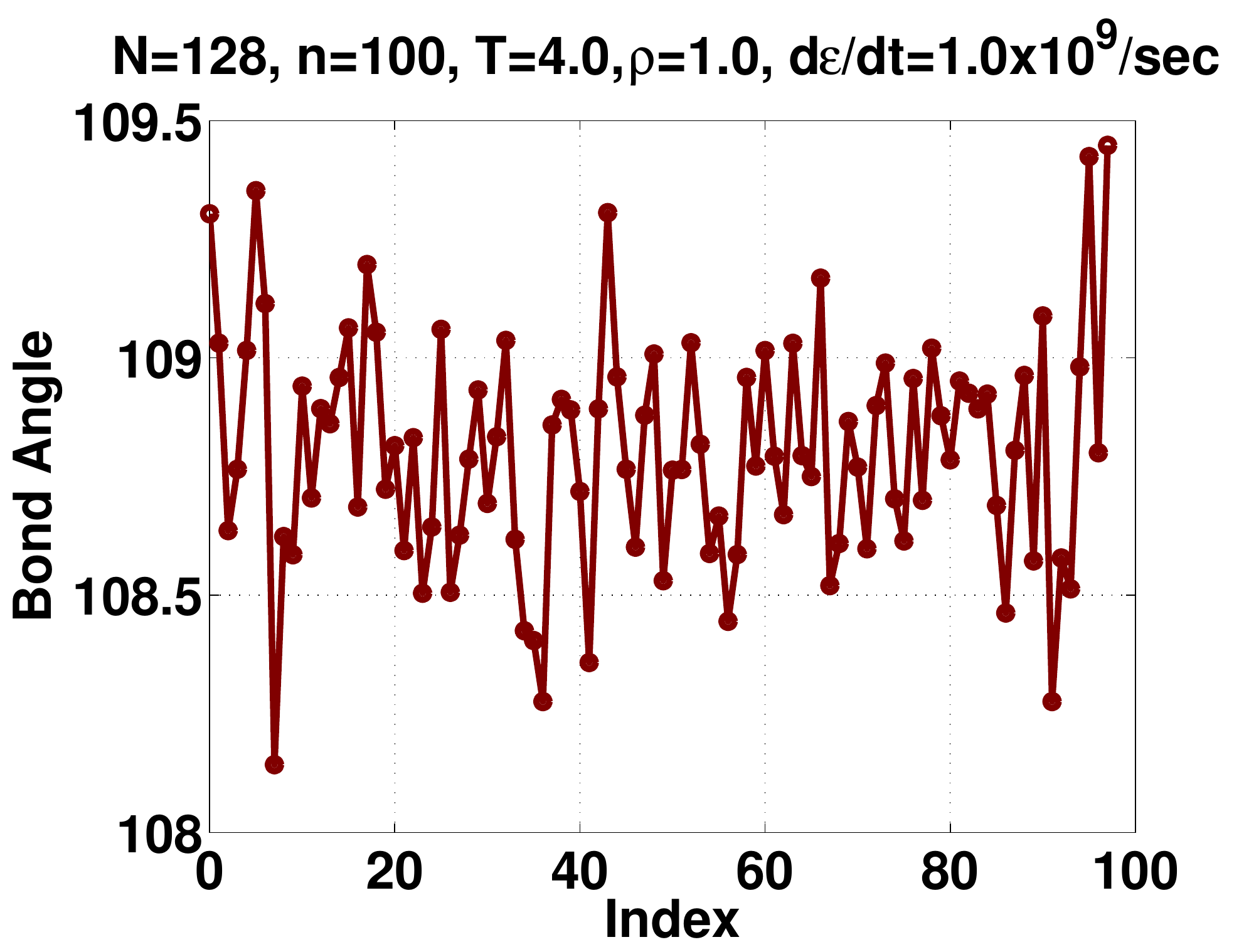}}\\
 \subfloat[Dihedral angle]{\label{fig:LenDihedAng}\includegraphics[width=0.5\textwidth]{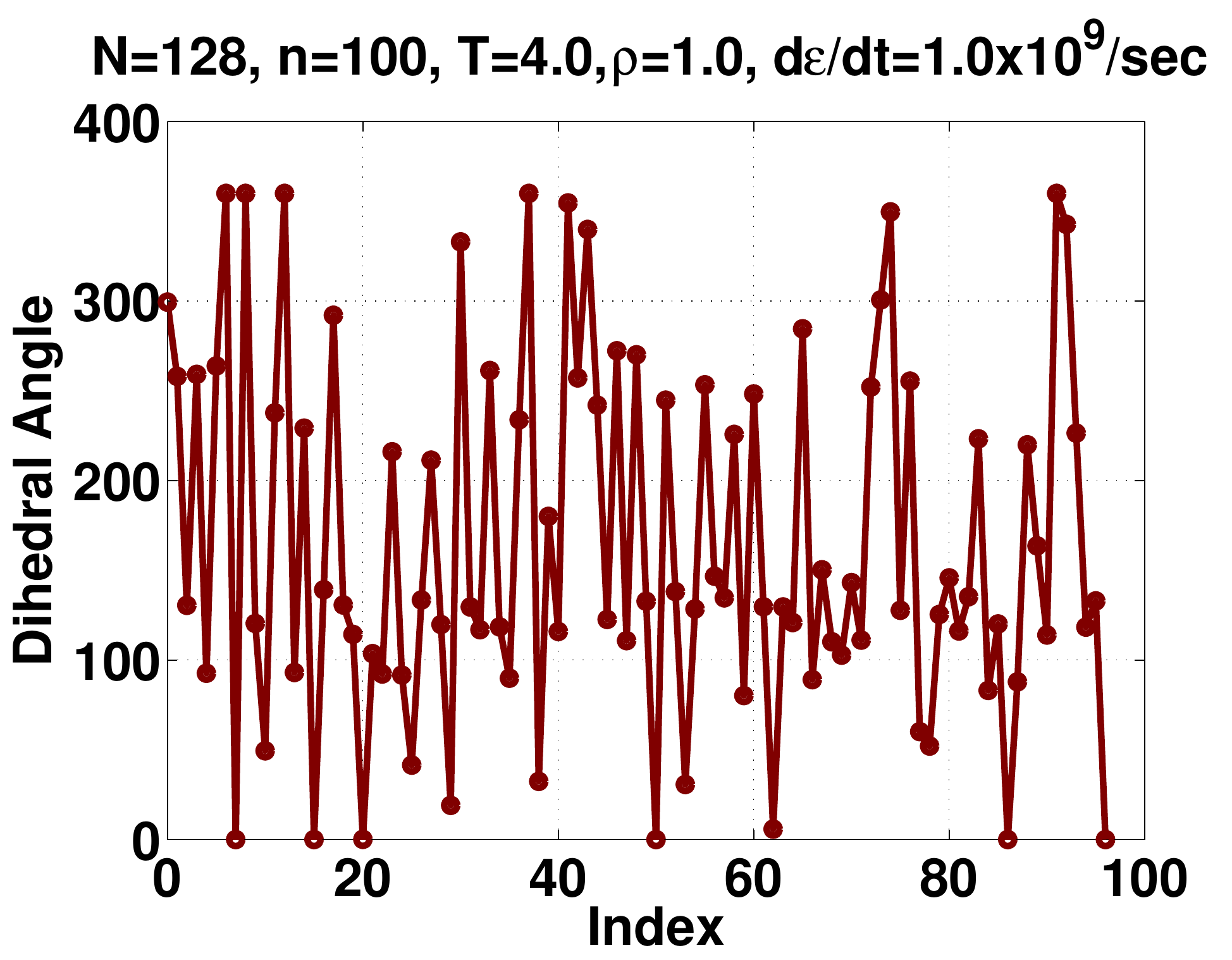}}
 \subfloat[Kinetic energy]{\label{fig:ke}\includegraphics[width=0.5\textwidth]{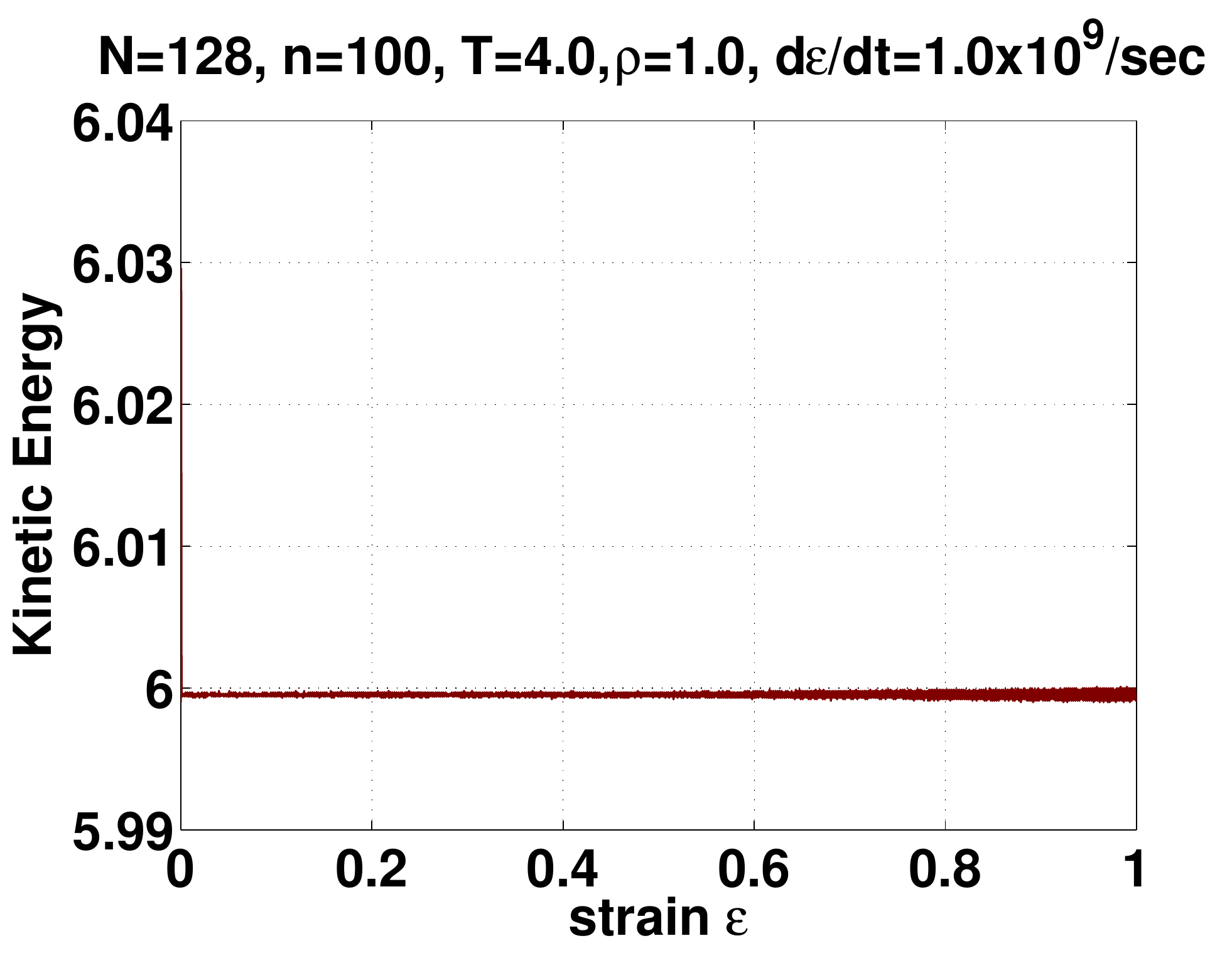}}\\
 \caption{Distribution of structural properties along the length of the chain for constant strain rate loading}
 \label{fig:LenDist}
\end{figure}

In the discussion above, we presented the variation of the micro-structure parameters as a function of strain. We now present variation of some of these micro-structure properties along the chain length. In order do so, at a given segment on the chain, we take the average value of the structural property over all the chains. Subsequently, we take the time-average of this ensemble average. We repeat this for other segments on the polymer chain, For instance, the mean bond length along the length of the chain is obtained by taking the time average of the average value of the bond-length of corresponding bonds in all the chains.

We observe in Figure \ref{fig:LenBondLen}, that the bond length has fluctuations along the chain. But these fluctuations are uniform throughout the chain as opposed to the observation of \citet{Saitta} wherein they report a jump in the bond length at the ends of the chain. This difference is due to the fact that Saitta and Klein's study considers a single short polymer chain loaded by an external force. As reported in their study, the mean value of the bond length along the length of the chain is higher than the equilibrium bond length value indicative of chains being in a stretched state. 

The bond angle variation along the length of the chain is shown in Figure~\ref{fig:LenBondAng}. We observe that the bond angles do not deviate much from the mean value and are equally scattered on both sides from the mean value. The variation of the dihedral angle along the length of the chain, for the constant strain rate loading, is shown in Figure~\ref{fig:LenDihedAng}. The set of dihedral angles can be divided into four groups, namely $\{0,\, 113^\circ,\, 247^\circ,\, 360^\circ\}$. This set of dihedral angles correspond to the minimum potential energy of the polymeric system. We observe that the number of bonds falling in each segment is almost the same.

Figure~\ref{fig:ke} shows the variation of the kinetic energy during the simulation. Throughout this simulation, the temperature was kept constant at a non-dimensional temperature $T=4.0$. As a result of which the kinetic energy takes a value corresponding to the specified temperature. In the course of the simulation, the temperature keeps fluctuating about this mean value with a very small standard deviation.

\section{Control parameters, stress-strain, and micro-structure}
\label{sec:control_para_stress_strain_resp}

The control variables for the polymeric system subject to dynamic strain are the parameters that are externally set or imposed such as strain rate, density, temperature, and chain length. These parameters control the behavior of the polymeric system. On a macroscopic scale, they affect the stress-strain characteristics. Here we study the effect of the control parameters on the stress response and correlate it with the polymer micro-structure response. Not only can we now explain well known mechanical behavior of polymers and elastomer \cite{Rajagopal}, but certain anomalous behavior can be explained too. As before, we consider a constant strain rate loading imposed on the polymer.

\subsection{Strain rate}
\label{subsec:control_para_strain_rate}

\begin{figure}
\centering
\subfloat[Stress vs. strain]{\label{fig:EffectStrainRate}\includegraphics[width=0.5\textwidth]{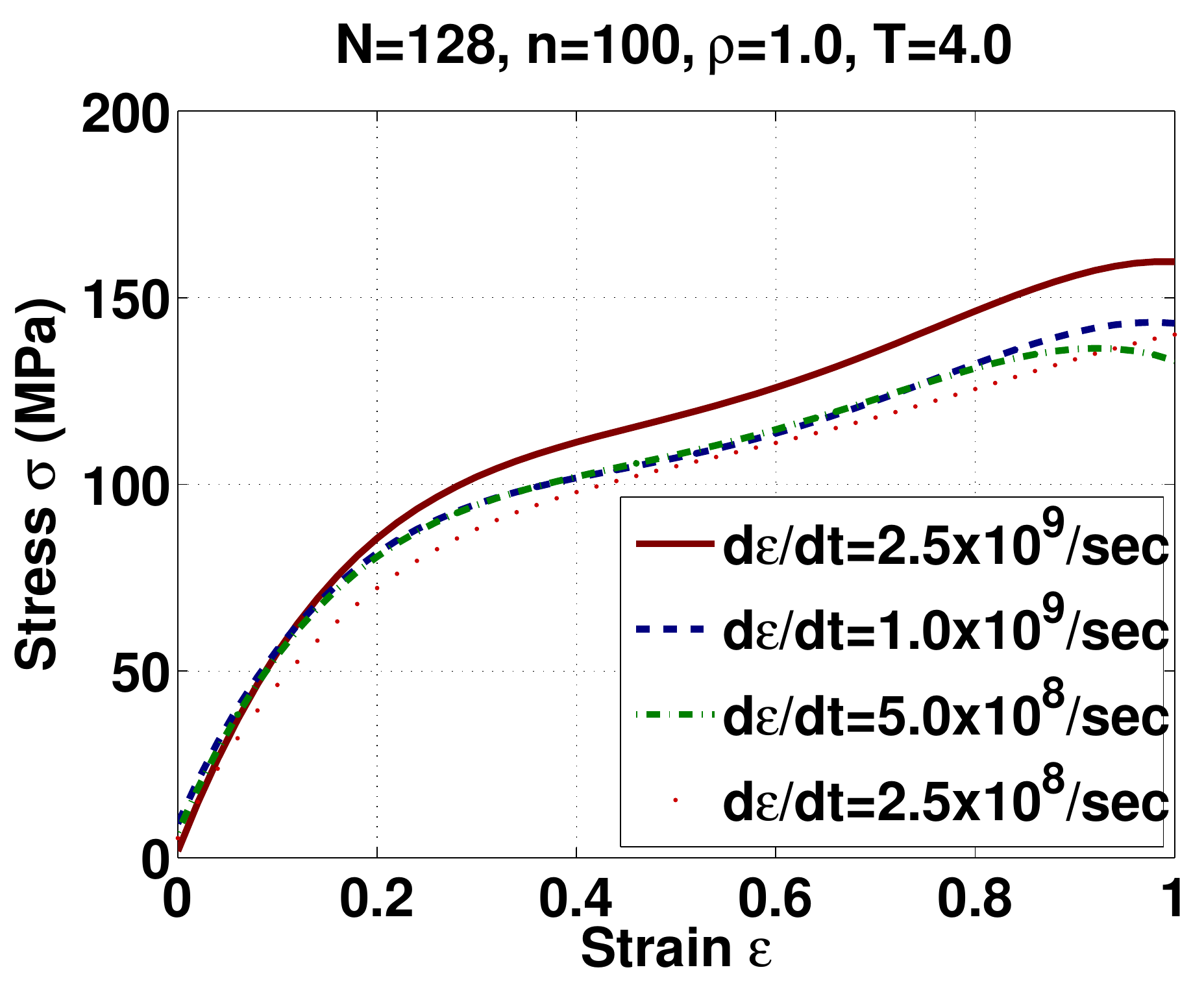}}
 \subfloat[Modulus vs strain]{\label{fig:EffectStrainRateModulus}\includegraphics[width=0.5\textwidth]{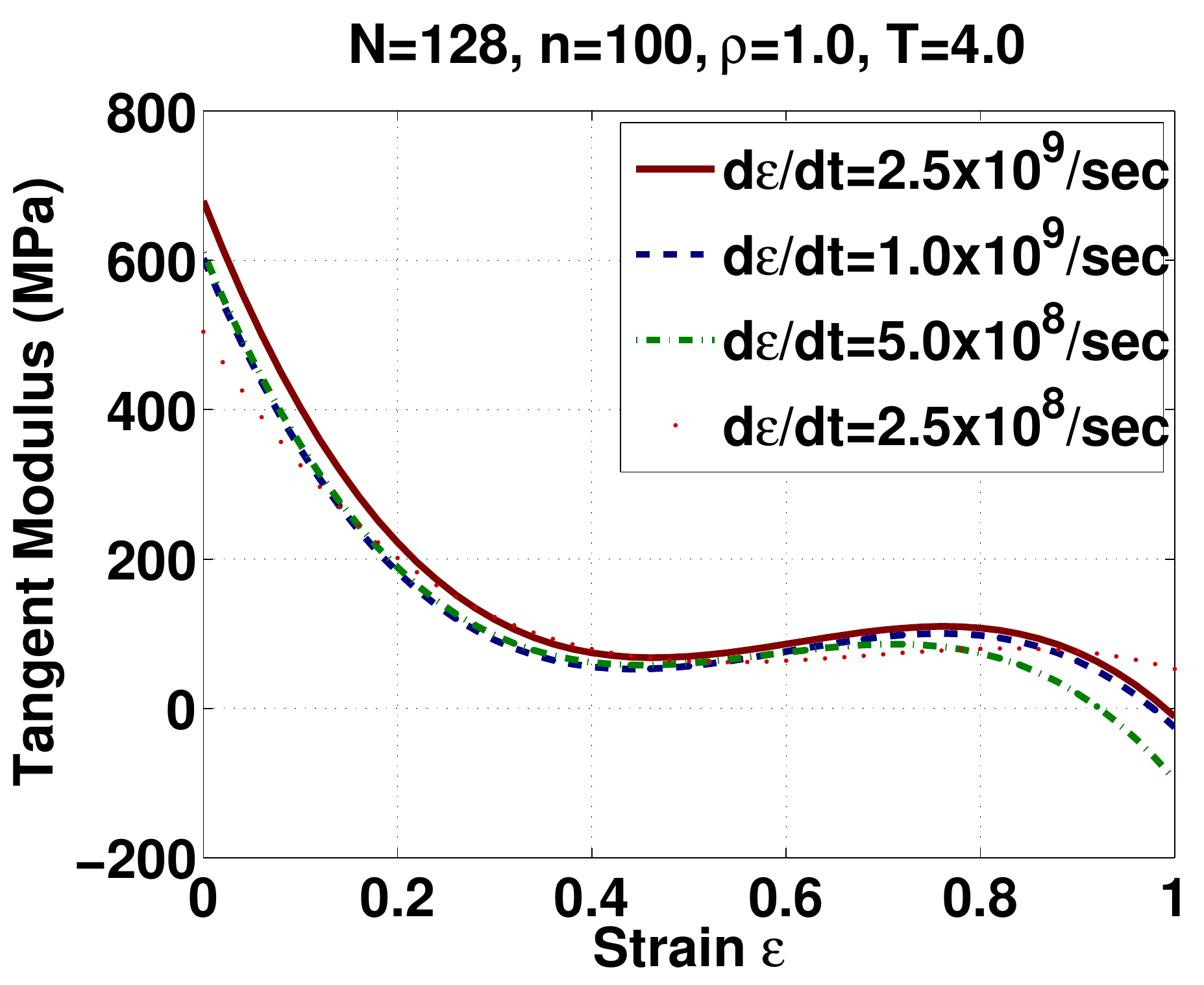}}
\caption{Effect of strain rate on stress response}
\label{fig:EffStrainRateLinear}
\end{figure}

Figure~\ref{fig:EffectStrainRate} shows the effect of different strain rates on the tensile behavior of the polymer. For low values of strain, the stress is the same for all strain rates. Only for strains $\epsilon > 0.2$ that we see an appreciable effect of strain rate on the stress response. Note however that when the strains reach a high value, the stress almost reaches a constant value. This is true across all the strain rates. Figure~\ref{fig:EffectStrainRateModulus} is obtained by taking the derivative of the above curve with respect to strain. From this figure it is clear that the material is stiffer at higher strain rates. Now if we consider any one of these curves, there is a plateau region where stiffness does not vary much with strain. This is because the deformation in this phase is mostly due to the uncoiling of the chain. At large values of strain, the modulus decreases by a small amount that could be due to structural relaxation. We now show below how the loading rate influences the evolution of the micro-structure.

\begin{figure}
\centering
\includegraphics[width=0.5\textwidth]{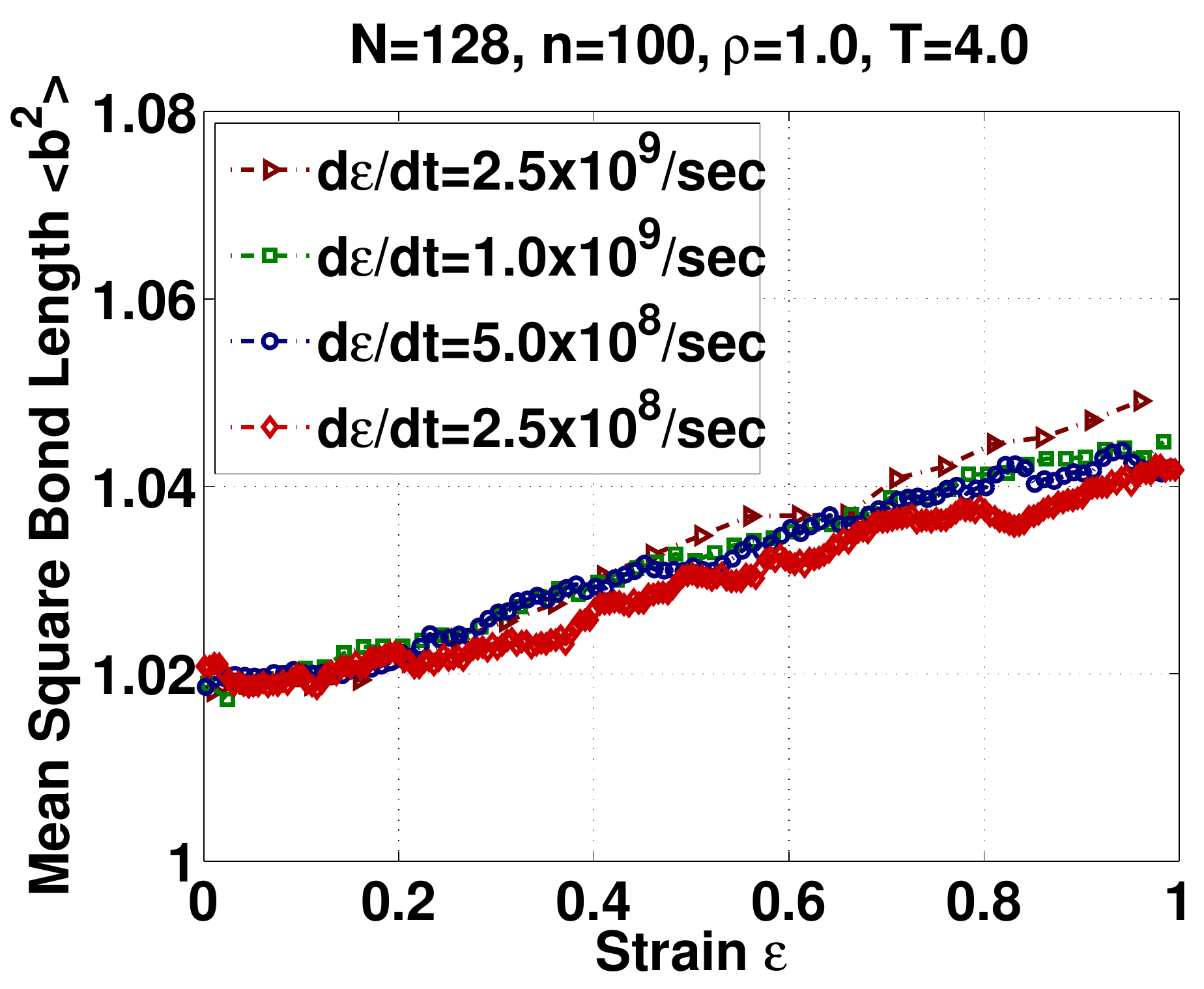}
\caption{Effect of loading rate on mean-square bond length}
\label{fig:EffRateMeanBondLenLin}
\end{figure}

We first consider the influence of strain rate on the evolution of the mean-square bond length with strain. This is shown in \ref{fig:EffRateMeanBondLenLin}. Mean-square bond length grows with strain for all strain rates and we find that at higher rates of loading the deformation in the bond length is more. Also note that during the initial part of loading, the bond length does not increase significantly. But after a strain value of $\epsilon = 0.2$, it increases uniformly for all strain rates. 

\begin{figure}
\centering
\includegraphics[width=0.5\textwidth]{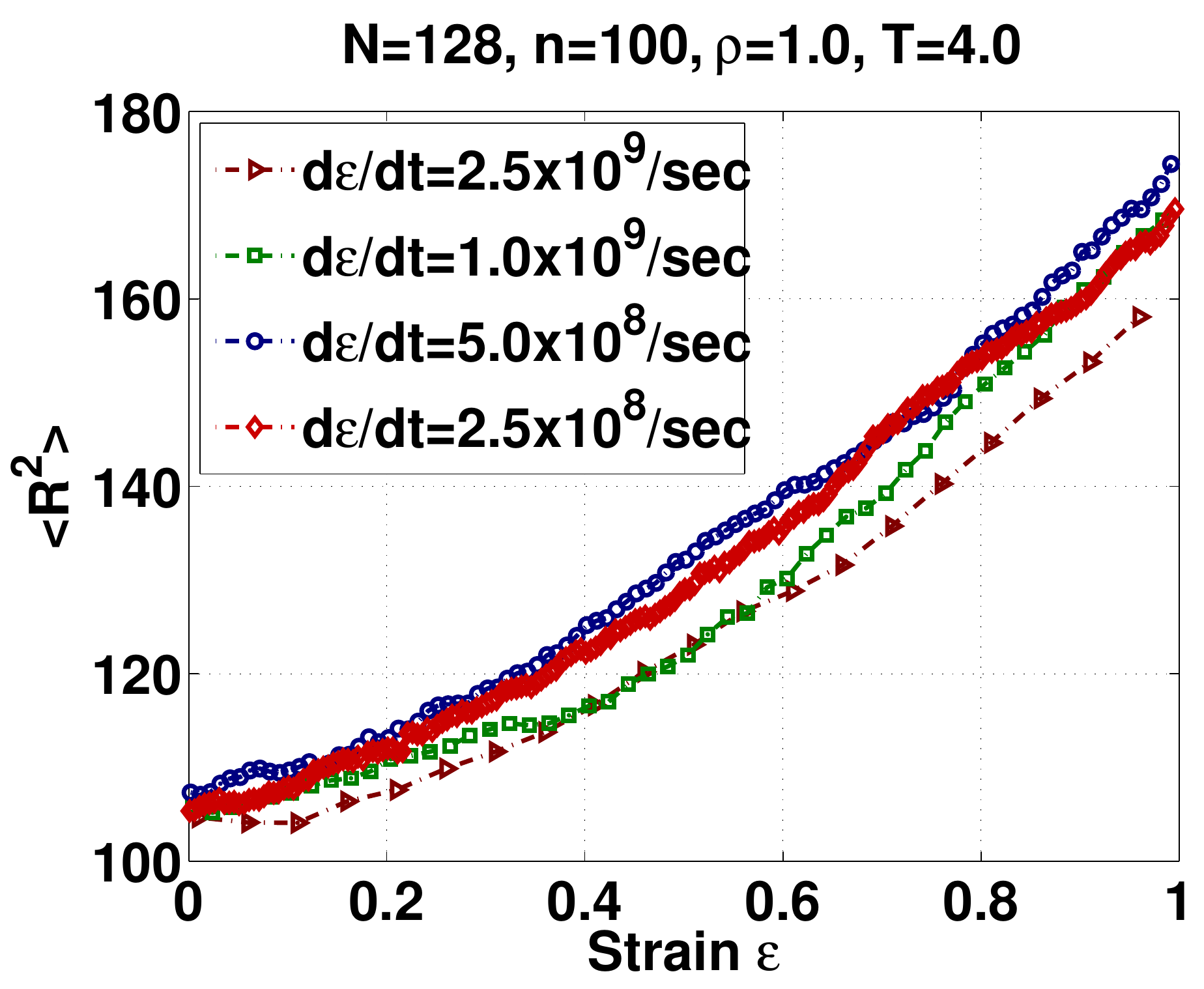}
\caption{Effect of loading rate on end-to-end length}
\label{fig:EffRateEndToEndLin}
\end{figure}

Effect of the strain rate on the mean-square end-to-end length for constant strain rate loading is shown in \ref{fig:EffRateEndToEndLin}. We observe that the mean-square end-to-end length increases in a parabolic manner. Also, as opposed to the case of mean-square bond length, we find that the mean-square end-to-end length is lower at high strain rates. This indicates that at high strain rates internal deformation of the chains is dominant in the over-all deformation of the chains. 

\begin{figure}
\centering
\includegraphics[width=0.5\textwidth]{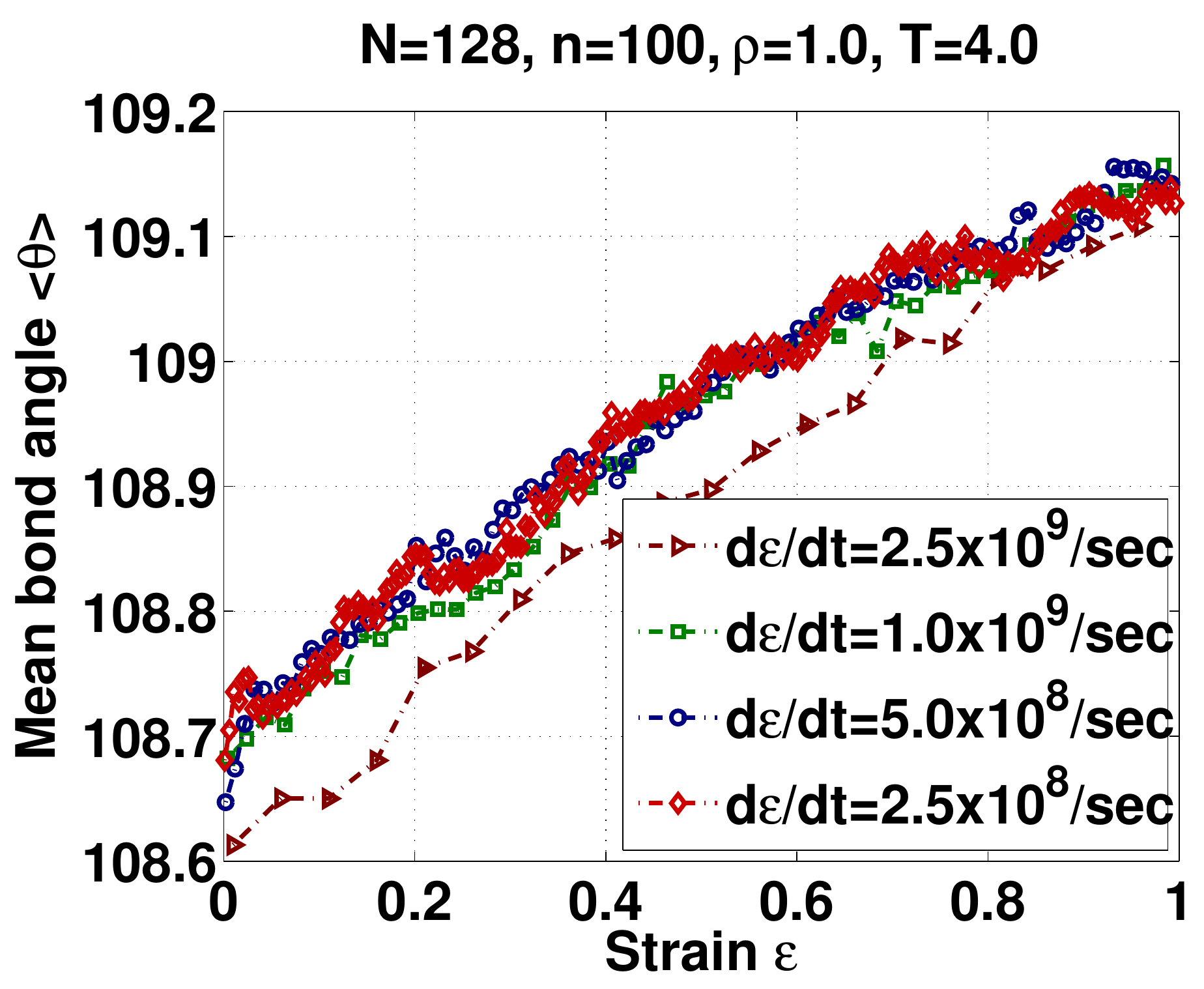}
\caption{Effect of loading rate on mean bond angle}
\label{fig:EffRateBondAngLin}
\end{figure}

The variation of the mean bond angle when the system is subjected to uniaxial constant strain rate loading is shown in \ref{fig:EffRateBondAngLin} at various strain rates. We observe that the change in mean bond angle is more at strain rate $\dot{\epsilon} = 2.5\times 10^{9}/\mathrm{sec}$ whereas for other strain rates this change is almost same. Also, the variation of the mean bond angle with strain is very similar for all strain rates except $\dot{\epsilon} = 2.5\times 10^9/\mathrm{sec}$.

\begin{figure}
\centering
\includegraphics[width=0.5\textwidth]{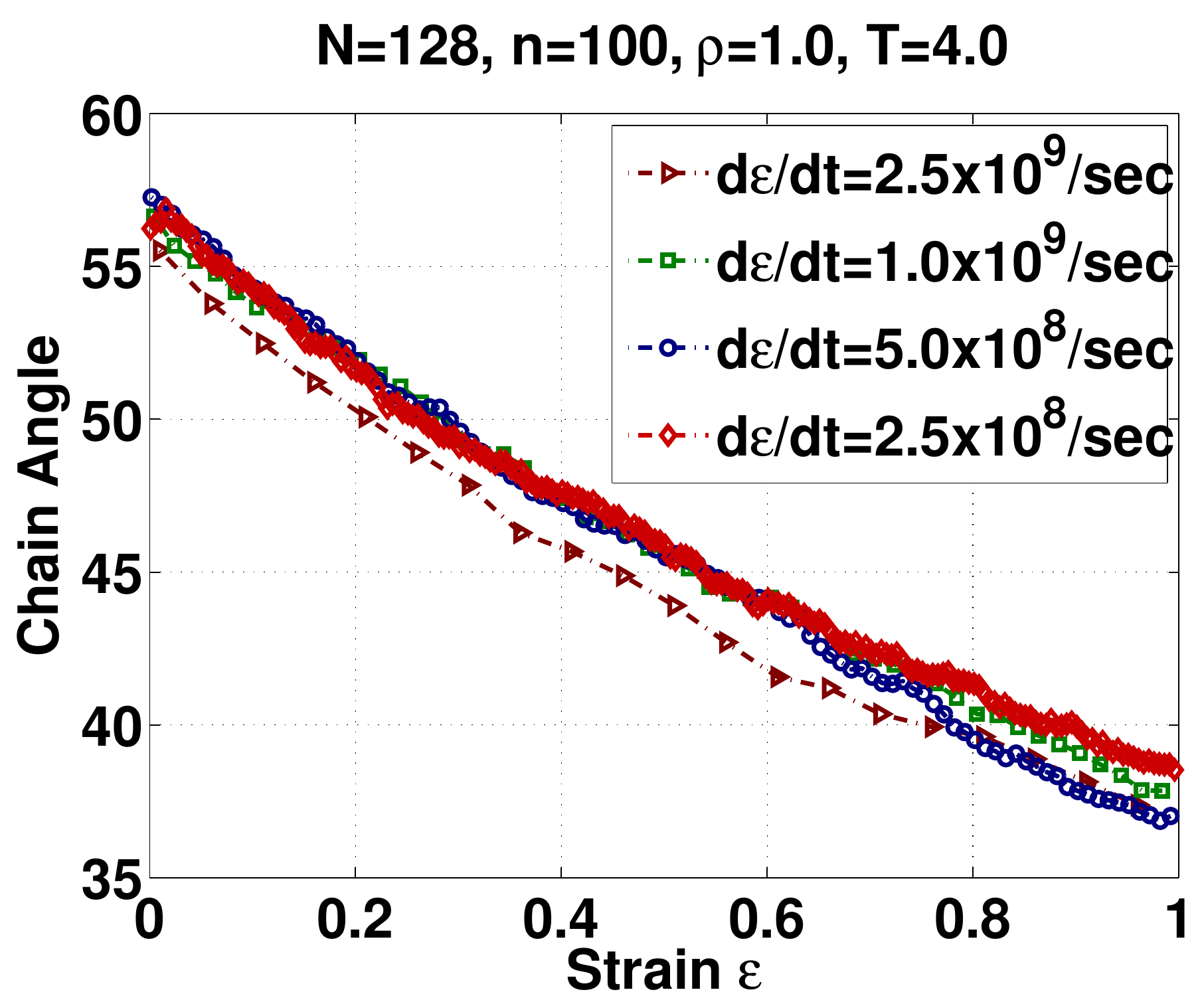}
\caption{Effect of loading rate on mean chain angle}
\label{fig:EffRateChainAngLin}
\end{figure}

\ref{fig:EffRateChainAngLin} shows the effect of strain rate on chain angle for constant strain rate uniaxial loading. We find that, for all strain rates, the chain angle decreases at almost constant rate. Furthermore, we notice that alignment of the chain in the loading direction is more at $\dot\epsilon = 2.5\times 10^9/\mathrm{sec}$ where as this alignment is similar at other strain rates.

\begin{figure}
\centering
\includegraphics[width=0.5\textwidth]{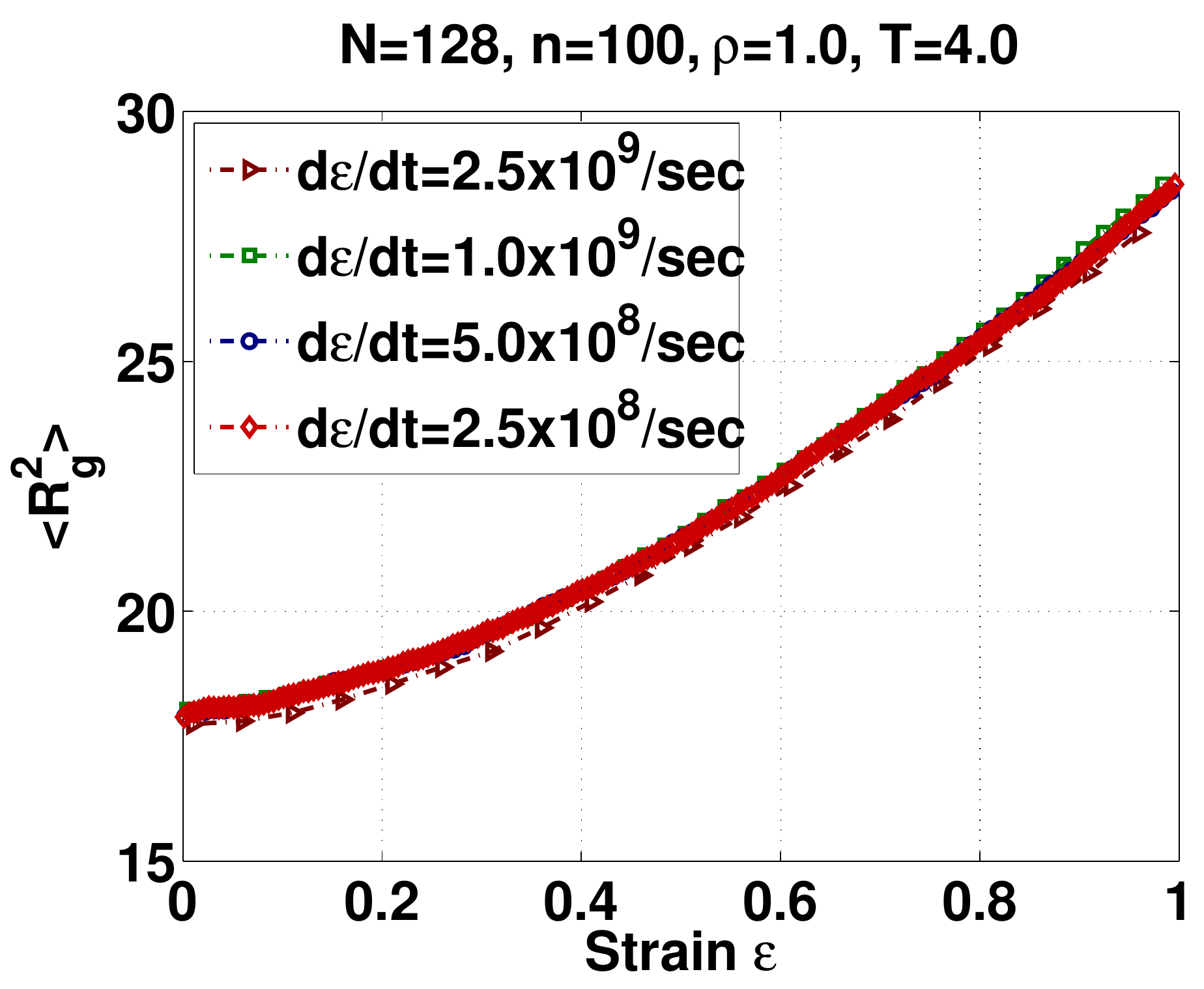}
\caption{Effect of loading rate on radius of gyration}
\label{fig:EffRateRadGyrLin}
\end{figure}

The variation of mean-square radius of gyration for constant strain rate loading at various strain rates is shown in \ref{fig:EffRateRadGyrLin}. We observe that the mean-square radius of gyration increases with strain in a nonlinear sense. There is no significant change in the variation of radius of gyration with strain at different strain rates. 

\begin{figure}
\centering
\subfloat[$g2/g1$]{\label{fig:EffRateG2Lin}\includegraphics[width=0.5\textwidth] {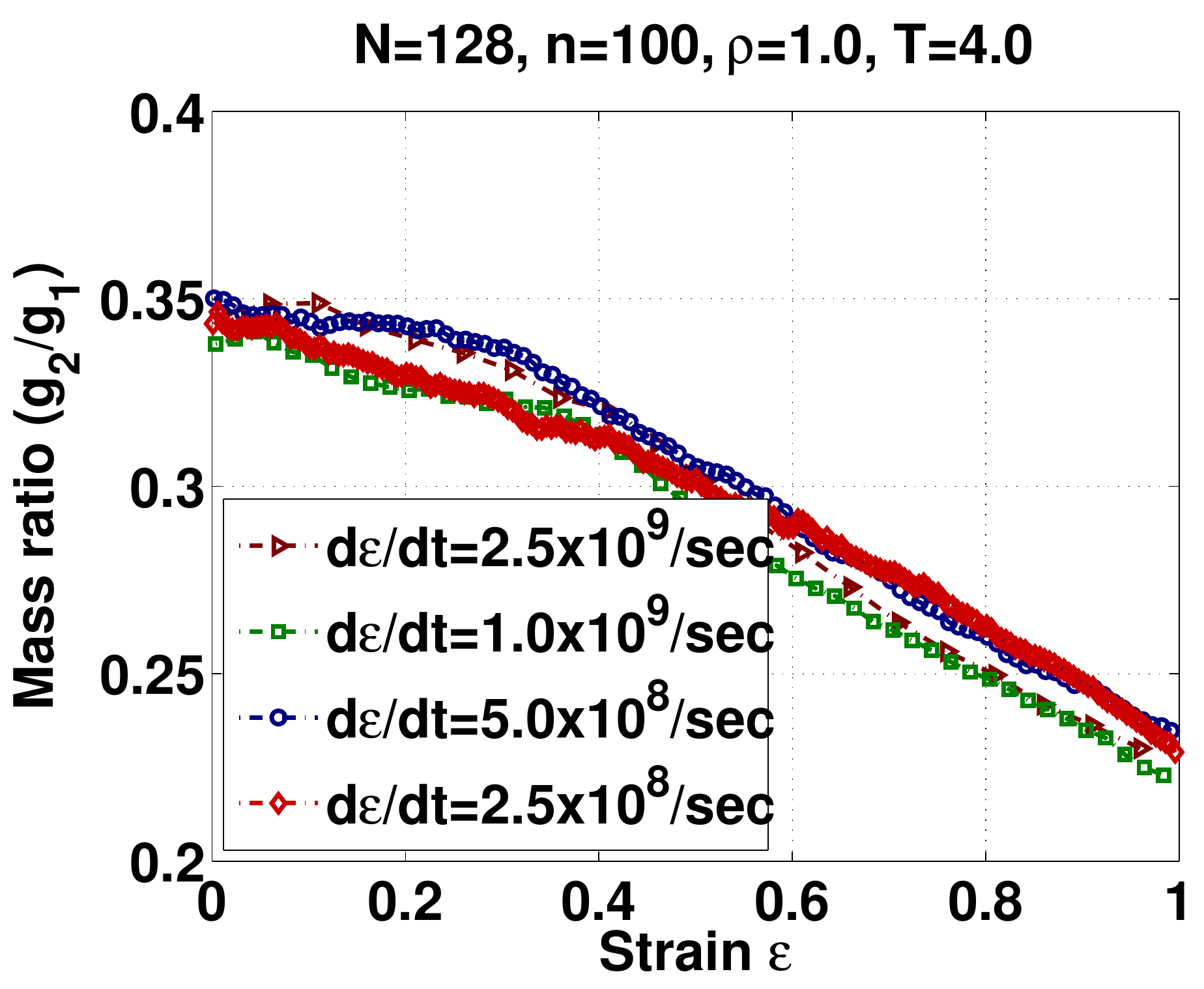}}
\subfloat[$g3/g1$]{\label{fig:EffRateG3Lin}\includegraphics[width=0.5\textwidth]{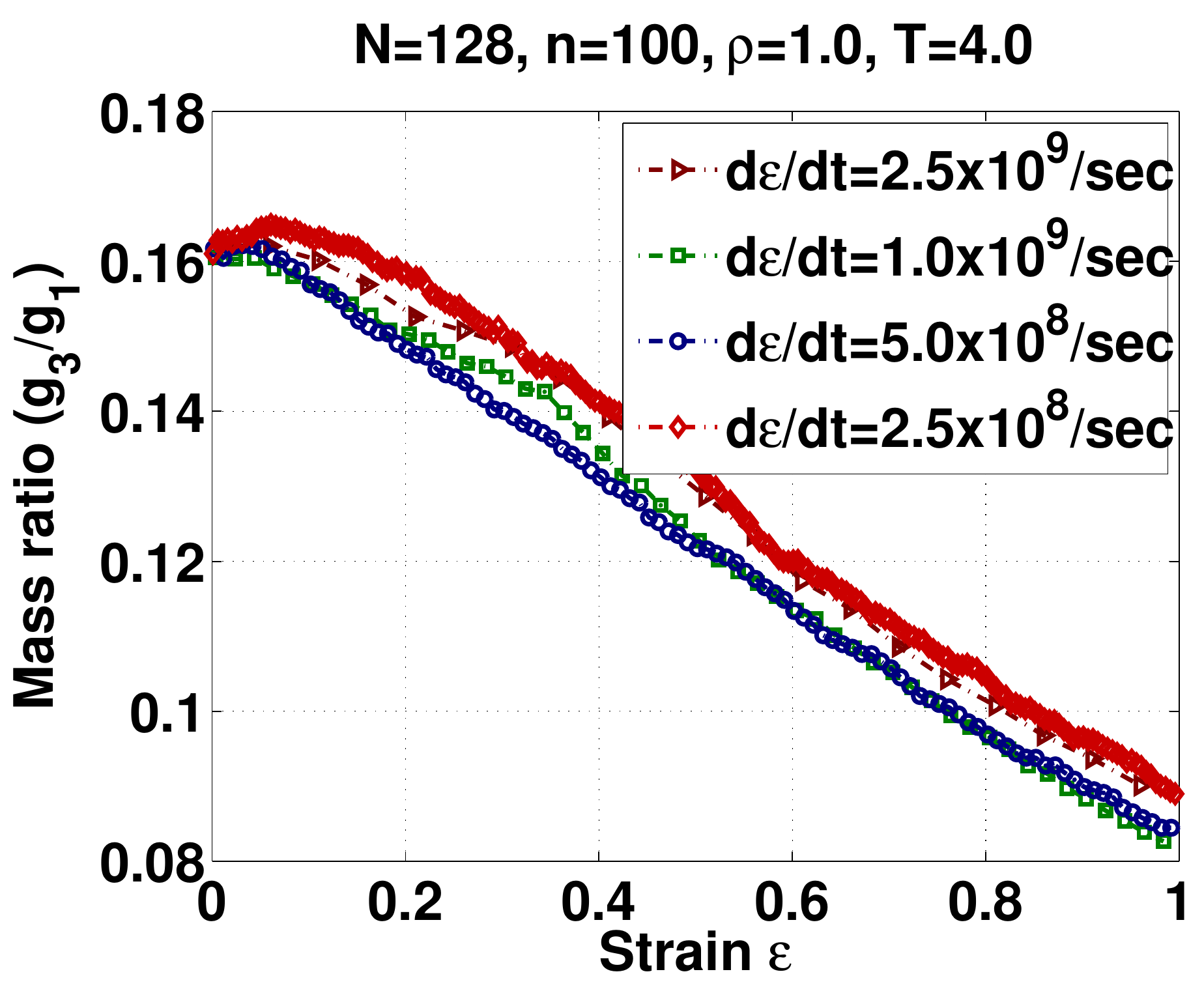}}
\caption{Effect of strain rate on mass ratios}
\label{fig:EffectRateMassRatioLin}
\end{figure}

\ref{fig:EffectRateMassRatioLin} shows the variation of the mass ratios for various strain rates under constant strain rate loading. Mass ratios initially decrease very slowly but subsequently decrease with a larger rate. Beyond a strain value of $\epsilon = 0.2$ we observe that the rate of change of the mass ratios with strain is almost constant. With increasing strain in the system, we observe that the mass distributions decrease linearly. This is indicative of the fact that chains align themselves in one direction. Further, the mass ratio $g_3/g_1$ is very small which points to the fact that in the third direction the monomer distribution is very low. Hence, chains become like a flattened cigar and the distribution of the monomers is primarily in a plane.  The rate of change of mass ratio $g_2/g_1$ is low at low strain rates indicating that alignment of the chains in the loading direction is low at low strain rates. 

We find that at a high value of strain rate the change in the mean-square bond length and mean bond angle is more, where-as the change in the mean-square end-to-end length is small and mean-square radius of gyration remains same at all strain rates. This indicates that internal deformation in the chain dominates over the uncoiling at higher strain rates.  

\subsection{Density}
\label{subsec:control_para_density}

\begin{figure}
\centering
\subfloat[Stress versus strain]{\label{fig:EffectDensity}\includegraphics[width=0.5\textwidth]{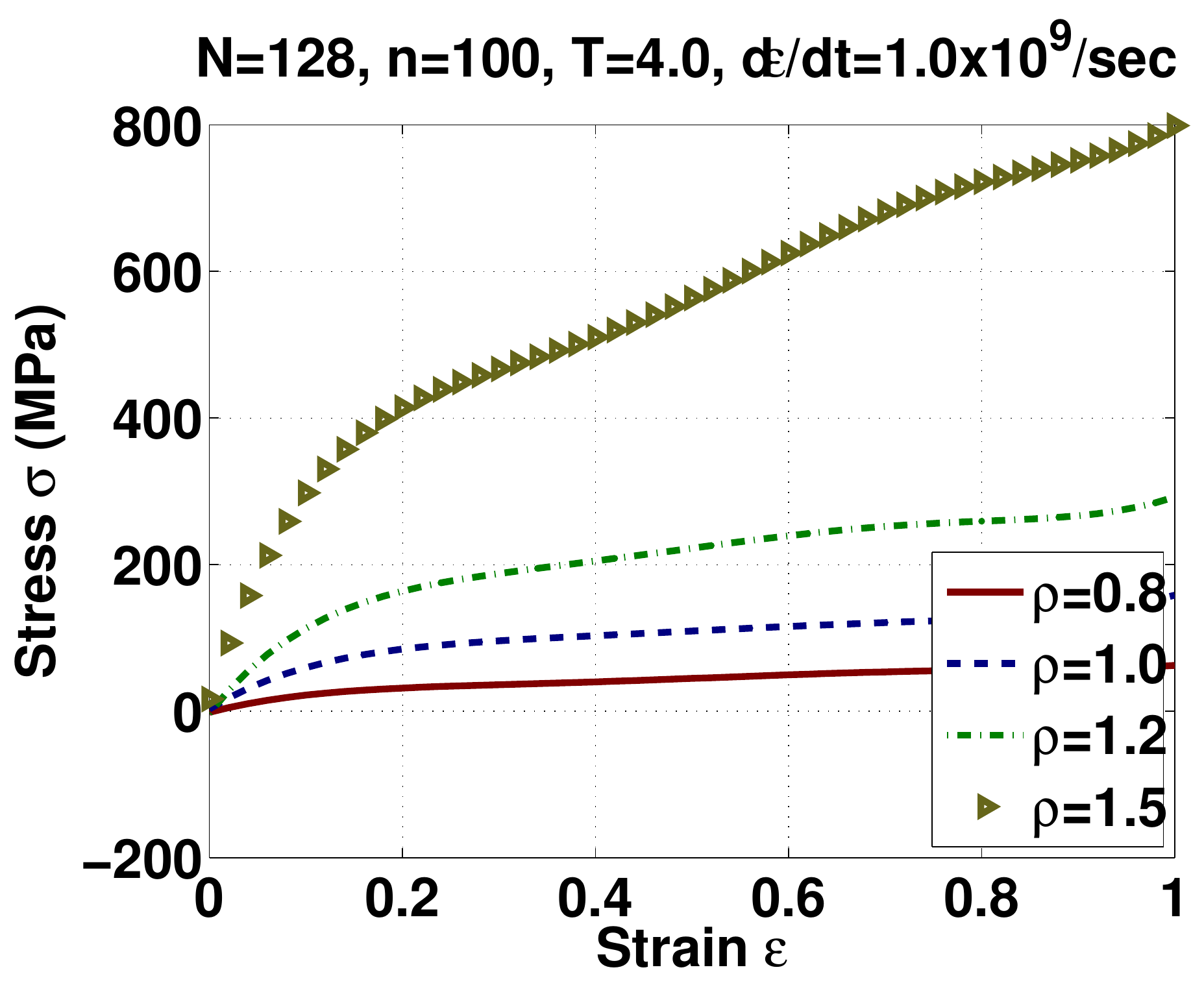}}
\subfloat[Modulus versus strain]{\label{fig:EffectDensityModulus}\includegraphics[width=0.5\textwidth]{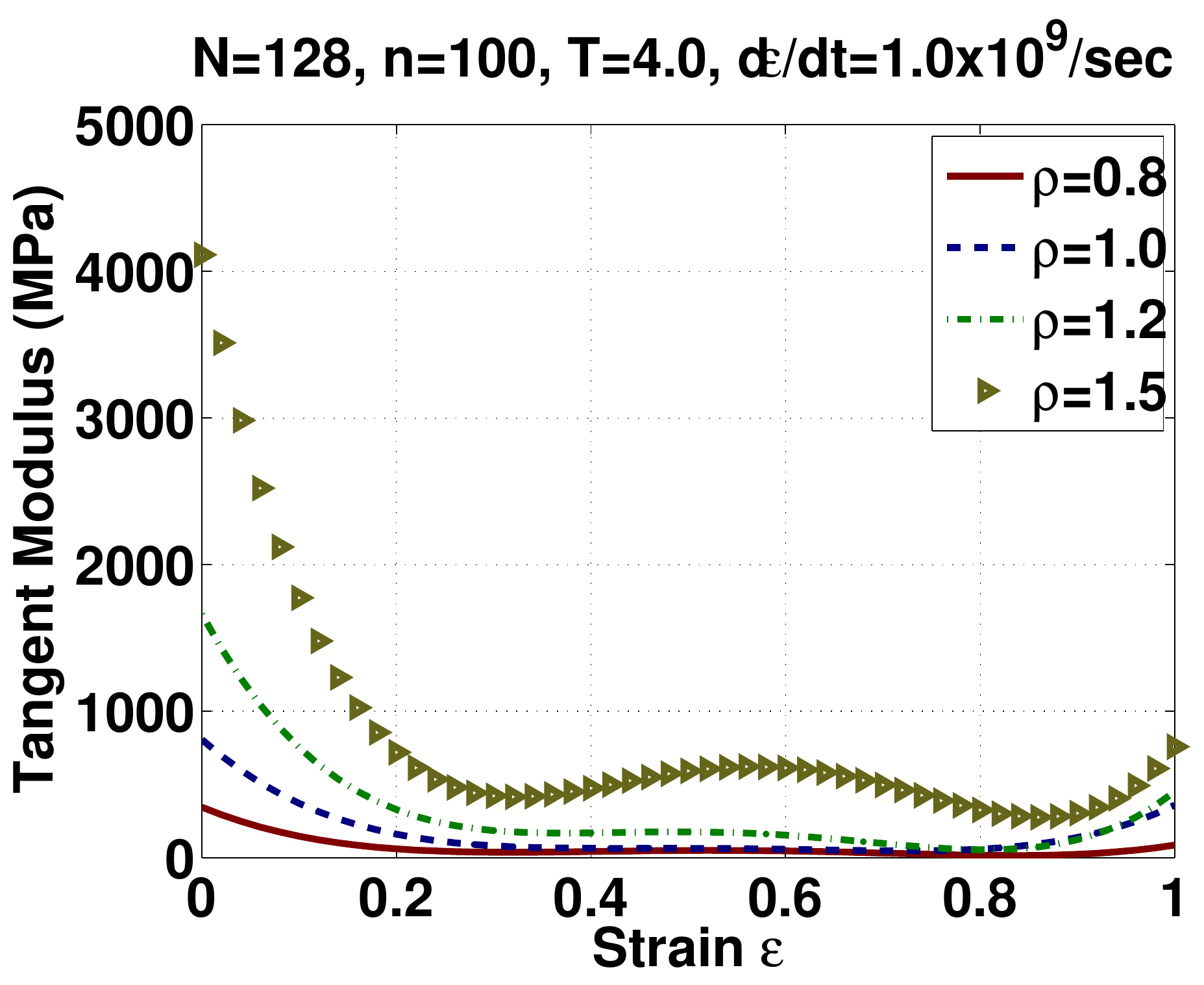}}
\caption{Effect of density on stress response}
\label{fig:EffDensityLinear}
\end{figure}

The effect of density on the stress-strain behavior of the polymer for a constant strain rate is shown in Figure~\ref{fig:EffectDensity}. Higher the density, greater is the stress in the polymer. Also, we find a very huge jump in the stress from $\rho=0.8$ to $\rho=1.5$. In fact, jump in the stress level from $\rho=1.2$ to $\rho=1.5$ itself is very significant. Similar behavior was also observed by \citet{Bower2006}. \citet{Bower2004} report glass transition takes place at $\rho=1.2$ and $T = 2.0$ in which they mention about presence of an additional energy component in the difference stress.  This component increases with increment in density and decrement in temperature. This could be one of the possible reasons for a huge jump in stress from $\rho=1.2$ to $\rho=1.5$. 

Variation of the modulus is shown in Figure~\ref{fig:EffectDensityModulus}. One observes three phases of deformation in the material. During the initial loading phase the polymer with higher density is more stiffer. In the intermediate range of strain values, for the polymer with density $\rho=1.5$, there is an increase in stress and even the modulus increases too. This is due to the fact that for a dense polymer it takes more time to relax. This is not true in the modulus versus strain curves for lower values of strain. This intermediate or plateau region is followed by a third region where the modulus again increases. This is due to the chains being sufficiently uncoiled and the imposed strain resulting in the deformation of bonds that are stiff. Therefore, the stress and the modulus in the polymer starts increasing during this phase.

\begin{figure}
\centering
\includegraphics[width=0.5\textwidth]{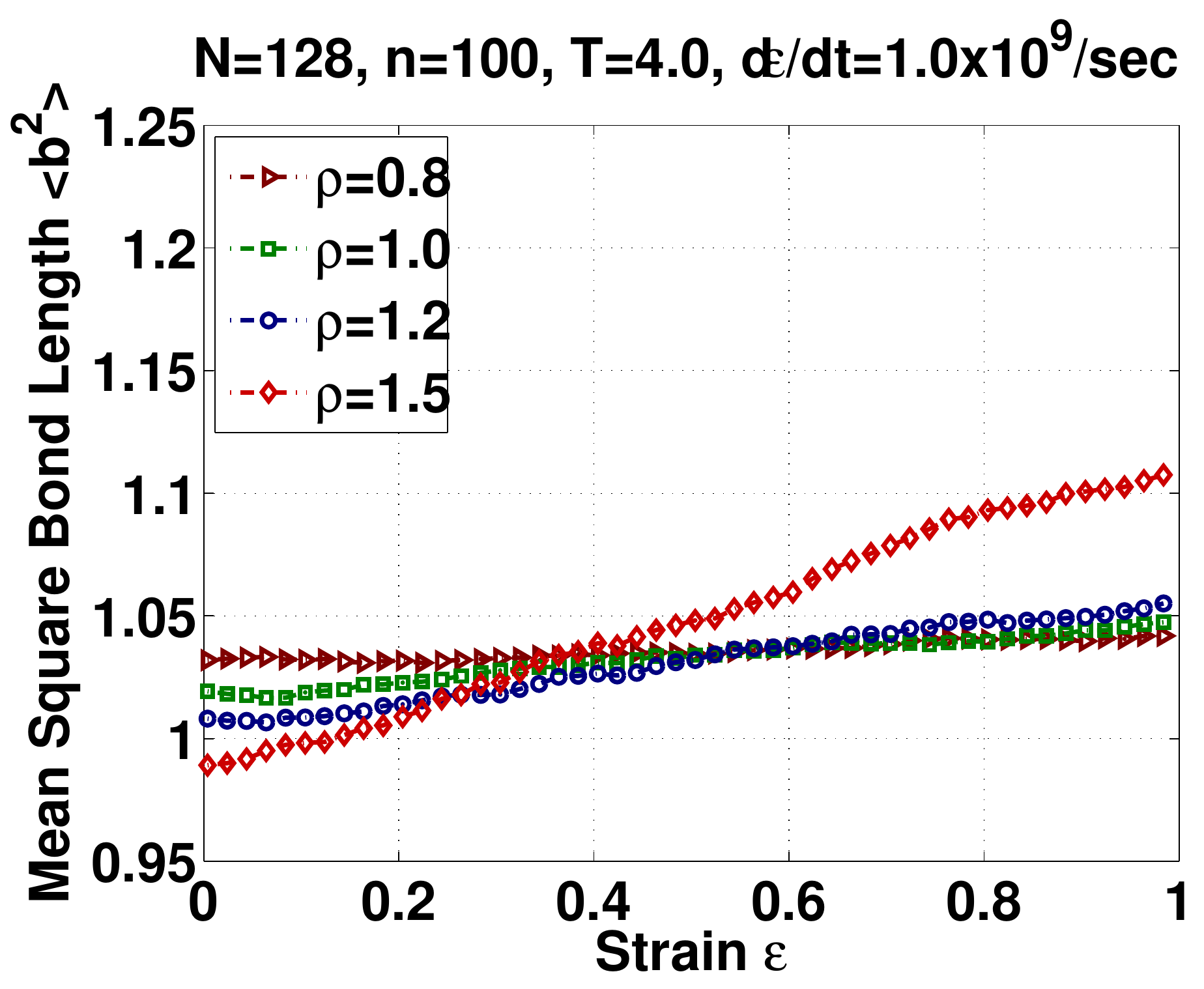}
\caption{Effect of density on mean-square bond length}
\label{fig:EffDenMeanBondLenLin}
\end{figure}

We now study the effect of the density on the mean-square bond length as the polymer is strained. The effect is shown in \ref{fig:EffDenMeanBondLenLin}. There is a critical value of density, $\rho=1.5$, above which mean-square bond length increases with strain. At values of densities below this, the mean-square bond length shows a rather slow increase with strain. This implies that deformation of the bonds at higher densities is more. Also, at high density values, the chain is tightly packed and sliding and uncoiling are not easy because of the constraints from the neighboring molecules. As a result, bond deformation is the only mechanism to provide the appropriate strain in the material.  When there is no strain in the system, we observe that the mean-square bond length takes a lower value at higher densities to give a compact configuration.

\begin{figure}
\centering
\includegraphics[width=0.5\textwidth]{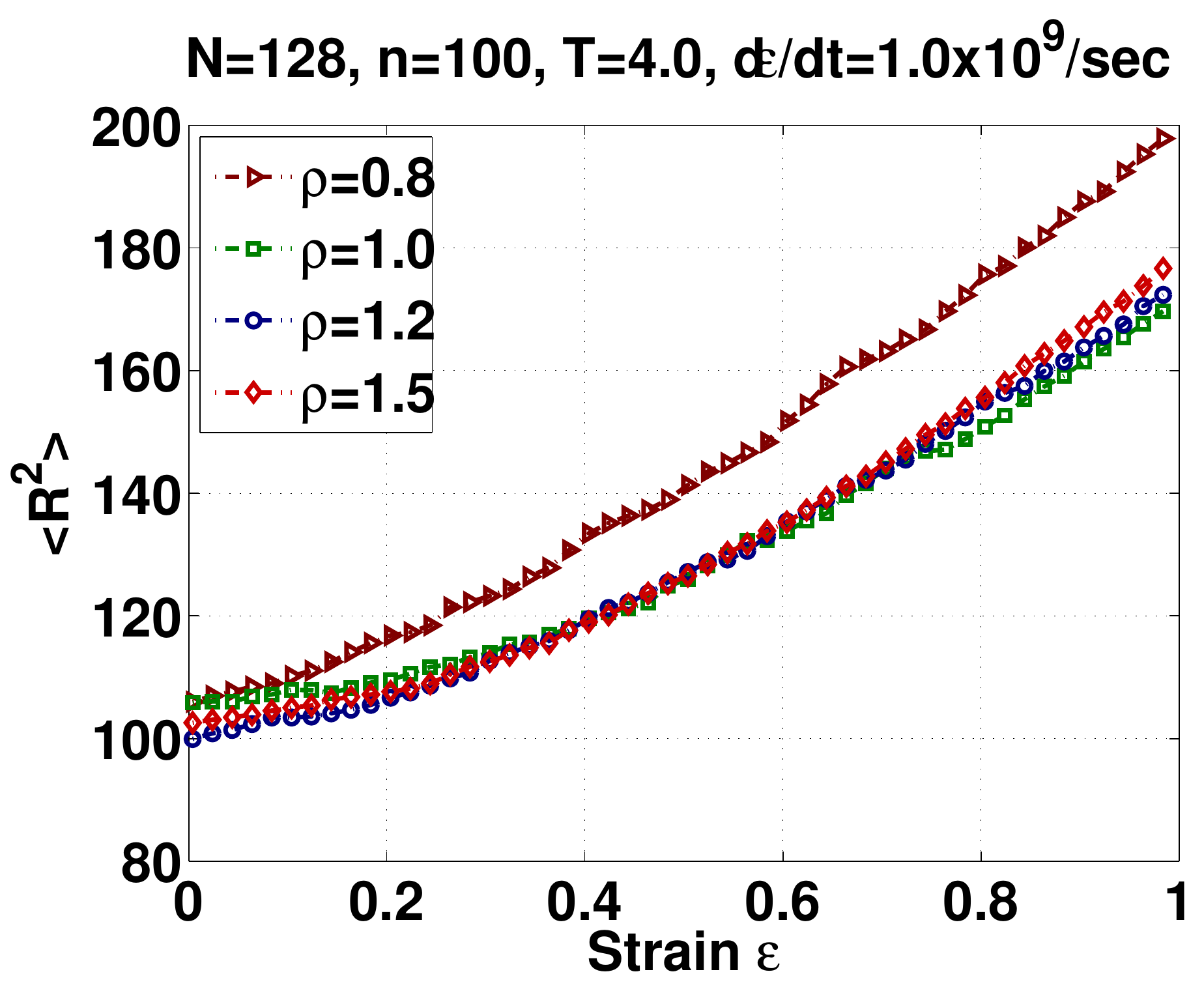}
\caption{Effect of density on end-to-end length}
\label{f15}
\end{figure}

The effect of density on the mean-square end-to-end length was also studied for constant strain rate loading. The results are shown in \ref{f15}. At a lower value of density, the mean-square end-to-end length is larger than that at a higher value of density. This is since at lower densities the chains get enough space to uncoil themselves when subjected to strain. This is not possible at higher densities. Also, we observe there is not much difference in the mean-square end-to-end length beyond $\rho = 1.0$. The only significant difference is observed for a density value of $\rho = 0.8$ in which case the chains have enough space in between them to open up.

\begin{figure}
\centering
\includegraphics[width=0.5\textwidth]{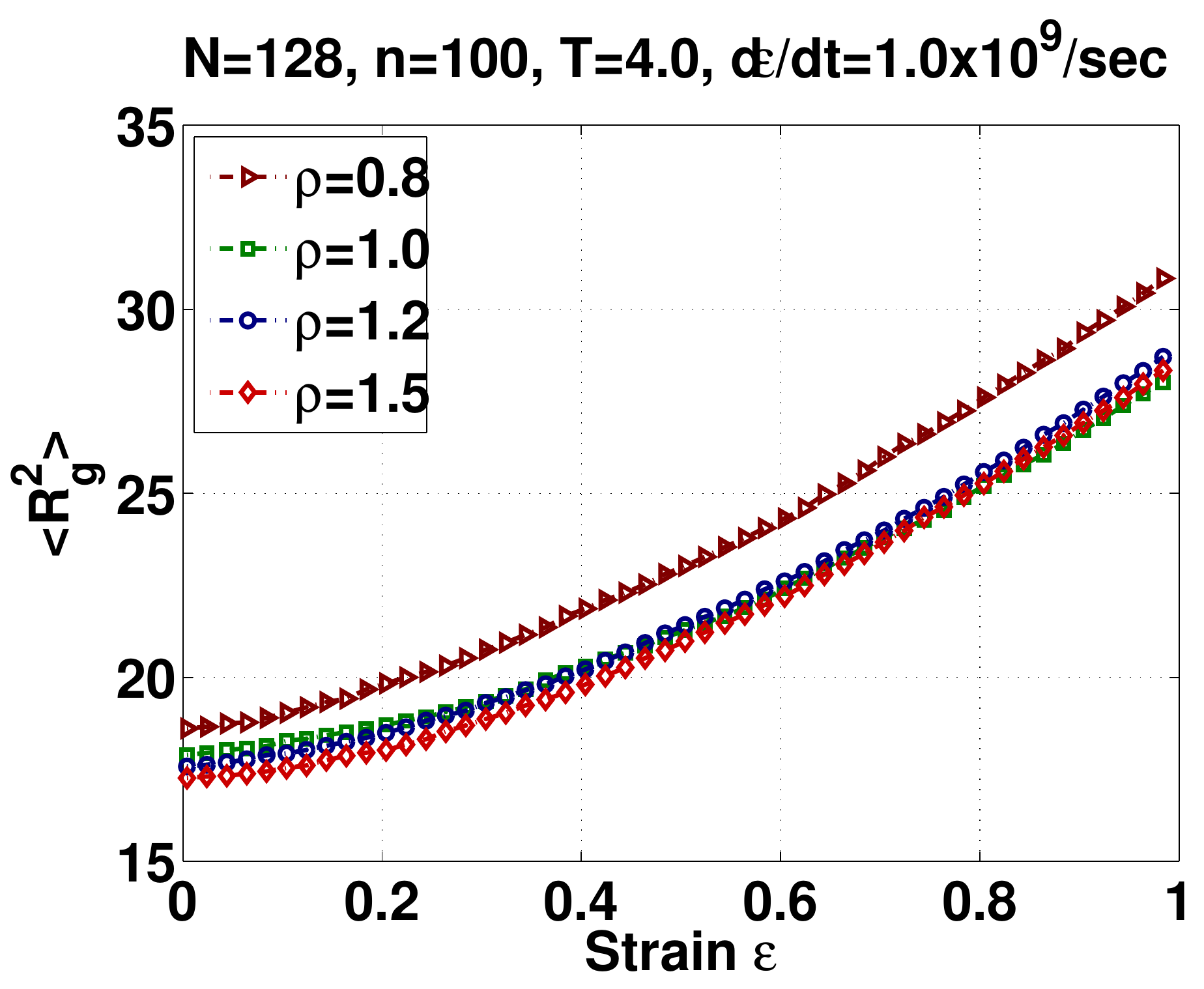}
\caption{Effect of density on radius of gyration}
\label{f19}
\end{figure}

\ref{f19} shows the effect of density on the radius of gyration under the constant strain rate loading. At lower densities, the radius of gyration is larger than that at higher density for the same reason as with the mean-square end-to-end length discussed above. But, this effect is more dominant for density $\rho = 0.8$. For other values of densities, this effect is not very significant. The radius of gyration initially increases slowly with strain but subsequently increases at a faster rate because in this phase the chain uncoiling is predominant.

\begin{figure}
\centering
\subfloat[$g2/g1$]{\label{fig:EffDenG2Lin}\includegraphics[width=0.5\textwidth]{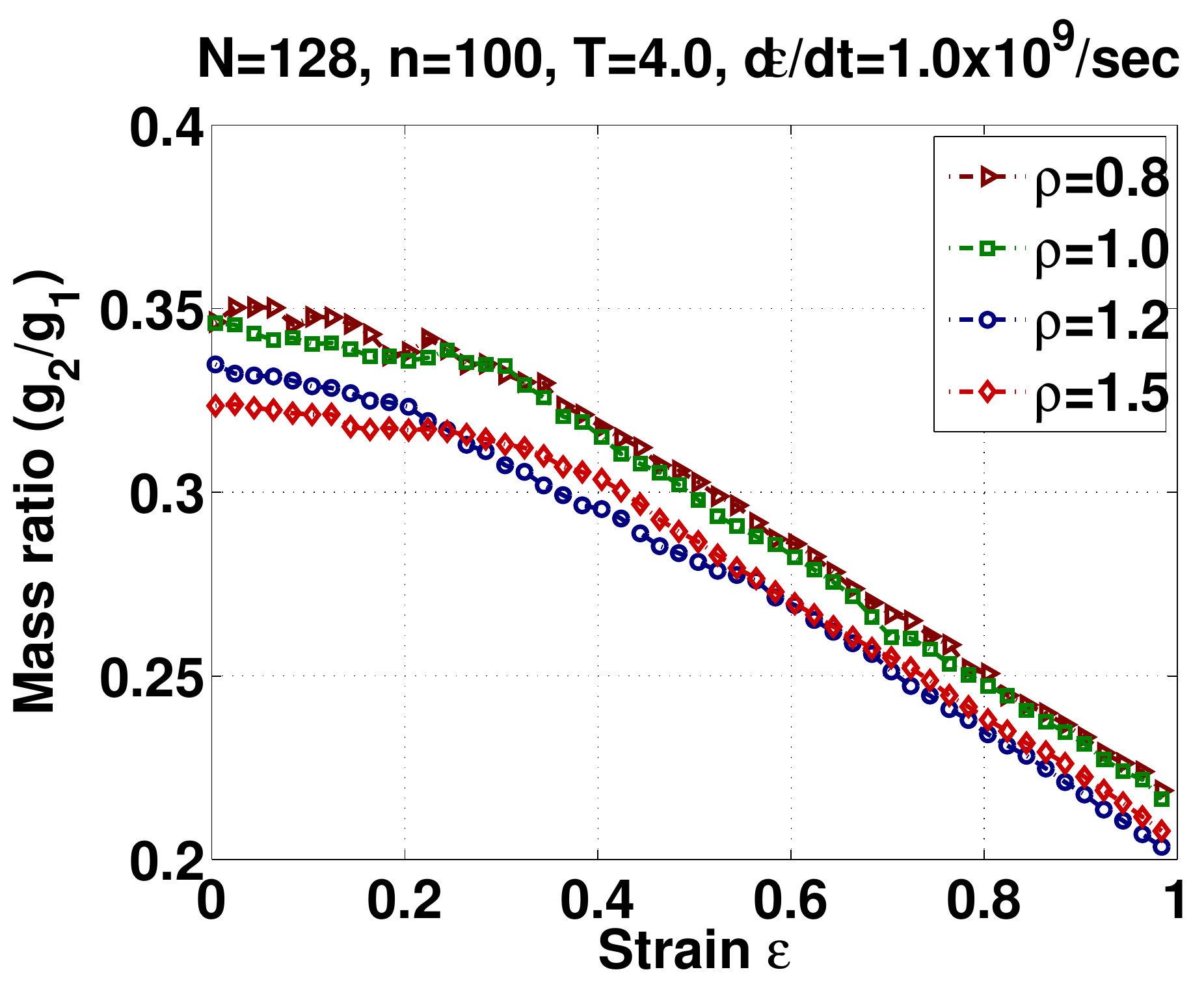}}
\subfloat[$g3/g1$]{\label{fig:EffDenG3Lin}\includegraphics[width=0.5\textwidth]{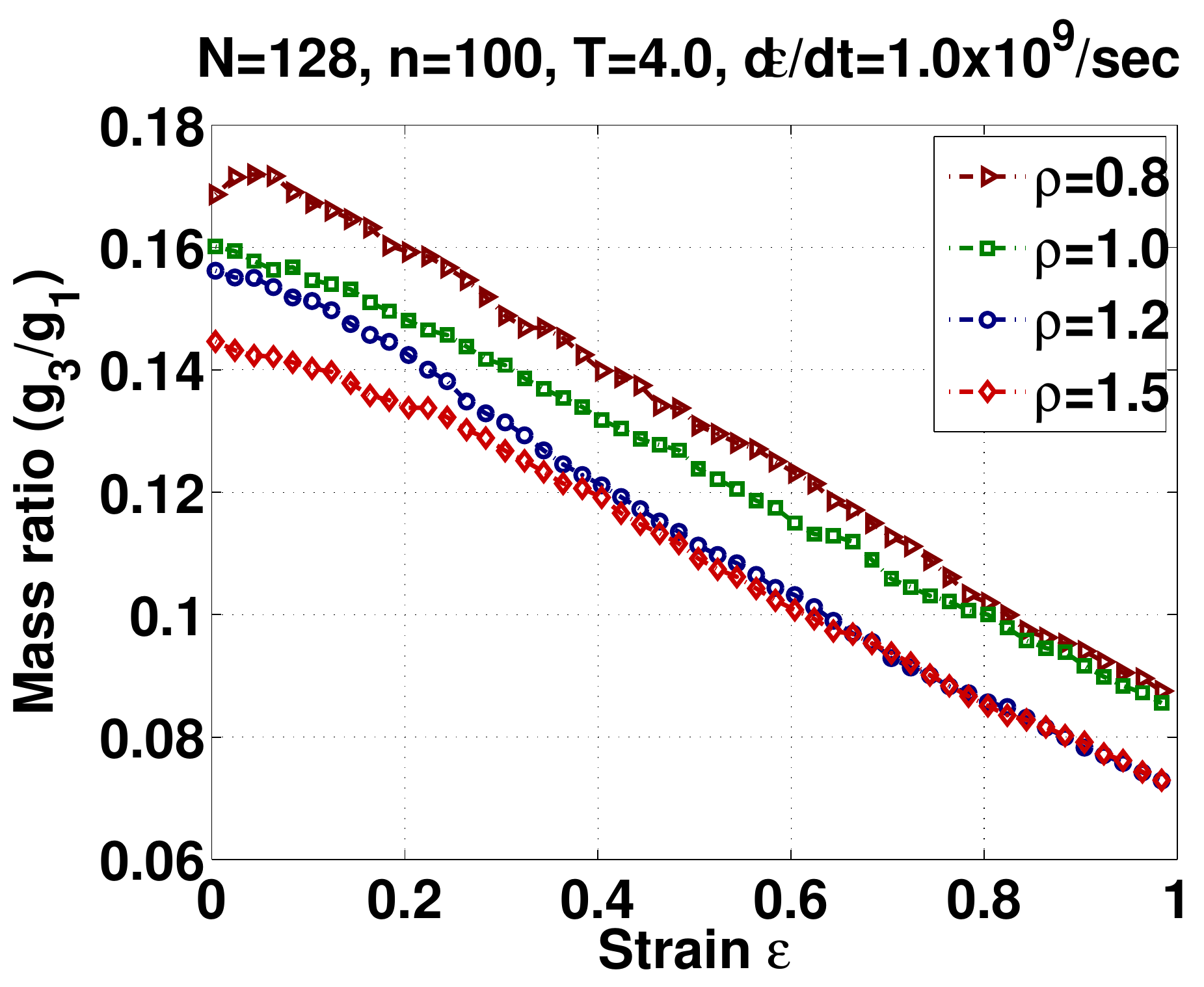}}
\caption{Effect of density on mass ratios}
\label{fig:EffectDenMassRatioLin}
\end{figure}

The variation of the mass ratios under constant strain rate loading at different densities is shown in \ref{fig:EffectDenMassRatioLin}. Higher the density, the initial configurations of the molecules are more flat to provide the compactness. This remains true approximately at all the values of strain during stretching. As the system is stretched, initially the change in the mass ratios is very small but subsequently decreases with constant rate. At a density of $\rho = 1.5$, this transition from a small change in the mass ratios with strain to a change at a constant rate is delayed and we see the cross over of this curve with the $\rho = 1.2$ curve. Mass ratios decrease  with strain due to alignment of the chains in one direction.

\begin{figure}
\centering
\includegraphics[width=0.5\textwidth]{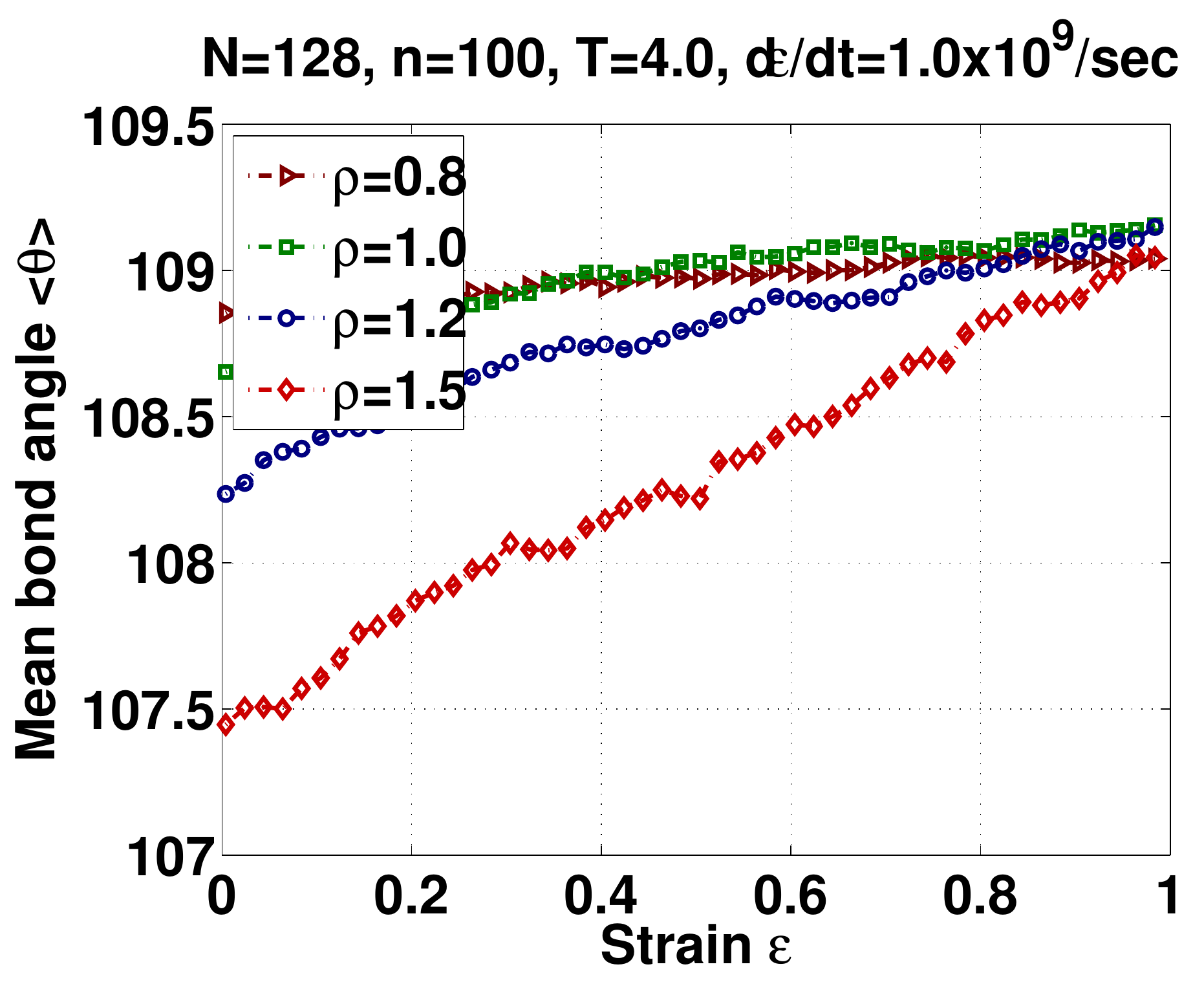}
\caption{Effect of density on mean bond angle}
\label{fig:EffDenBondAngLin}
\end{figure}

The variation of the mean bond angle at different densities for uniaxial constant strain loading is shown in \ref{fig:EffDenBondAngLin}. We find that before the loading is applied, the mean bond angle is lower at higher densities to give a compact configuration. As the system is stretched, mean bond angle increases at all the densities. But, the rate at which the mean bond angle increases is greater at higher values of densities. At low densities, the change in the mean bond angle is very small. Note that irrespective of the density of the polymer, in all the cases, the mean bond angle converges to the same value at high values of strain. 

\begin{figure}
\centering
\includegraphics[width=0.5\textwidth]{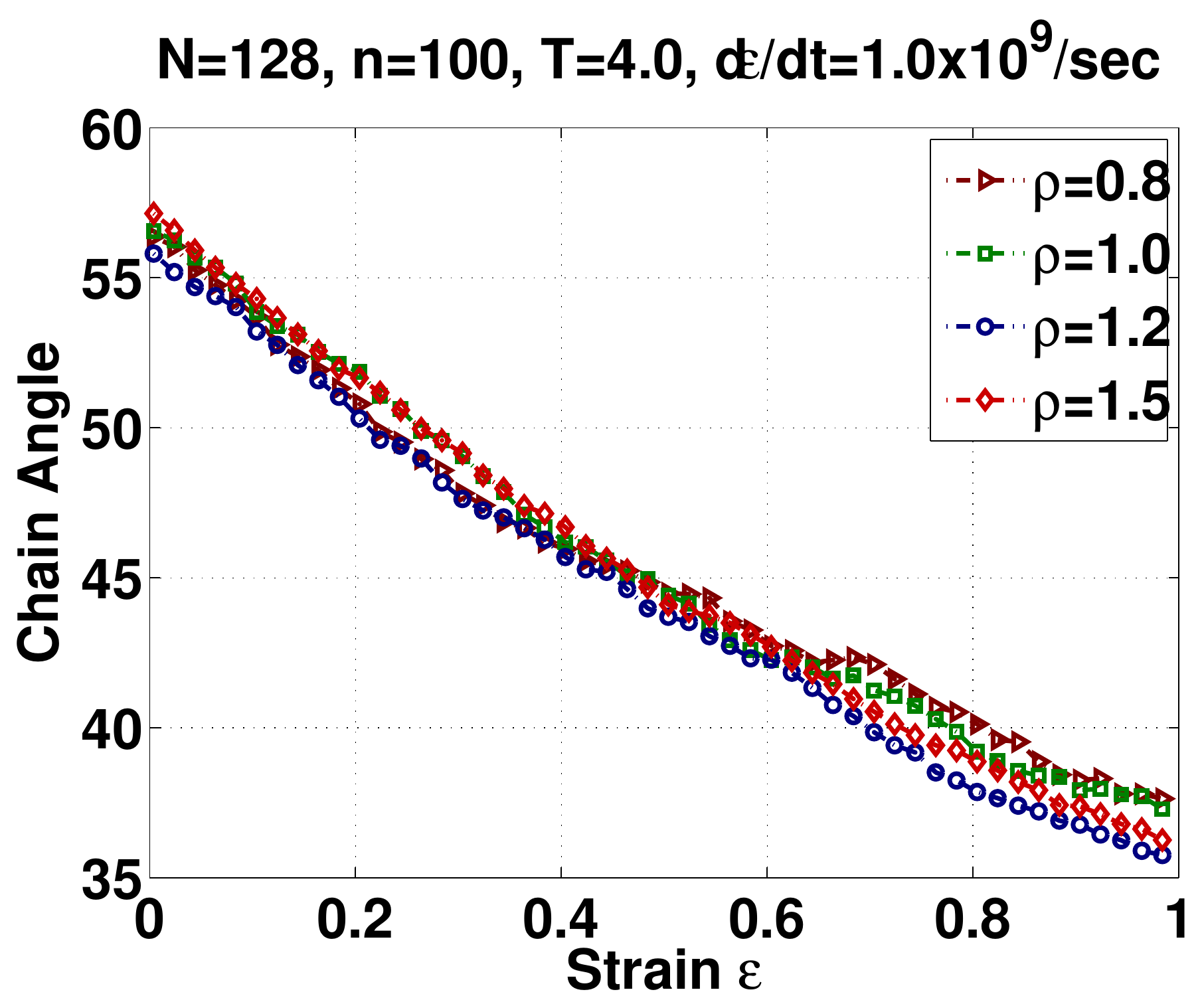}
\caption{Effect of density on mean chain angle}
\label{fig:EffDensChainAngLin}
\end{figure}

The effect of density on mean chain angle for uniaxial constant strain rate loading is shown in \ref{fig:EffDensChainAngLin}.  There is no significant effect of density on the behavior of mean chain angle and it decreases with loading in similar manner at all densities.

\subsection{Temperature}
\label{subsec:control_para_temp}

\begin{figure}
\centering
\subfloat[Stress vs. strain]{\label{fig:EffectTemp}\includegraphics[width=0.5\textwidth]{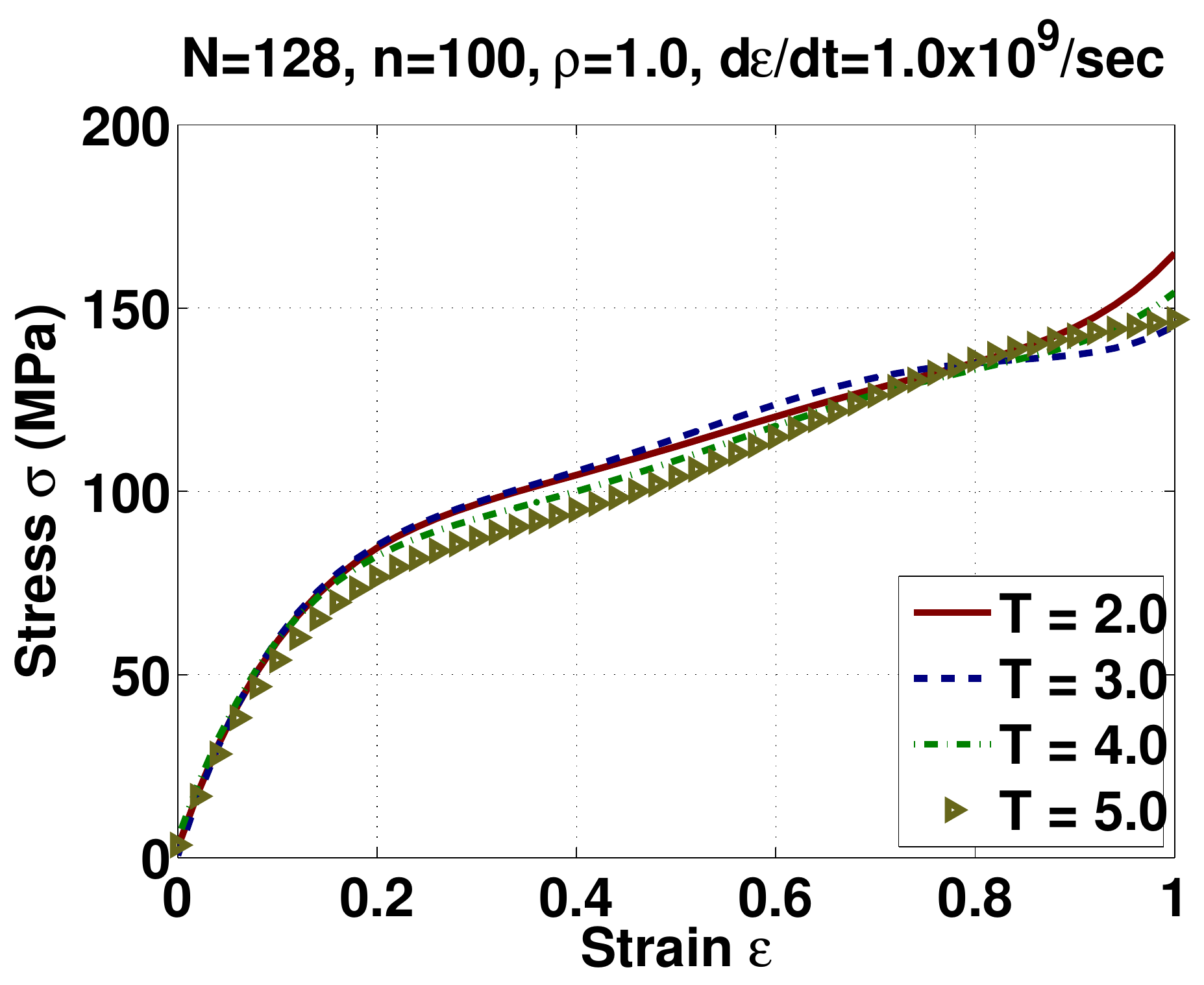}}
  \subfloat[Modulus vs strain]{\label{fig:EffectTempModulus}\includegraphics[width=0.5\textwidth]{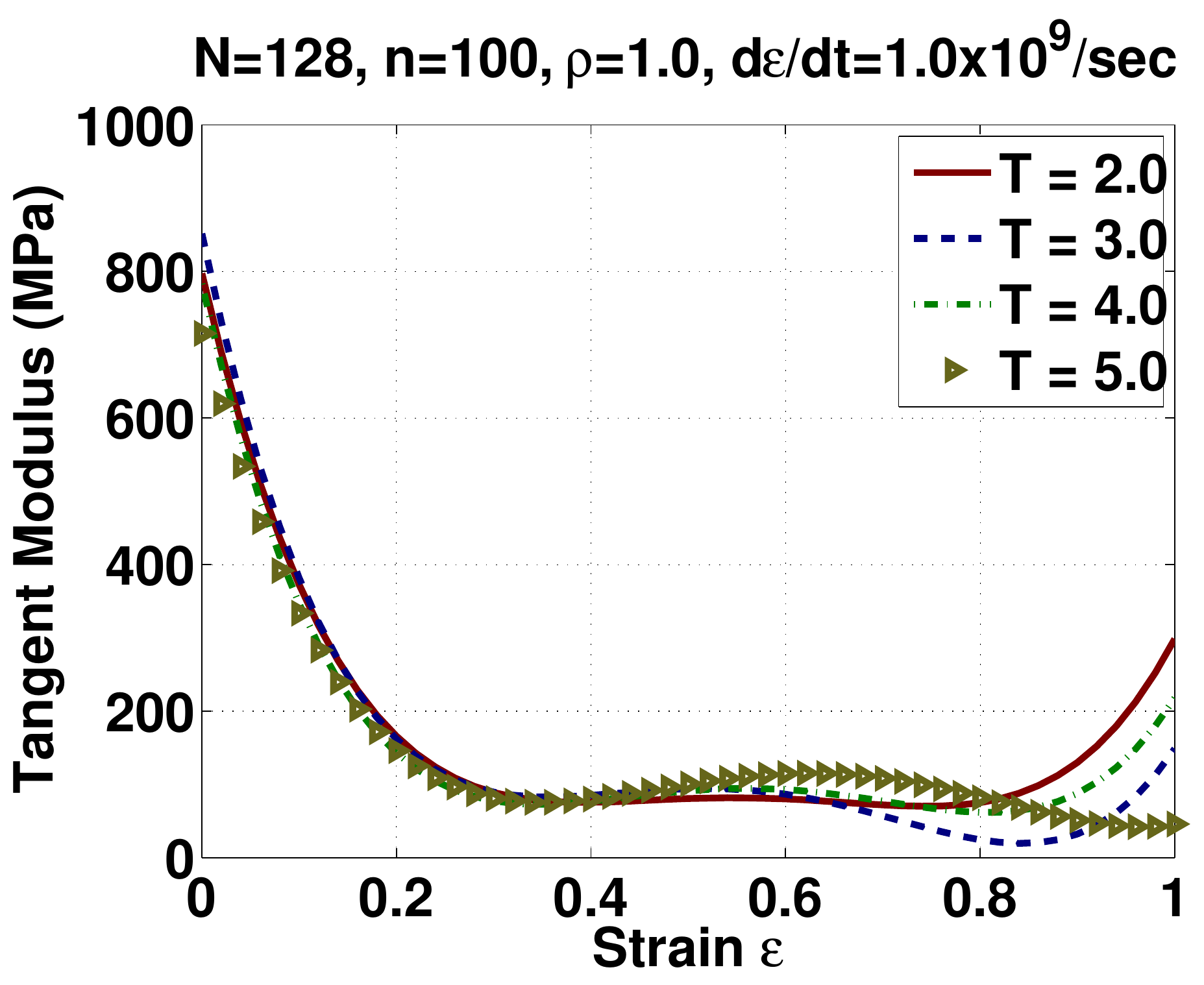}}
\caption{Effect of temperature on stress response}
\label{fig:EffTempLinear}
\end{figure}

\begin{figure}
\centering
\subfloat[$g2/g1$]{\label{fig:EffTempG2Lin}\includegraphics[width=0.5\textwidth]{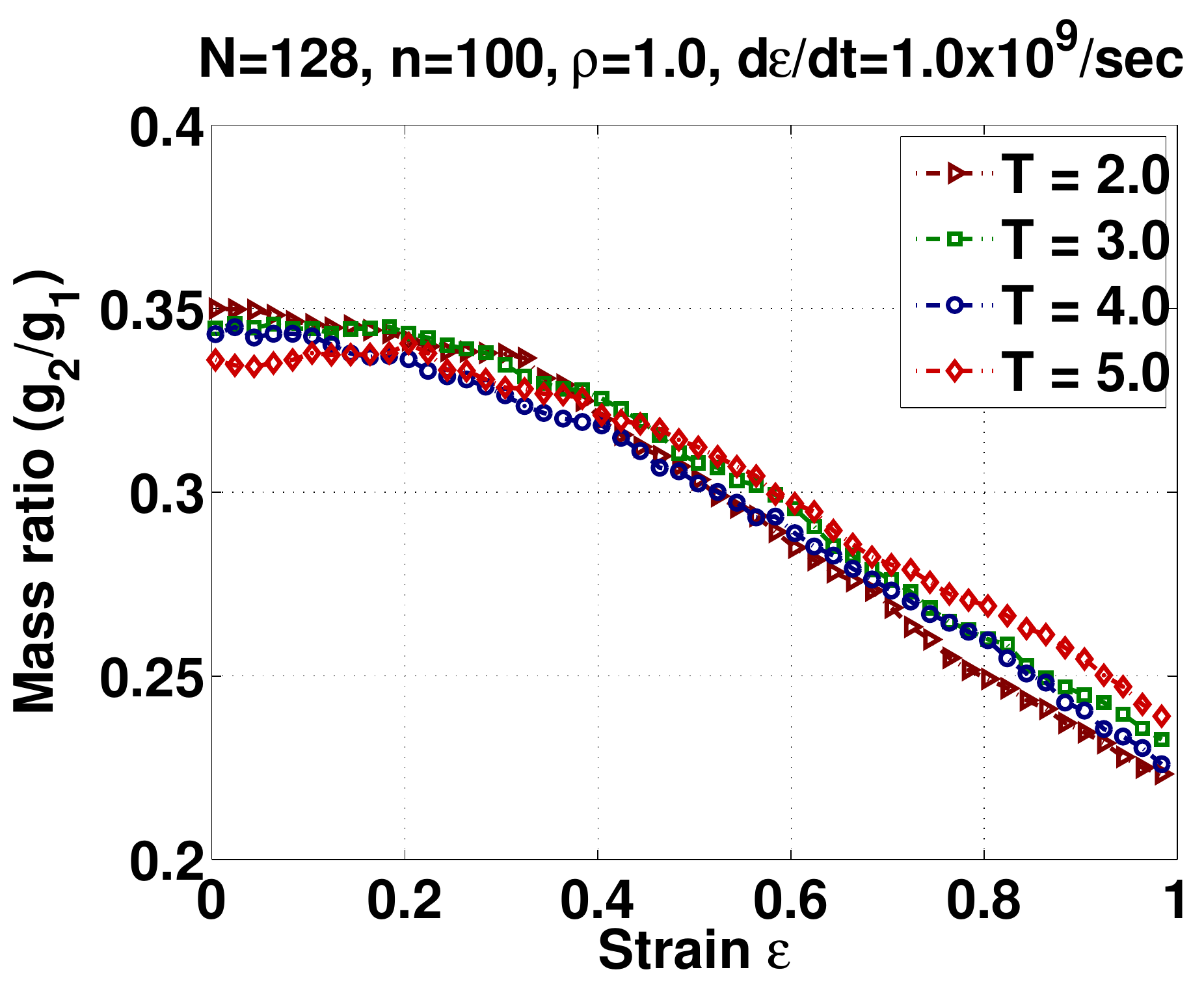}}
\subfloat[$g3/g1$]{\label{fig:EffTempG3Lin}\includegraphics[width=0.5\textwidth]{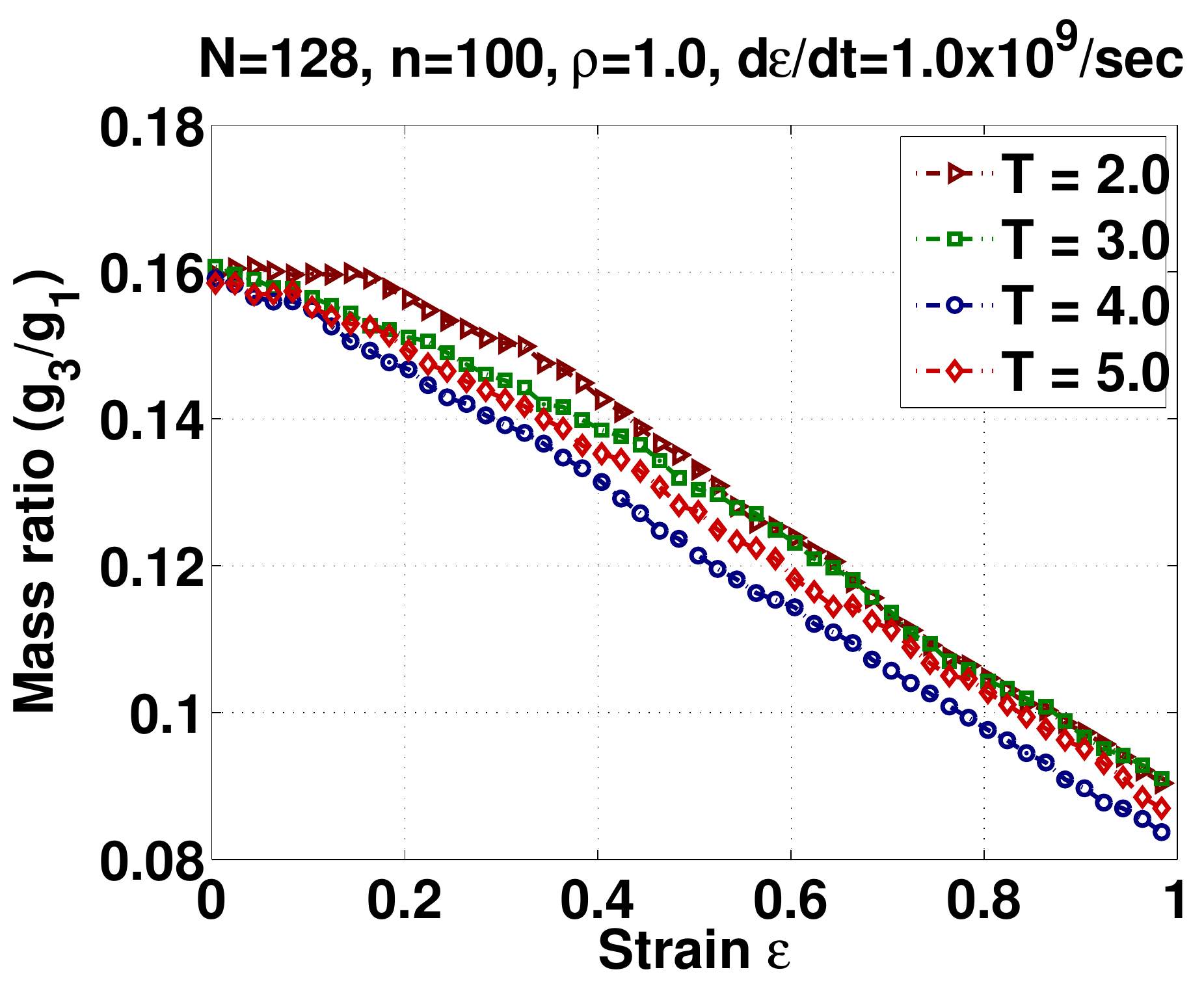}}
\caption{Effect of temperature on mass ratios}
\label{fig:EffectTempMassRatioLin}
\end{figure}

The stress response of the system at different temperatures is shown in Figure~\ref{fig:EffectTemp}. One observes that lower the temperature, higher is the stress in the polymer. This was also observed by \citet{Chui} who performed a Monte-Carlo simulation of a polymer in a compression test. At lower values of temperature, the kinetic energy of the united atoms is very small and hence it takes more time for uncoiling and relaxation. Whereas at higher temperatures, the monomers have high kinetic energy and hence relax faster. Therefore, we find the stress developed in the polymer at higher temperatures is less as it attains a more relaxed structure very soon. Also, at higher temperatures, distribution of the monomers in the chains change from spherical to a less spherical shape slowly under the stretch. In contrast, at low temperatures, this change is slow initially but subsequently becomes faster. This is also corroborated by the mass ratio variation with strain at different temperatures shown in \ref{fig:EffectTempMassRatioLin}. This indicates that at high temperatures anisotropy in the stress is low resulting in lower stress. The modulus curve for different temperatures is shown in Figure~\ref{fig:EffectTempModulus}. This figure clearly shows that at low temperature the polymer exhibits a higher modulus at higher strains but at low strains there is very small difference in modulus.

\begin{figure}
\centering
\includegraphics[width=0.5\textwidth]{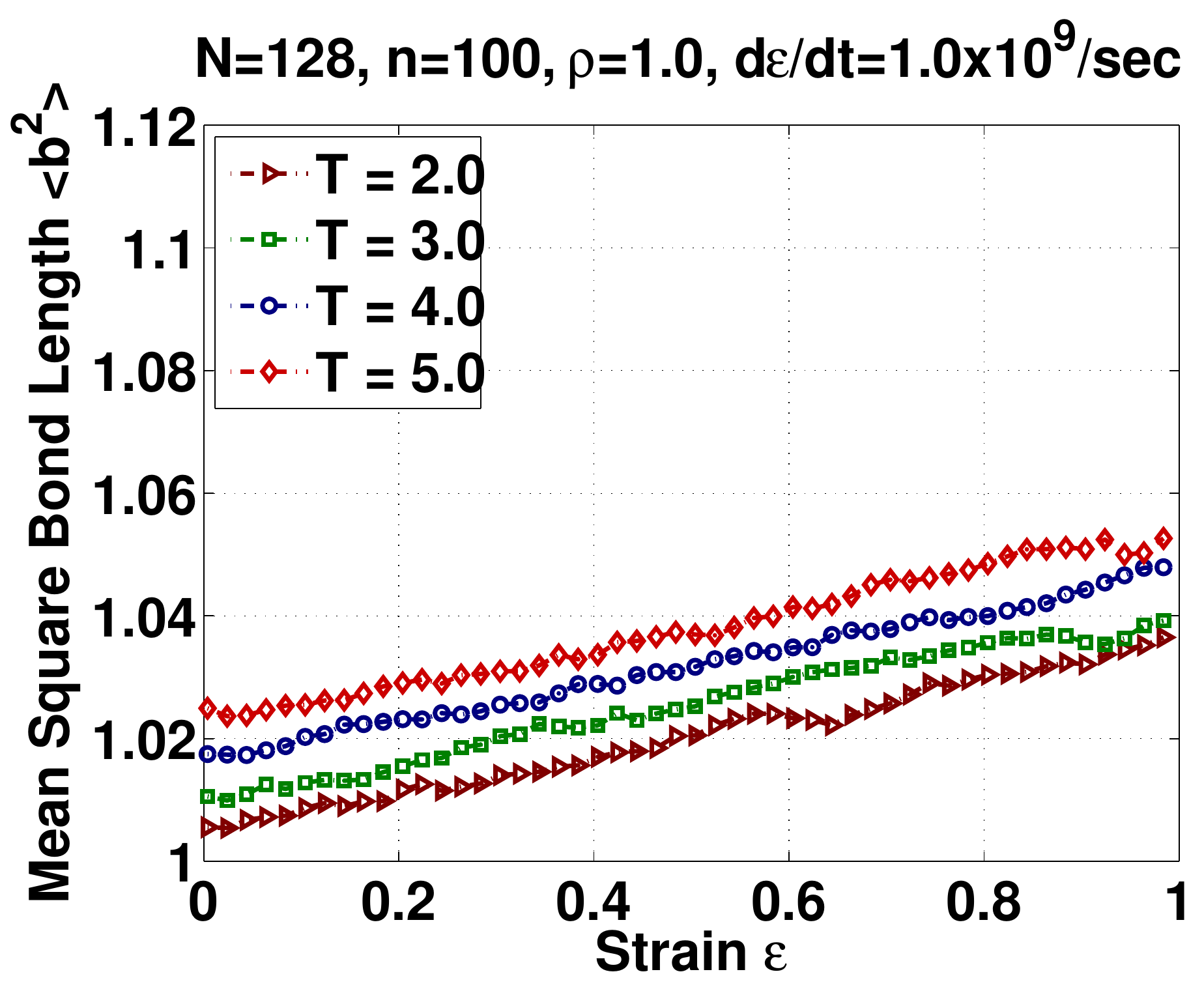}
\caption{Effect of temperature on mean-square bond length}
\label{fig:EffTempMeanBondLenTri}
\end{figure}

We begin our investigation of the influence of temperature on the micro-structure parameters that determine stress response with the effect of temperature on the mean-square bond length. The result is shown in \ref{fig:EffTempMeanBondLenTri}. We find that the mean-square bond length is larger at higher temperatures. Also, we note that as the system is loaded, the mean-square bond length at different temperature values increases with the same rate. 

\begin{figure}. 
\centering
\includegraphics[width=0.5\textwidth]{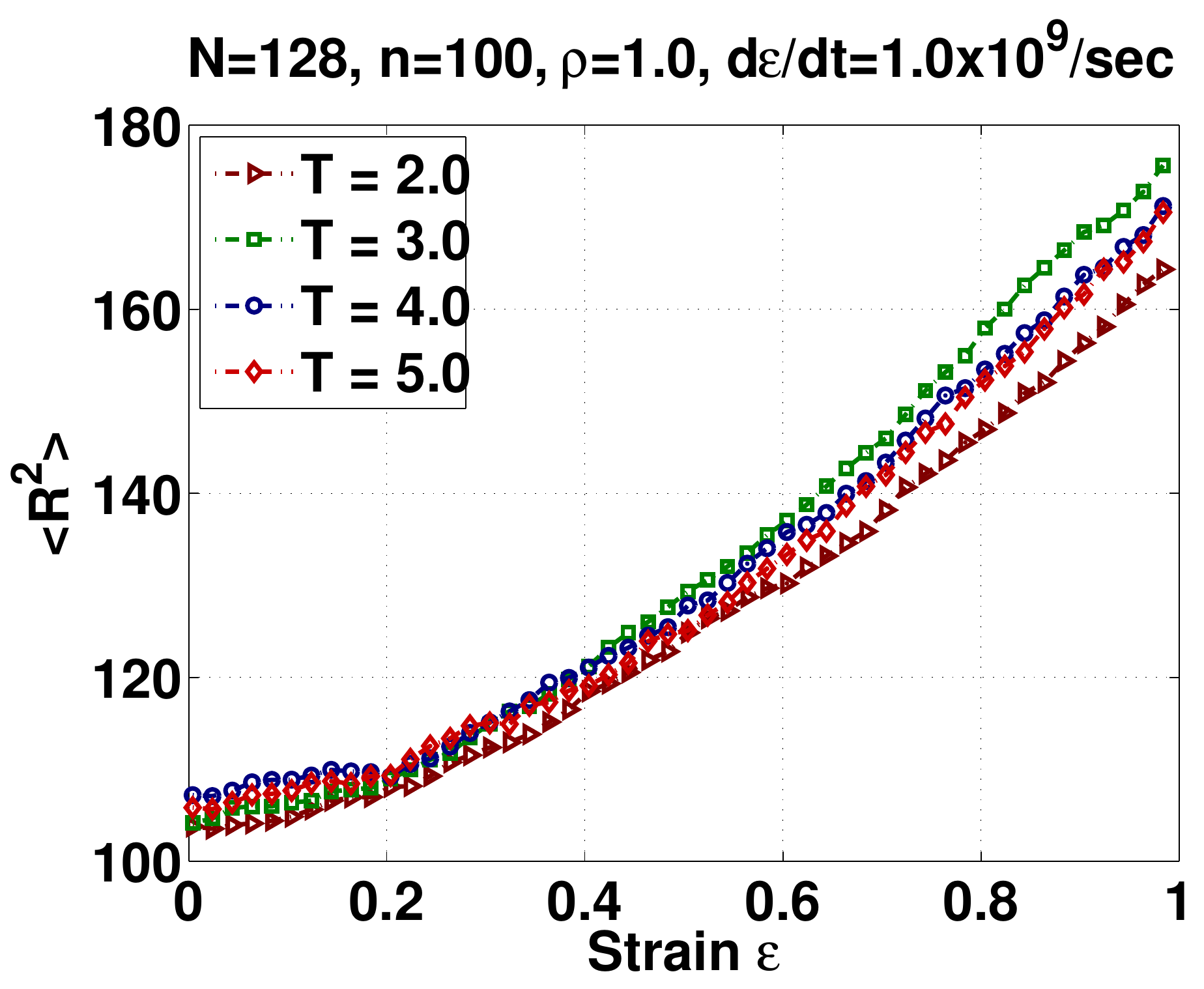}
\caption{Effect of temperature on end-to-end length}
\label{fig:EffTempEndToEnd}
\end{figure}

The effect of temperature on the mean-square end-to-end length under constant strain rate loading is shown in \ref{fig:EffTempEndToEnd}. At higher temperatures the end-to-end length is larger because the kinetic energy dominates over the potential energy and the system relaxes faster. 

However, note that though the mean-square bond length and the mean-square end-to-end length variations with strain show significant change at different temperatures, the axial stress versus strain variation at different temperatures, shown in Figure~\ref{fig:EffectTemp}, does not show any significant change. This is primarily because though the temperature does affect the bond length as well as the end-to-end length significantly, the orientation of the bonds and chains being random, stresses generated are isotropic and therefore increase the pressure in the polymer leaving the anisotropic axial stress largely unaffected.

\begin{figure}
\centering
\includegraphics[width=0.5\textwidth]{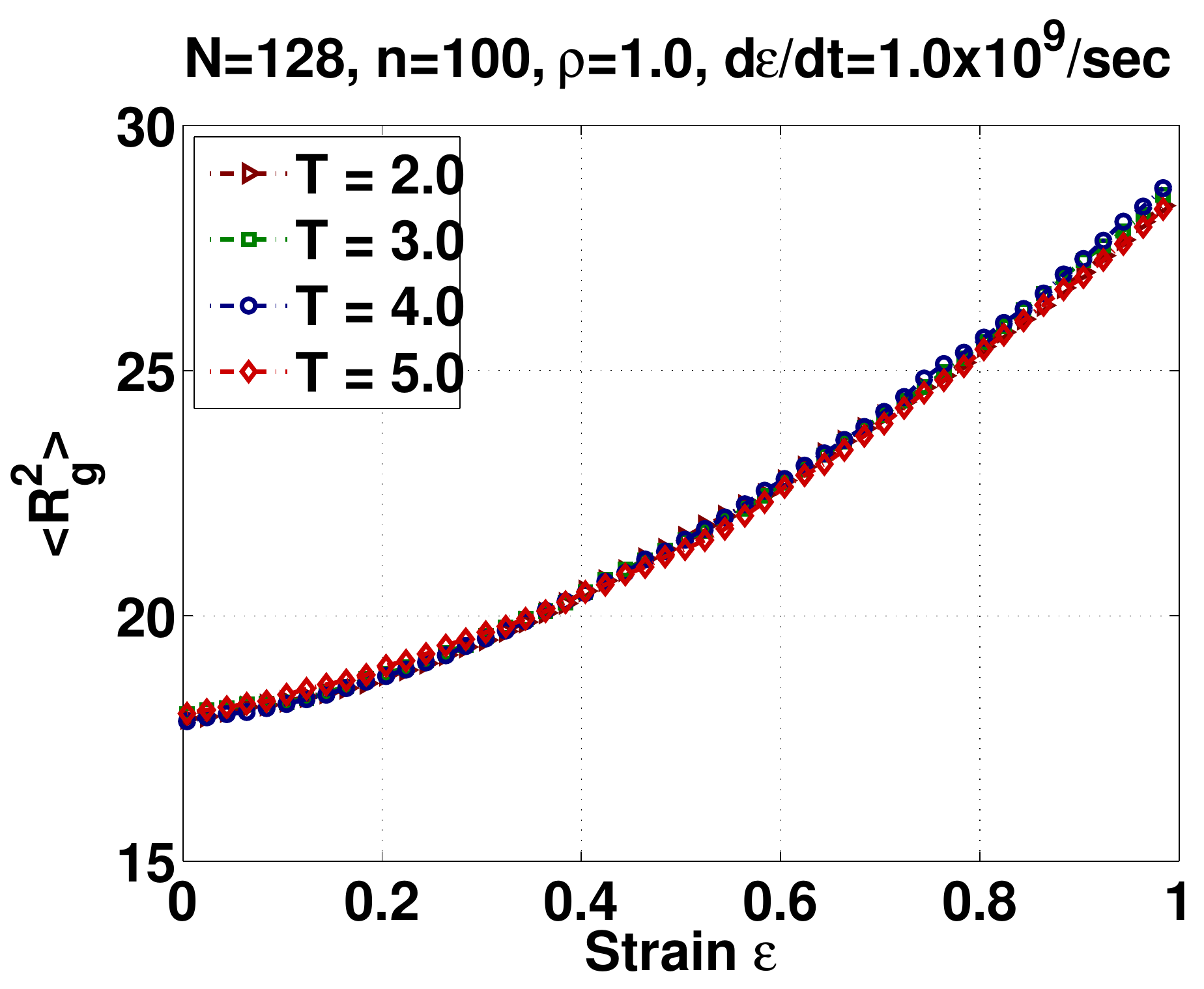}
\caption{Effect of temperature on radius of gyration}
\label{fig:EffTempRadGyrTri}
\end{figure}

The effect of temperature on mean-square radius of gyration in constant strain rate loading is shown in \ref{fig:EffTempRadGyrTri}.  We observe that there is no significant effect of temperature on mean-square radius of gyration and its variation remains similar. This is also reflected in the stress-strain behavior where we do not find any significant variation at different temperatures.

The variation of mass ratio under uniaxial constant strain loading at various temperatures is shown in \ref{fig:EffectTempMassRatioLin}. There is no significant difference in the variation in the mass ratio with temperature except that it decreases under tensile strain at all temperatures. We observe that at high temperatures, chains are more flat initially but as the system gets stretched, initially change in mass ratio $g_2/g_1$ is very slow at higher temperatures and it continues till higher values of strain. Subsequently, mass ratio at all the temperatures decreases at constant rate. As a result we find the cross-over of the $g_2/g_1$ variation with strain at different temperatures. $g_3/g_1$ is lower at higher temperatures, the only exception being at $T=5.0$ where this curve falls between $T=3.0$ and $T=4.0$.

\begin{figure}
\centering
\includegraphics[width=0.5\textwidth]{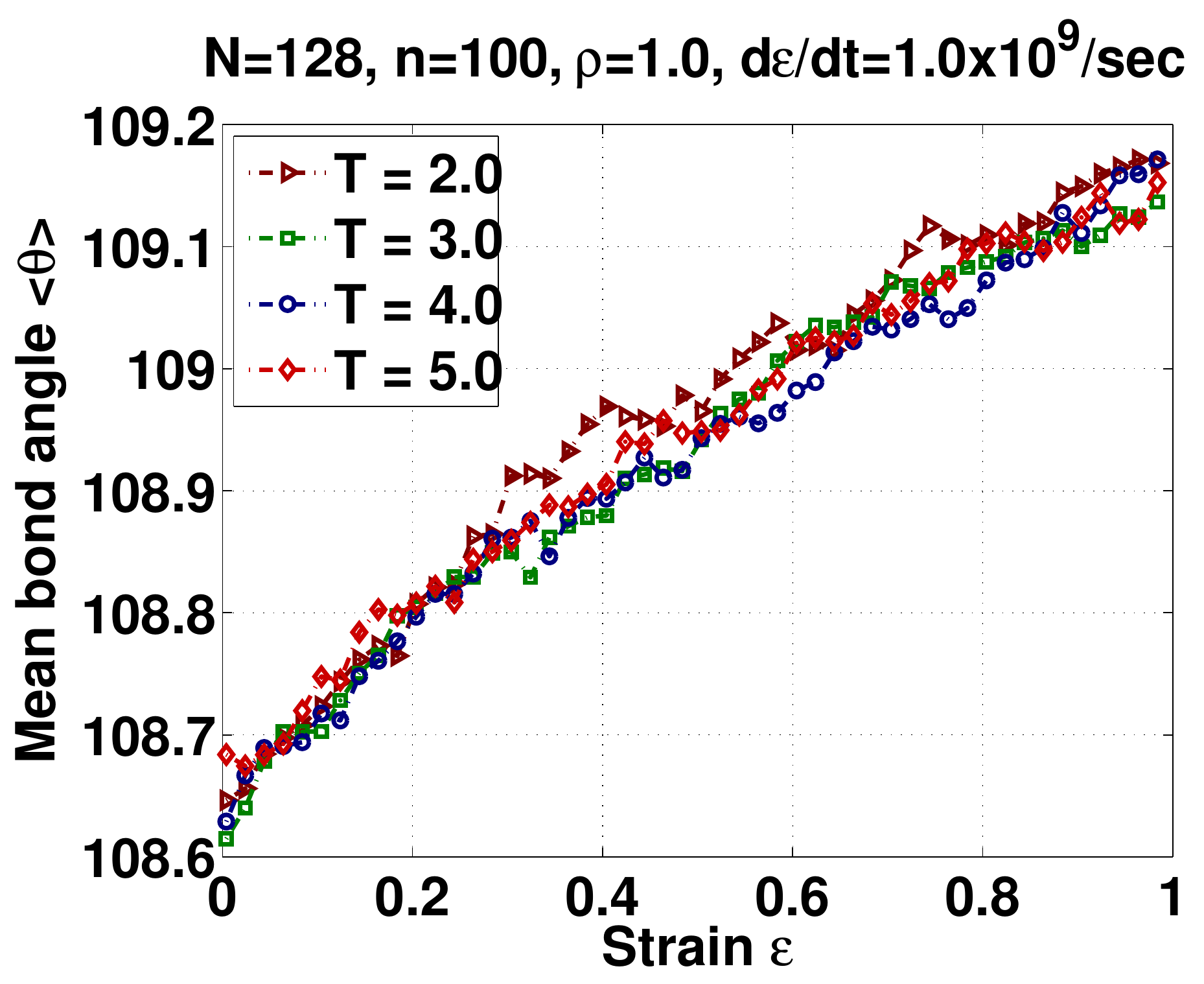}
\caption{Effect of temperature on mean bond angle}
\label{fig:EffTempBondAngLin}
\end{figure}

The effect of temperature on mean bond angle is shown in \ref{fig:EffTempBondAngLin} for uniaxial constant strain rate loading. We observe that the mean bond angle increases with the loading in all the cases, but, there is no significant difference in mean bond angle at various temperatures. 

\begin{figure}
\centering
\includegraphics[width=0.5\textwidth]{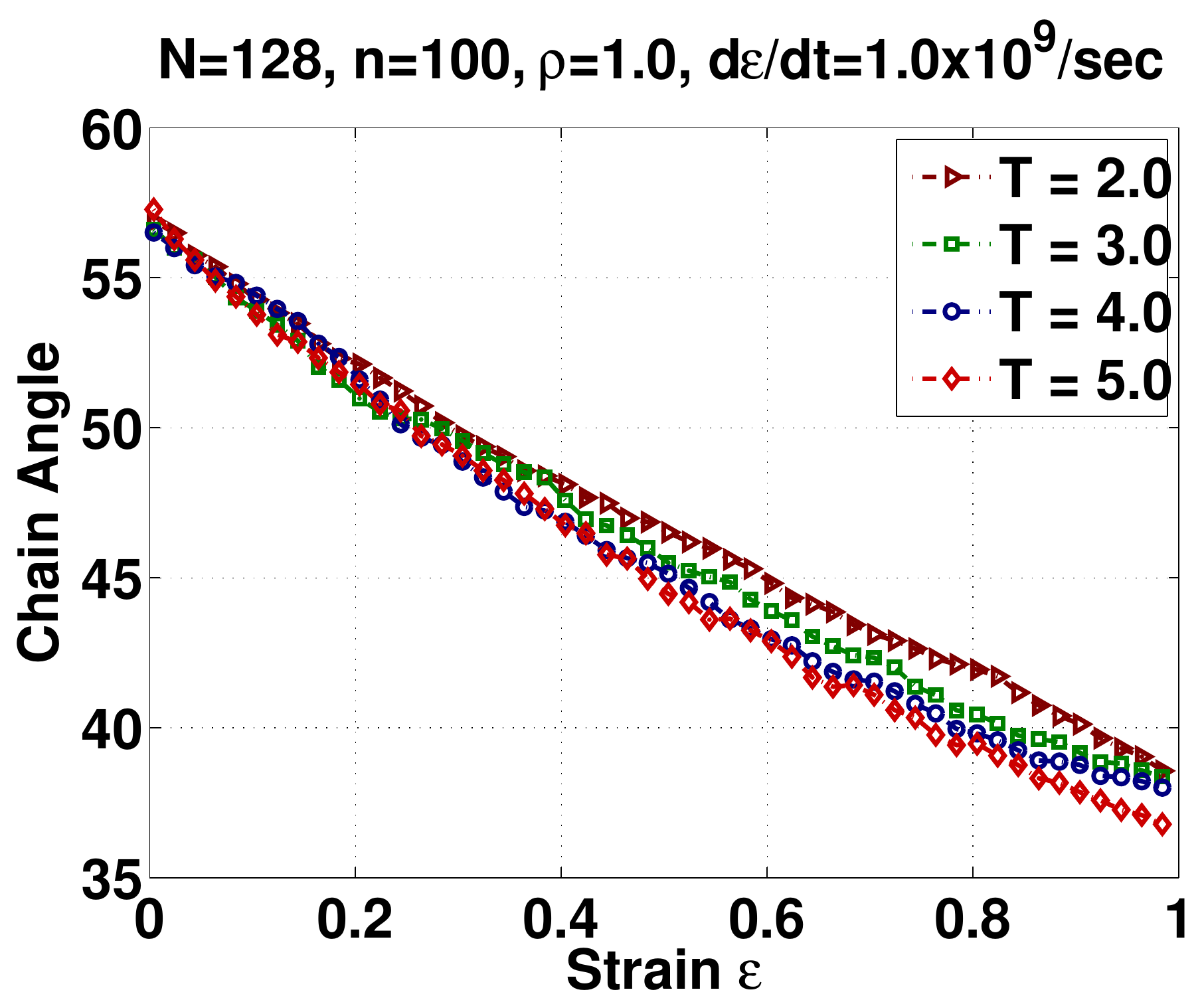}
\caption{Effect of temperature on mean chain angle}
\label{fig:EffTempChainAngLin}
\end{figure}

\ref{fig:EffTempChainAngLin} compares the variation in mean chain angle at various temperatures when the system is subjected to uniaxially constant strain rate load. We find that the mean chain angle again decreases with strain at all temperatures with more alignment observed at higher temperature. But, this change is again not very significant leading us to state that temperature does not affect the behavior of the mean chain angle.

\subsection{Chain length}
\label{subsec:control_para_chainlength}

\begin{figure}
\centering
\subfloat[Stress vs. strain]{\label{fig:EffectChainLenStress}\includegraphics[width=0.5\textwidth]{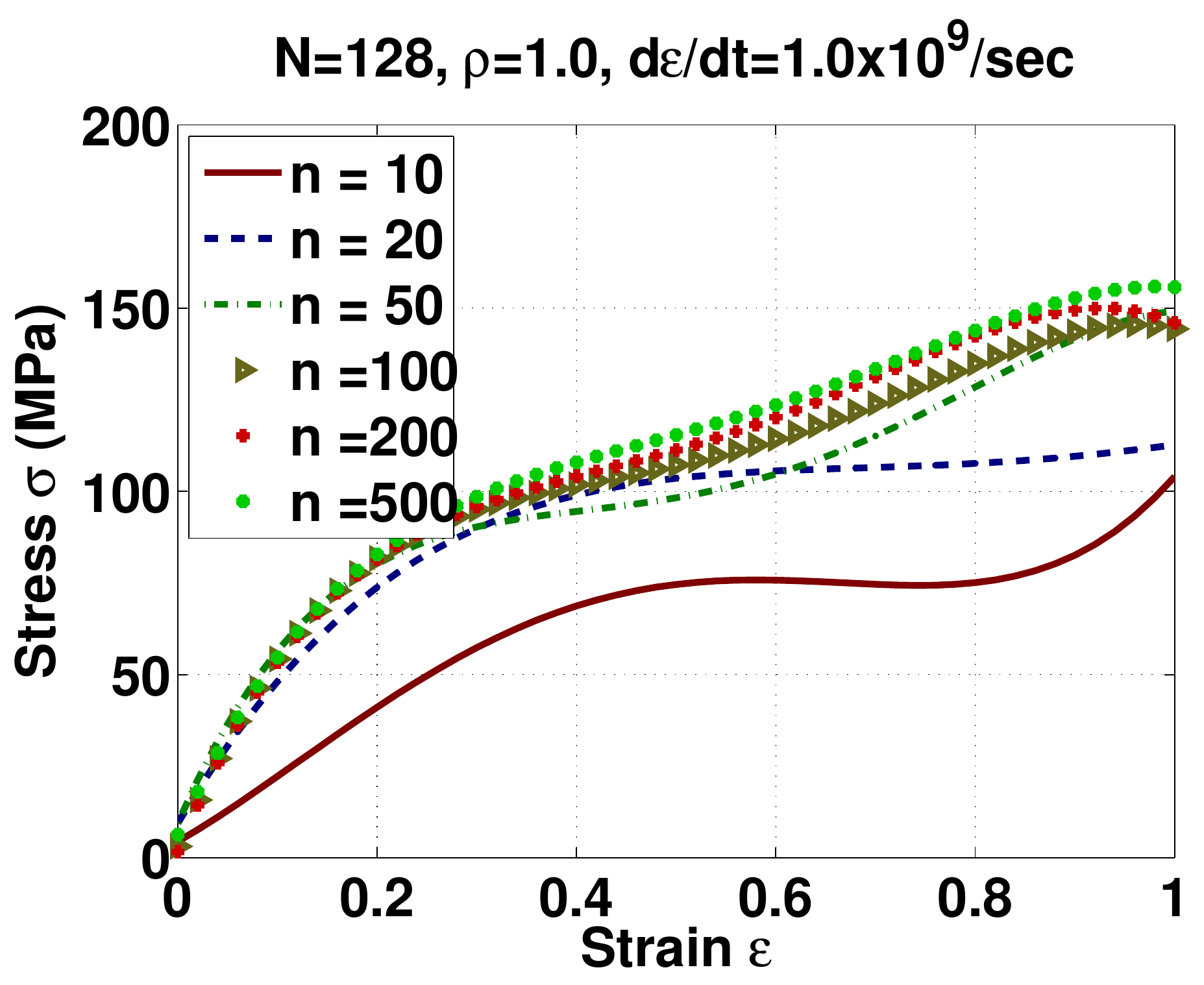}}
\subfloat[Modulus vs strain]{\label{fig:EffectChainLenMod}\includegraphics[width=0.5\textwidth]{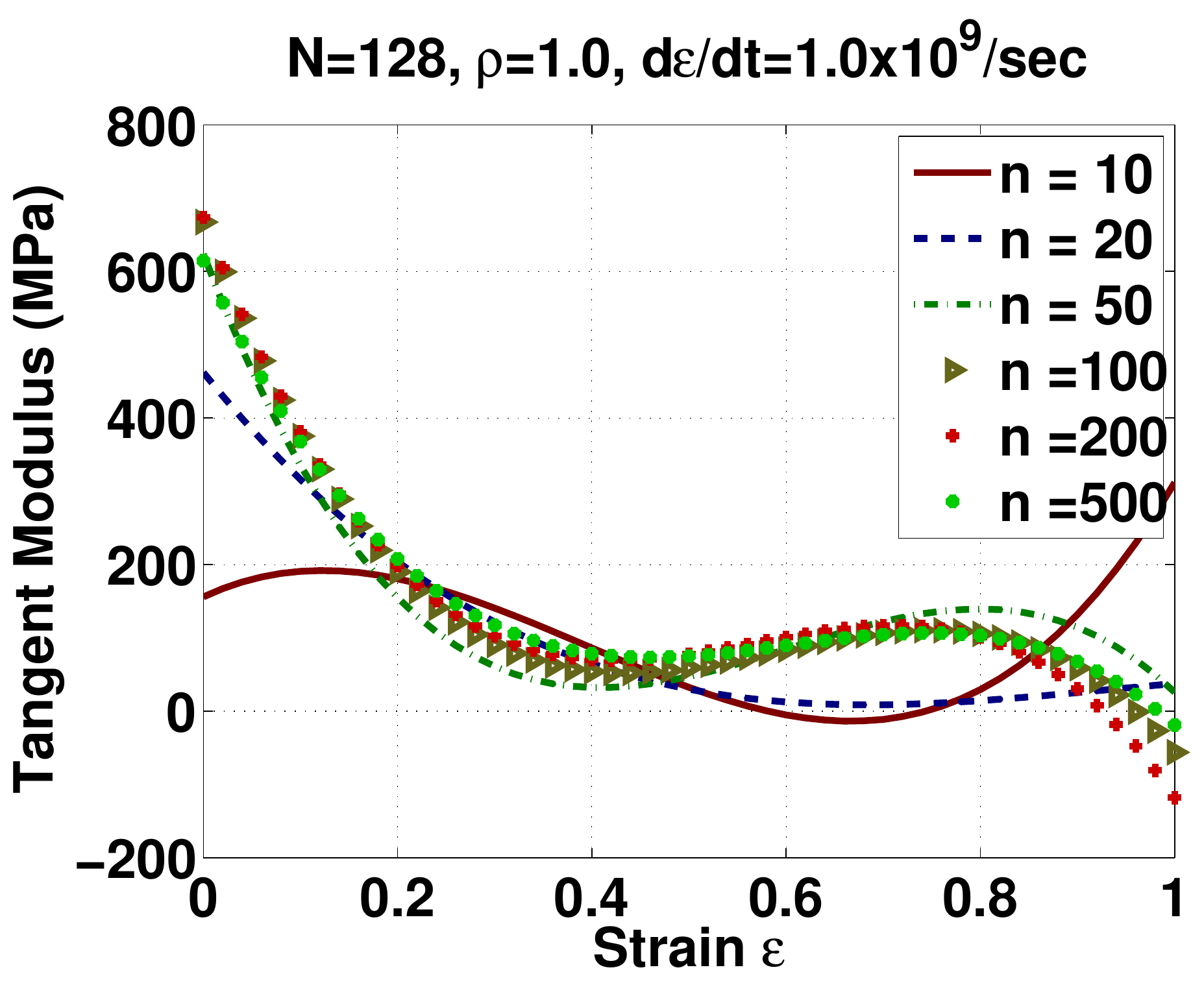}}
\caption{Effect of chain length on stress response}
\label{fig:EffChainLen}
\end{figure}

We consider different chain lengths of the polymer and subject the system to constant strain rate. The chain lengths are varied by varying the number of monomers in a polymer chain. The density is kept constant by controlling the simulation cell dimensions. The stress response is shown in \ref{fig:EffChainLen}. Short length polymers, $n=10$, $20$, and below, show distinctly different stress response. Longer polymers develop more stress and we observe that their modulus or stiffness is greater. A study by \citet{kremer1990} shows that below a chain length $N_e=35$ there is no entanglement in the polymer chain. This length is called the entanglement length. Our observation of increased stress in long chain polymers, that is polymers whose chain length in greater than $n=50$, can be attributed to entanglement effects. We also observe that beyond a particular value of chain length, $n = 100$, there is no significant difference in the stress-strain behavior.

\begin{figure}
\centering
\includegraphics[width=0.5\textwidth]{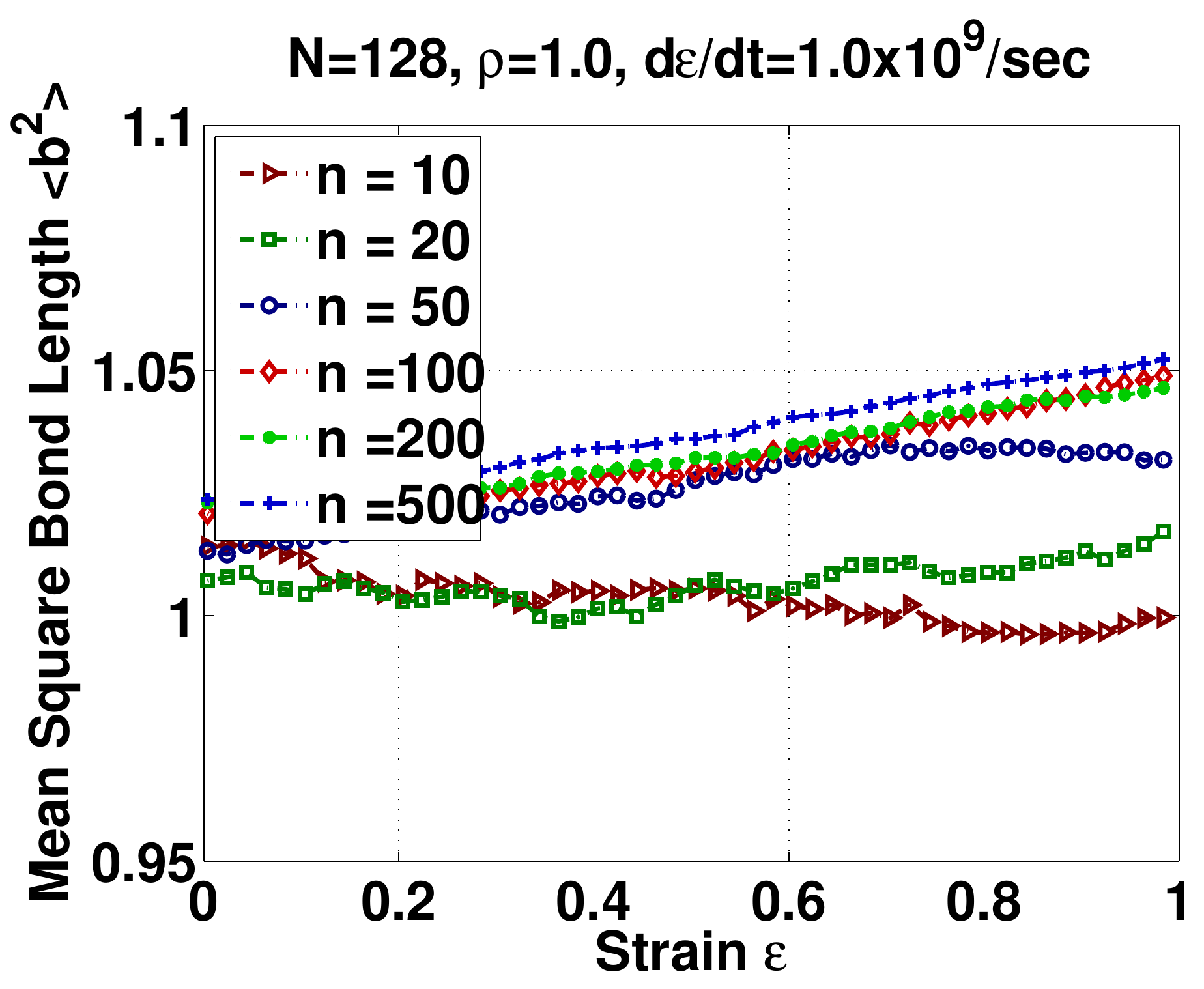}
\caption{Effect of chain length on mean-square bond length}
\label{fig:EffChainLenMeanBondLenLin}
\end{figure}

The effect of chain length on the mean-square bond length is illustrated in \ref{fig:EffChainLenMeanBondLenLin}. When the chains are of smaller length, there is no significant change in the mean-square bond length and the deformation is mostly from the sliding of the chains on one another. Short chain polymers can slide over each other very freely as possibility of entanglement is rare. Also, since the chains are shorter, there is not much scope for uncoiling of chains. Further, we observe that at high strain values, in the range $0.9\le\epsilon\le1.0$, the mean-square bond length decreases for shorter polymers. In fact, this is the main reason for the drastic change in their stress-strain behavior as observed in \ref{fig:EffChainLen}. In the case of long chain polymers, the mean-square bond length grows with strain since chains cannot slide so freely. In such a case, the total elastic deformation has contributions from the partial sliding of the chains, uncoiling, and the deformation due to stretching of the bonds. Also, beyond $n=100$ there is no significant change in the variation of the mean-square bond length. This is one of the reasons for similar stress-strain behavior of longer chain polymers as seen in \ref{fig:EffChainLen}.

\begin{figure}
\centering
\includegraphics[width=0.5\textwidth]{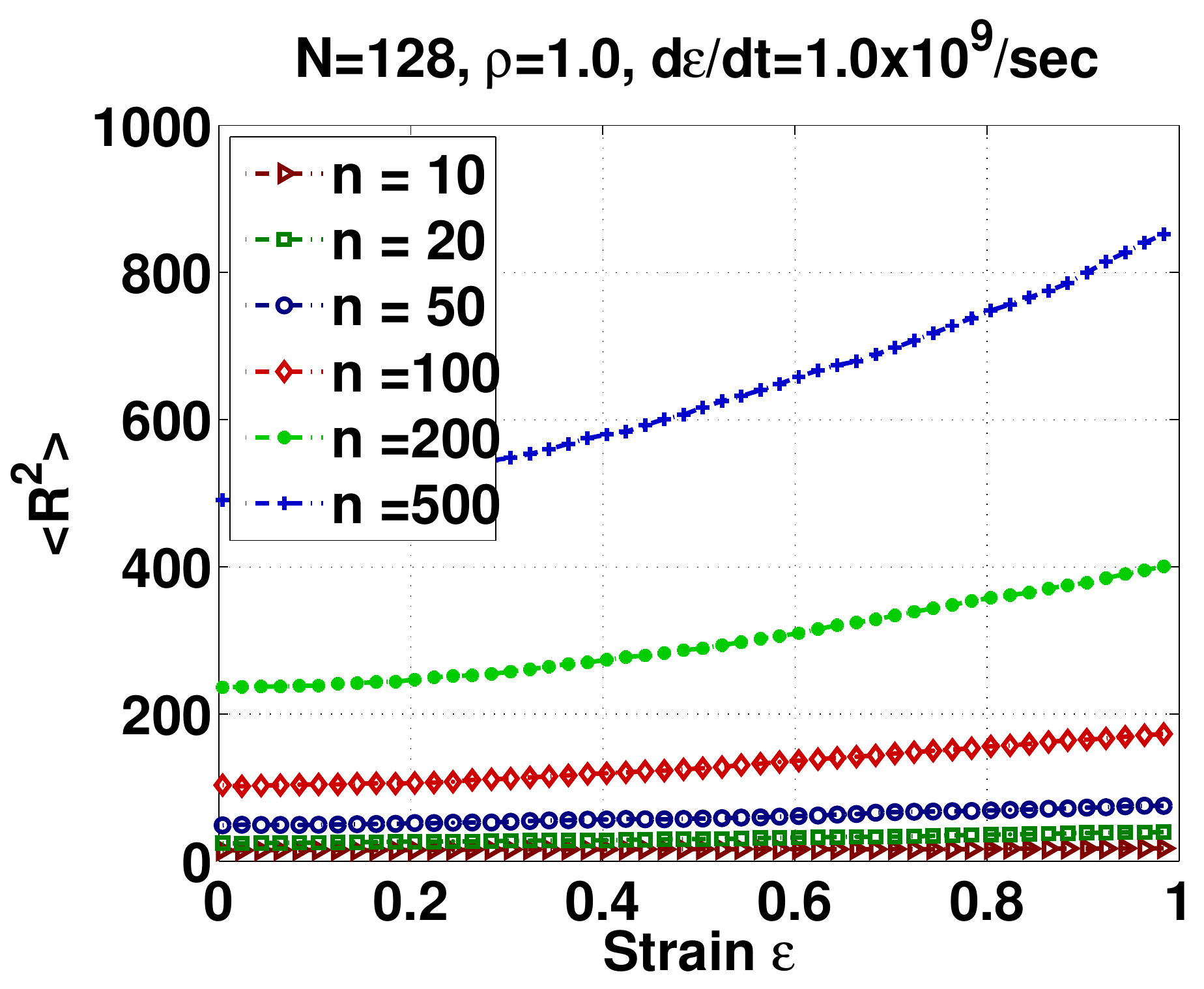}
\caption{Effect of chain length on end-to-end length}
\label{fig:EffChainLenEndToEndLin}
\end{figure}

\ref{fig:EffChainLenEndToEndLin} compares the effect of chain length on the mean-square end-to-end length under constant strain rate loading. The mean-square end-to-end length shows very little variation with strain for chain lengths $n=50$ and below. This is since for small chain lengths there is not much scope of uncoiling. For longer chain lengths, uncoiling is possible and in fact at higher values of strain there is a nonlinear increase in the end-to-end length.

\begin{figure}
\centering
\includegraphics[width=0.5\textwidth]{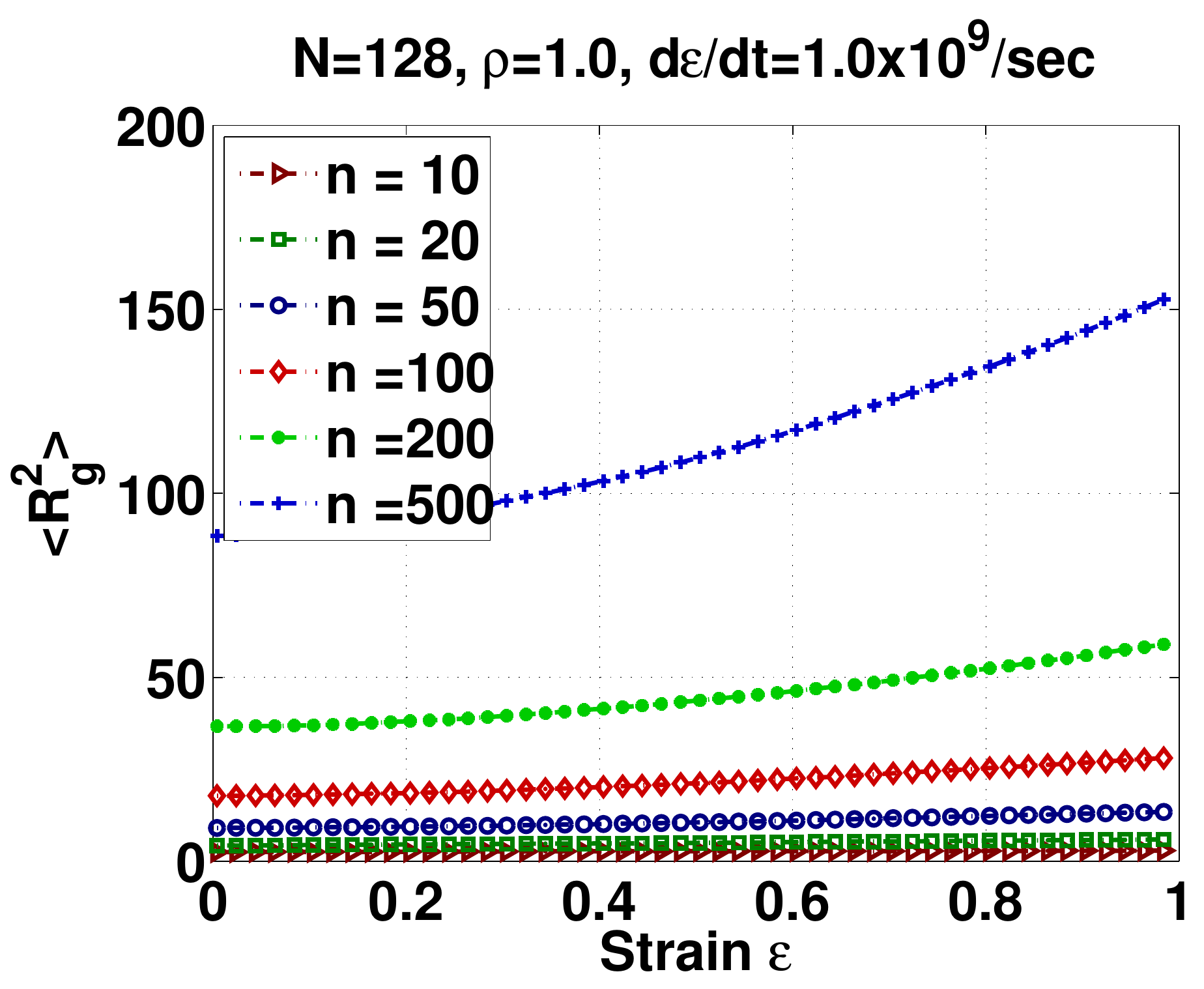}
\caption{Effect of chain length on radius of gyration}
\label{f21}
\end{figure}

The effect of chain length on the radius of gyration is presented in \ref{f21} for constant strain rate loading. For shorter chains, as expected, there is no significant change in the radius of gyration. But, longer chains show a larger increase of radius of gyration with strain and also a increasing rate of change with strain. In short chain polymers, the overall deformation is due to the alignment of the chains in the loading direction. Further, there is slipping over one another rather than actual deformation of the chains.

\begin{figure}
\centering
\subfloat[$g2/g1$]{\label{fig:EffChainLenG2Lin}\includegraphics[width=0.5\textwidth]{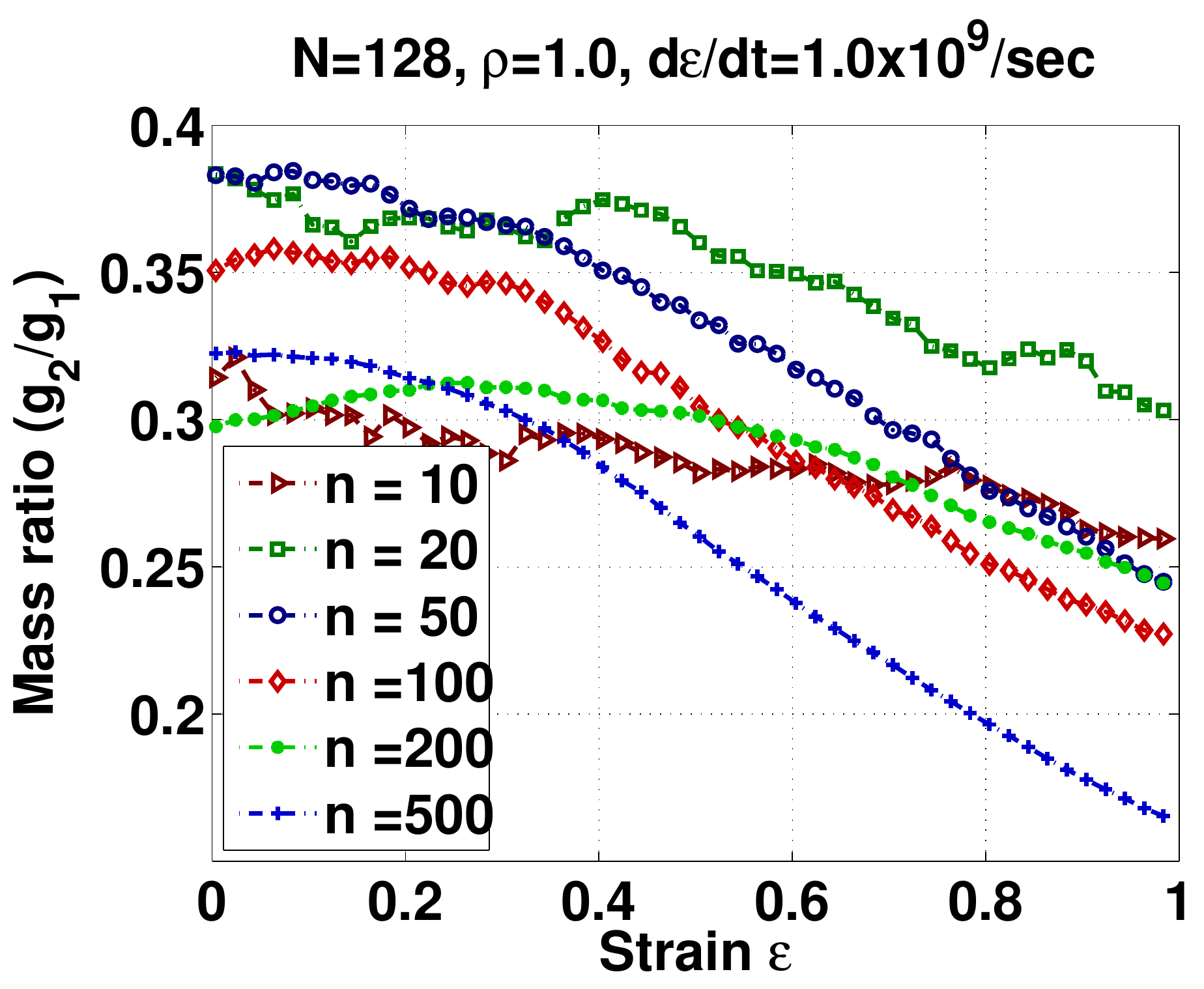}}
\subfloat[$g3/g1$]{\label{fig:EffChainLenG3Lin}\includegraphics[width=0.5\textwidth]{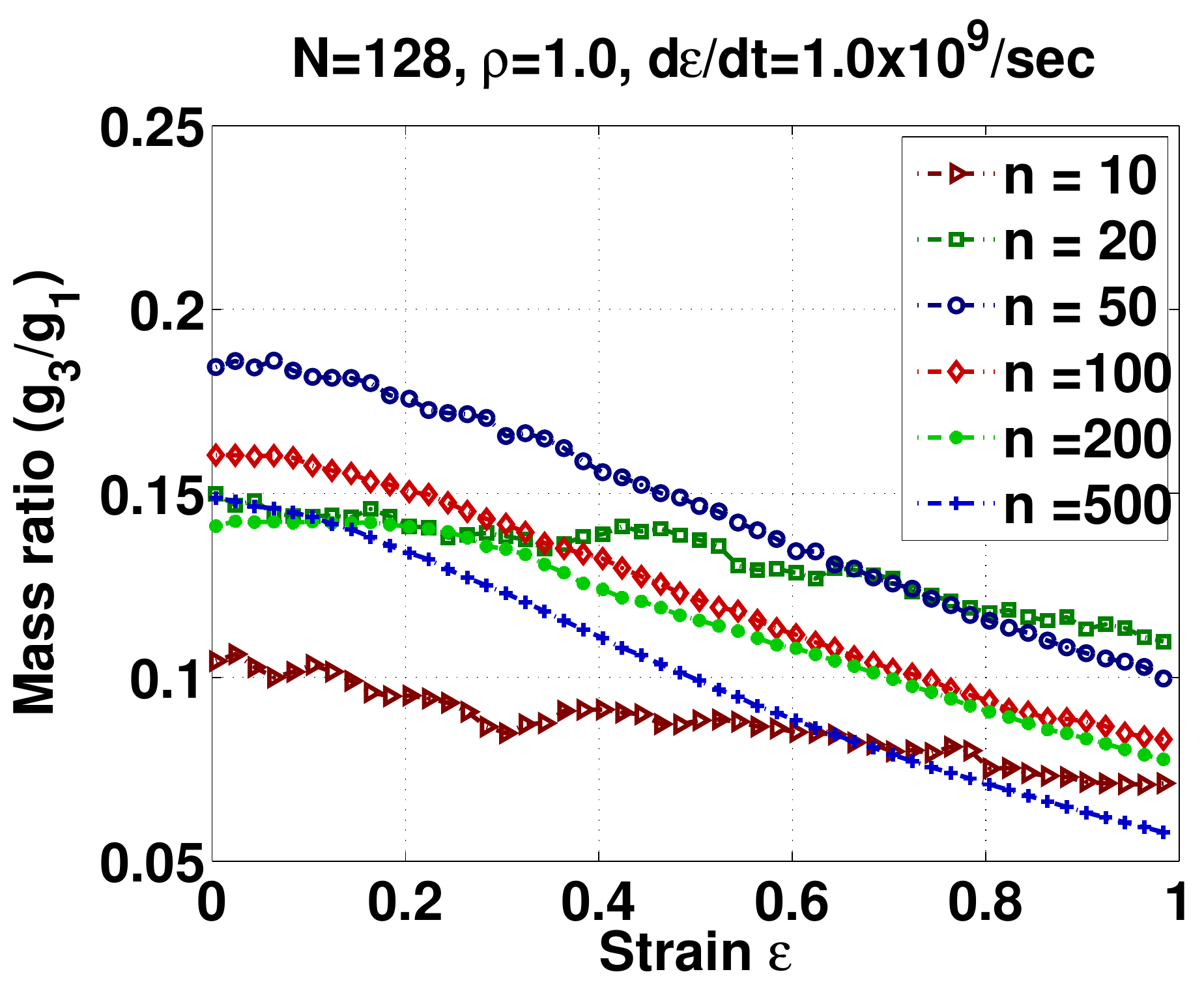}}
\caption{Effect of chain length on mass ratios}
\label{fig:EffectChainLenMassRatioLin}
\end{figure}

The effect of chain length on the mass ratios in constant strain rate loading condition is shown in \ref{fig:EffectChainLenMassRatioLin}. We observe that for very short chains, say chain lengths with $n=10$ or less number of monomers, the mass ratio is very small in the beginning of the loading. This implies that very short chains are flattened from the beginning. For larger chain polymers the mass ratios take higher value in the beginning of the loading. But, beyond $n=100$ we observe that mass ratio again take lower values as the chain length increases. This indicates that very longer chains are flat because of the constraints coming due to presence of neighboring molecules. Now, as the system is stretched, we find that in very short chain polymers, change in mass ratio is small as compared to the that in long chain polymers. As a result, in a long chain polymer, the mass is distributed more in the loading direction at high values of strains. For $n = 500$, we observe that the chains almost take a one dimensional structure.

\begin{figure}
\centering
\includegraphics[width=0.5\textwidth]{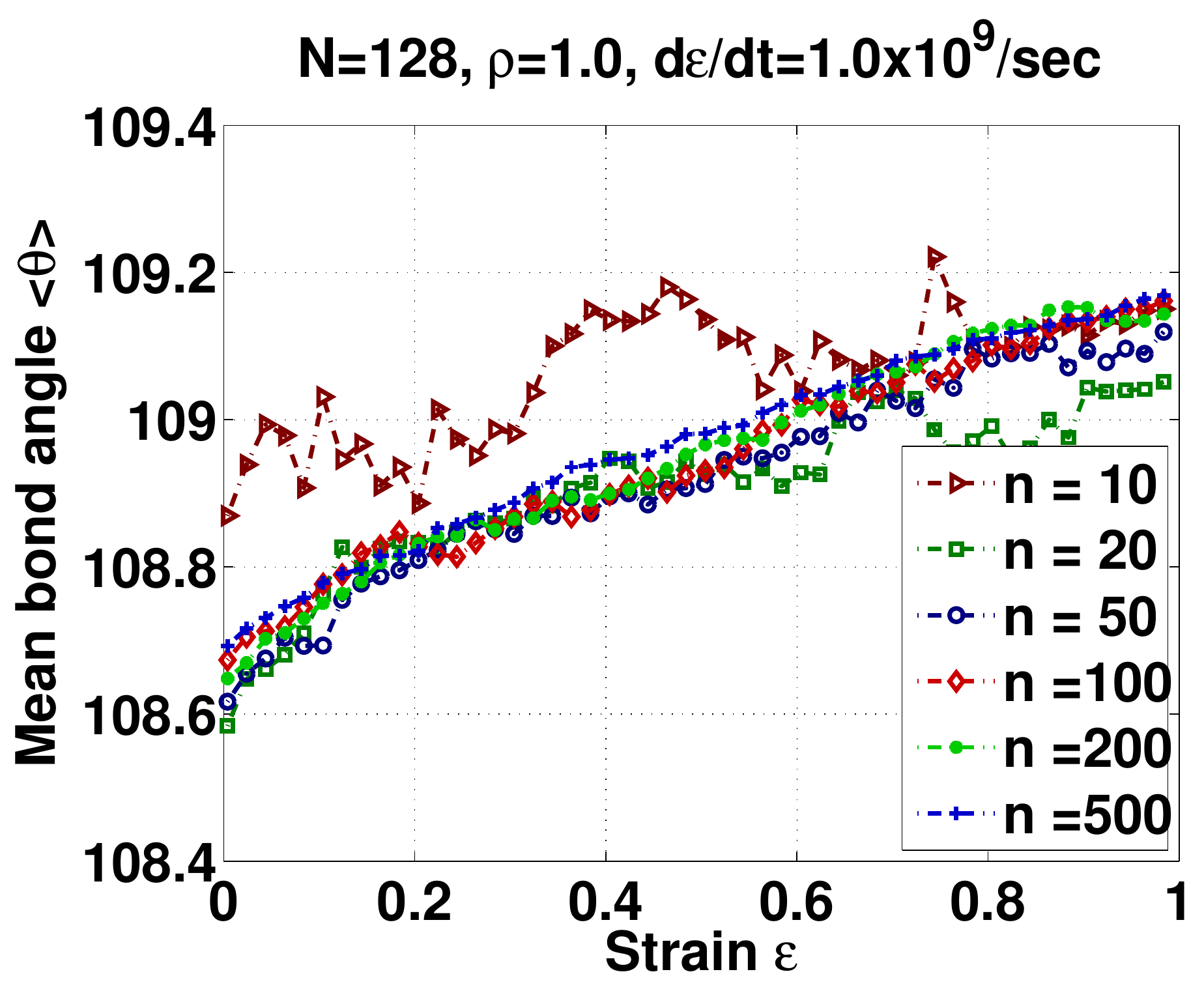}
\caption{Effect of chain length on mean bond angle}
\label{fig:EffChainLenBondAngLin}
\end{figure}

The effect of chain length on bond angle is shown in \ref{fig:EffChainLenBondAngLin} for constant strain rate loading. For very short chain polymers ($n = 10$) we find that there is a lot of fluctuations in the bond angle but the overall change in it is very small as compared to longer polymers. Even at $n = 20$ we observe fluctuations at high values of strain. The variation is similar for all chain lengths and it increases as the system is pulled. Small chain length polymers can easily slide over each other and hence the deformation is mostly due to slipping rather than actual internal deformation, a fact noticed while discussing \ref{fig:EffChainLenMeanBondLenLin}, \ref{fig:EffChainLenEndToEndLin} and \ref{f21}. Long polymers cannot slip over each other so easily and encounter entanglement constraints. Hence, internal deformation of the chains is dominant in this case and we see uniform increment in the mean bond angle with strain. 

\begin{figure}
\centering
\includegraphics[width=0.5\textwidth]{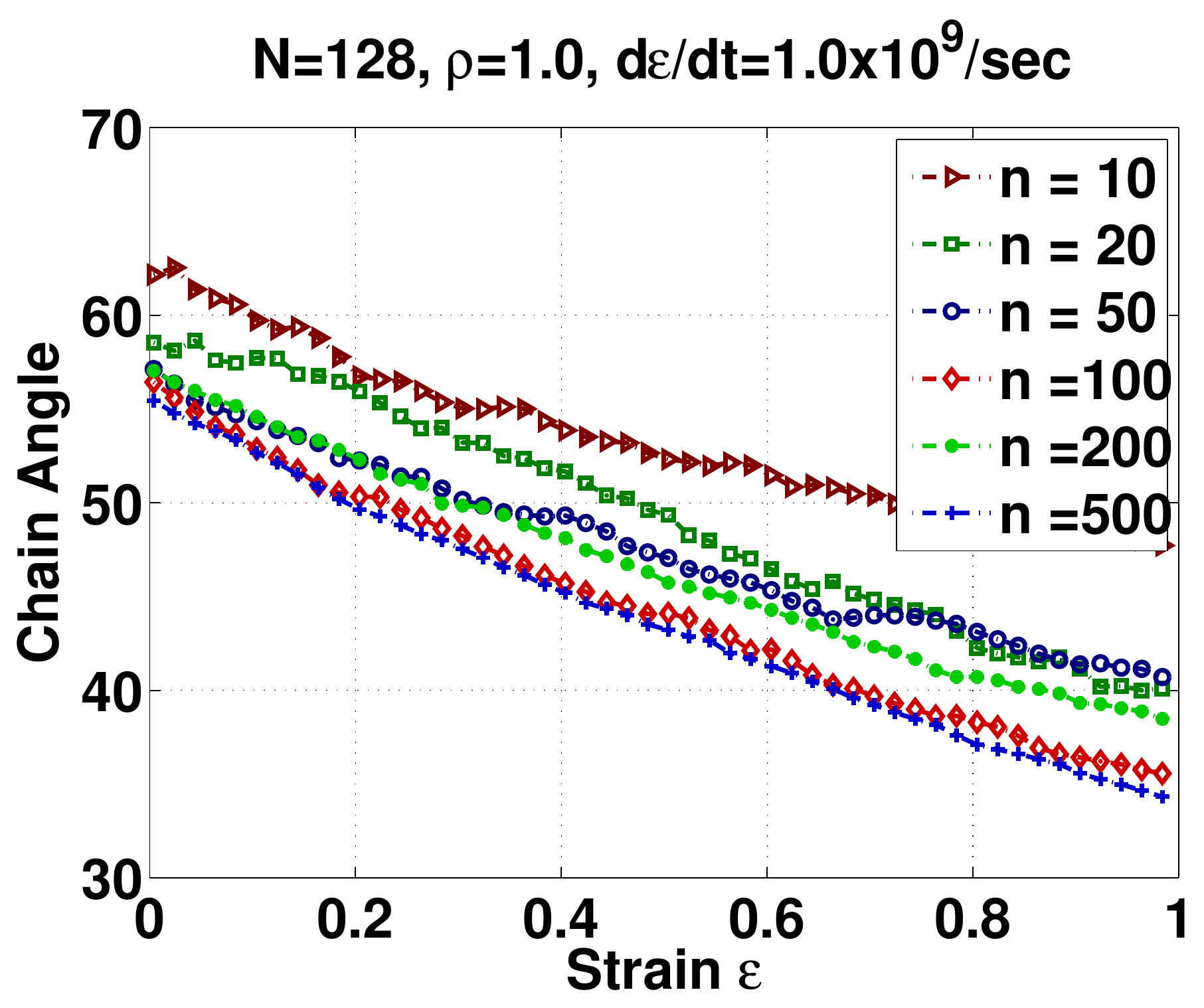}
\caption{Effect of chain length on mean chain angle}
\label{fig:EffChainLenChainAngLin}
\end{figure}

\ref{fig:EffChainLenChainAngLin} shows the effect of the polymer size on the mean chain angle. We find that at the initiation of loading itself, the mean chain angles are lower for longer polymers. As the system is pulled, the alignment of the chains in the loading direction is observed for all the polymer chain lengths considered. But note thought that the alignment is typically more for longer polymer chains. This is also supported by the initial alignment of the chains in case of long polymers. Except for $n = 200$ we find that as the system size increases, the alignment of the chains throughout the loading is more for long polymers.

\section{Conclusions}
\label{sec:conclusion}

We have presented a study of the stress response of a linear polymer subjected to uniaxial constant strain rate loading. The tool used for the study was molecular dynamics simulation. The distribution of micro-structure parameters along the chain length as well as the time-evolution of the micro-structure parameters is computed. These are then used to understand the effect of the micro-structure parameters on the stress response. We observe that the bond length has fluctuations along the chain. But these fluctuations are uniform throughout the chain as opposed to the fact that a jump occurs in the bond length at the ends of the chain as in a single short polymer chain loaded by an external force. While investigating the effect of externally defined parameters such as temperature, density, chain length, and strain rate, we observe that density has a very strong effect on the stress-strain behavior of the polymer and there is a large jump in the stress levels as the density is increased. At higher densities the deformation is mostly dominated by bond deformation, and change in overall size  and shape of the polymer chain is less significant. Characteristics of the polymer are very different for different chain lengths. Short chain polymers more or less behave like rigid molecules. There is no significant change in their internal structure when loaded. On the other hand, long chain polymers are very sensitive to the loading and the external environment. In the case of short chain polymers, those below the entanglement length, since there is no significant change the internal structure, the deformation is mostly due to slipping of the chains over one another. In long chain polymer this is not the case as they get entangled. Deformation in long chain polymers is due to deformation of the internal structure as well as their uncoiling. Temperature does not have a very significant effect on the stress strain behavior, but, it has significant effect on the mean-square bond length of the polymer and we find that relaxation in the polymer is faster at higher temperatures. Further, the end-to-end length, radius of gyration, and other parameters that characterize the micro-structure are not affected much by temperature. But, we find that chains loose their coiled structure very slowly at high temperature. Rate of the loading mostly affects the micro-structure very mildly and thereby stress-strain response. We observe that parameters associated with the covalent bond increase with the rate of loading where as parameters representing the overall size of the polymer chains decrease. This indicates that internal deformation of the chain dominates at higher rates of loading. 

The united atom model used for the molecular dynamics simulation is a powerful tool to study the statistics of polymer dynamics and correlate it to mechanical behavior. This can then be useful to design polymers and elastomers with desired mechanical characteristics.

\bibliographystyle{jfm_ase}
\bibliography{myrefjfmase}

\begin{thebibliography}{32}
\expandafter\ifx\csname natexlab\endcsname\relax\def\natexlab#1{#1}\fi

\bibitem[Bergstr{\"o}m \& Boyce(2001)]{Bergstrom}
{\sc Bergstr{\"o}m, J.~S. \& Boyce, M.C.} 2001 Deformation of elastomeric
  networks: Relation between molecular level deformation and classical
  statistical mechanics models of rubber elasticity. {\em Macromolecules\/}
  {\bf 34}, 616--626.

\bibitem[Boggs(1952)]{Boggs}
{\sc Boggs, F.~W.} 1952 Statistical mechanics of rubber. {\em Journal of
  Chemical Physics\/} {\bf 20}~(4), 632--636.

\bibitem[Bower \& Weiner(2004)]{Bower2004}
{\sc Bower, A.~F. \& Weiner, J.~H.} 2004 The role of pressure in rubber
  elasticity. {\em Journal of Chemical Physics\/} {\bf 120}~(24), 11948--11964.

\bibitem[Bower \& Weiner(2006)]{Bower2006}
{\sc Bower, A.~F. \& Weiner, J.~H.} 2006 Role of monomer packing fraction in
  rubber elasticity. {\em Journal of Chemical Physics\/} {\bf 125},
  096101--1--2.

\bibitem[Brigadnov \& Dorfmann(2003)]{Brigadnov}
{\sc Brigadnov, I.~A. \& Dorfmann, A.} 2003 Mathematical modelling of
  magneto-sensitive elastomers. {\em International Journal of Solids and
  Structures\/} {\bf 40}, 4659--4674.

\bibitem[Chui \& Boyce(1999)]{Chui}
{\sc Chui, C. \& Boyce, M.~C.} 1999 Monte carlo modeling of amorphous polymer
  deformation: Evolution of stress with strain. {\em Macromolecules\/} {\bf
  32}, 3795--3808.

\bibitem[Doi \& Adwards(1986)]{Doi}
{\sc Doi, M. \& Adwards, S.~F.} 1986 {\em The Theory of Polymer Dynamics\/}.
  Clarendon Press, Oxford.

\bibitem[Fox \& Andersen(1984)]{Fox}
{\sc Fox, J.~R. \& Andersen, H.~C.} 1984 Molecular dynamics simulations of a
  supercooled monoatomic liquid and glass. {\em Journal of Chemical Physics\/}
  {\bf 88}~(18), 4019--4027.

\bibitem[Freed(1971)]{Freed}
{\sc Freed, K.~F.} 1971 Statistical mechanics of systems with internal
  constraints: Rubber elasticity. {\em Journal of Chemical Physics\/} {\bf
  55}~(12), 5588--5599.

\bibitem[Gao \& Weiner(1984)]{Gao1984}
{\sc Gao, J. \& Weiner, J.~H.} 1984 Excluded volume effect on stress
  transmission in rubber elasticity. {\em Journal of Chemical Physics\/} {\bf
  81}~(12), 6176--6185.

\bibitem[Gao \& Weiner(1987)]{Gao1987a}
{\sc Gao, J. \& Weiner, J.~H.} 1987 Excluded volume effects on force-length
  relations of long chain molecules. {\em Macromolecules\/} {\bf 20}~(1),
  142--148.

\bibitem[Gao \& Weiner(1989{\natexlab{{\em a\/}}})]{Gao1989a}
{\sc Gao, J. \& Weiner, J.~H.} 1989{\natexlab{{\em a\/}}} Contribution of
  covalent bond force to pressure in polymer melts. {\em Journal of Chemical
  Physics\/} {\bf 91}~(5), 3168--3173.

\bibitem[Gao \& Weiner(1989{\natexlab{{\em b\/}}})]{Gao1989b}
{\sc Gao, J. \& Weiner, J.~H.} 1989{\natexlab{{\em b\/}}} Excluded volume
  effects in rubber elasticity 4.nonhydostatic contribution to stress. {\em
  Macromolecules\/} {\bf 22}~(2), 979--984.

\bibitem[Gao \& Weiner(1991{\natexlab{{\em a\/}}})]{Gao1991a}
{\sc Gao, J. \& Weiner, J.~H.} 1991{\natexlab{{\em a\/}}} Anisotropic effects
  on chain-chain interactions in stretched rubber. {\em Macromolecules\/} {\bf
  24}~(7), 1519--1525.

\bibitem[Gao \& Weiner(1991{\natexlab{{\em b\/}}})]{Gao1991b}
{\sc Gao, J. \& Weiner, J.~H.} 1991{\natexlab{{\em b\/}}} Chain force concept
  in systems of interacting chains. {\em Macromolecules\/} {\bf 24}~(18),
  5179--5191.

\bibitem[Gao \& Weiner(1992)]{Gao1992c}
{\sc Gao, J. \& Weiner, J.~H.} 1992 Range of validity of entropic spring
  concept in polymer melt relaxation. {\em Macromolecules\/} {\bf 25}~(13),
  3462--3467.

\bibitem[Gao \& Weiner(1994)]{Gao1994a}
{\sc Gao, J. \& Weiner, J.~H.} 1994 Anisotropic stress in a confined chain:
  Excluded volume effects. {\em Journal of Chemical Physics\/} {\bf 100}~(1),
  682--686.

\bibitem[Greiner {\em et~al.\/}(1997)Greiner, Neise \& St{\"o}cker]{Greiner}
{\sc Greiner, W., Neise, L. \& St{\"o}cker, H.} 1997 {\em Thermodynamics and
  Statistical Mechanics\/}. Springer-Verlag, New York.

\bibitem[James(1947)]{James1947}
{\sc James, H.~M.} 1947 Statistical properties of networks of flexible chains.
  {\em Journal of Chemical Physics\/} {\bf 15}~(9), 651--668.

\bibitem[Kankanala \& Triantafyllidis(2003)]{Kankanala}
{\sc Kankanala, S.~V. \& Triantafyllidis, N.} 2003 On finitely strained
  magnetorheological elastomers. {\em Journal of Mechanics and Physics of
  Solids\/} {\bf 52}~(3), 2869--2908.

\bibitem[Kremer \& Grest(1990)]{kremer1990}
{\sc Kremer, K. \& Grest, G.~S.} 1990 Dynamics of entangled linear polymer
  melts: A molecular dynamics simulation. {\em Journal of Chemical Physics\/}
  {\bf 92}~(6), 5057--5086.

\bibitem[Kremer {\em et~al.\/}(1988)Kremer, Grest \& Carmesin]{Kremer1988}
{\sc Kremer, K., Grest, G.~S. \& Carmesin, I.} 1988 Crossover from
  \uppercase{R}ouse to reptation dynamics: A molecular dynamics simulation.
  {\em Physical Review Letters\/} {\bf 61}~(5), 566--569.

\bibitem[Nos{\'e}(1983)]{Nose}
{\sc Nos{\'e}, S.} 1983 A molecular dynamics method for simulation in the
  canonical ensemble. {\em Molecular Physics\/} {\bf 52}, 255--268.

\bibitem[Rapaport(2004)]{Rapaport}
{\sc Rapaport, D.~C.} 2004 {\em Art of Molecular Dynamics Simulation\/}.
  Cambridge University Press, Cambridge.

\bibitem[Rickayzen \& Powles(2003)]{Rickayzen}
{\sc Rickayzen, G. \& Powles, J.~G.} 2003 Viscoelasticity of fluids with
  steeply repulsive potentials. {\em Journal of Chemical Physics\/} {\bf
  118}~(24), 11048--11056.

\bibitem[Saitta \& Klein(1999)]{Saitta}
{\sc Saitta, A.~M. \& Klein, M.~L.} 1999 Polyethylene under tensile load:
  Strain energy storage and breaking of linear and knotted alkanes probed by
  first-principles molecular dynamics calculations. {\em Journal of Chemical
  Physics\/} {\bf 111}~(20), 9434--9440.

\bibitem[Sim{\~o}es {\em et~al.\/}(2004)Sim{\~o}es, Cunha \&
  Brostow]{Simoes2004}
{\sc Sim{\~o}es, R., Cunha, A.~M. \& Brostow, W.} 2004 Molecular deformation
  mechanisms and mechanical properties of polymers simulated by molecular
  dynamics. {\em e-Polymers\/} ~(067), 1--23.

\bibitem[Thien(2002)]{Thien}
{\sc Thien, N.~P.} 2002 {\em Understanding Viscoelasticity: Basics of
  Rheology\/}. Springer, Berlin.

\bibitem[Weiner \& Gao(1989)]{Weiner1989}
{\sc Weiner, J.~H. \& Gao, J.} 1989 Concept of intrinsic chain stress in rubber
  elasticity. {\em Macromolecules\/} {\bf 22}~(12), 4544--4549.

\bibitem[Weiner \& Gao(1990)]{Weiner1990}
{\sc Weiner, J.~H. \& Gao, J.} 1990 Intrinsic chain stress model for the mooney
  effect in swollen networks. {\em Macromolecules\/} {\bf 23}~(6), 1860--1865.

\bibitem[Weiner \& Gao(1994)]{Weiner1994}
{\sc Weiner, J.~H. \& Gao, J.} 1994 Simulation of viscoelasticity in polymer
  melts: Effect of torsional potential. {\em Modelling and Simulation in
  Materials Science and Engineering\/} {\bf 2}, 755--766.

\bibitem[Wineman \& Rajagopal(2000)]{Rajagopal}
{\sc Wineman, A.~S. \& Rajagopal, K.~R.} 2000 {\em Mechanical Response of
  Polymers: An Introduction\/}. Cambridge University Press, Cambridge.

\end{thebibliography}

\end{document}